%% file: acmart-primary/main.tex
\PassOptionsToPackage{table}{xcolor}
\documentclass[acmsmall]{acmart}

\usepackage{graphicx}
\usepackage{subcaption}
\usepackage{wrapfig}
\usepackage{float}
\usepackage{enumitem}
\usepackage{pifont}
\usepackage{xcolor}
\usepackage{multirow}
\usepackage{placeins}
\usepackage{textcomp}
\usepackage{booktabs}
\usepackage{array}
\usepackage[ruled,vlined]{algorithm2e}
\usepackage{tabularx}
\usepackage{xspace}

\SetKwInput{KwInput}{Input}
\SetKwInput{KwOutput}{Output}

\newif\ifrevblue
\newif\ifrevdarkblue

\revbluefalse
\revdarkbluefalse

\ifrevblue
    \newcommand{\rev}[1]{\textcolor{blue}{#1}}
\else
    \newcommand{\rev}[1]{#1}
\fi

\ifrevdarkblue
    \newcommand{\revdb}[1]{\textcolor{blue!60!black}{#1}}
\else
    \newcommand{\revdb}[1]{#1}
\fi

\newcommand{\system}{\textit{ErgoAssist}\xspace}

\title{ErgoAssist: Cognition-Aware Posture Feedback in Wearable Ergonomic Systems}

\author{Sarmistha Sarna Gomasta}
\orcid{0000-0003-1618-4708}
\affiliation{%
  \institution{University of Massachusetts Amherst}
  \country{United States}}
\email{sgomasta@umass.edu}

\author{Bhawana Chhaglani}
\orcid{0000-0002-4060-4883}
\affiliation{%
  \institution{University of Massachusetts Amherst}
  \country{United States}}
\email{bchhaglani@umass.edu}

\author{VP Nguyen}
\orcid{0000-0001-8078-4463}
\affiliation{%
  \institution{University of Massachusetts Amherst}
  \country{United States}}
\email{phuc@umass.edu}

\author{Prashant Shenoy}
\orcid{0000-0002-5435-1901}
\affiliation{%
  \institution{University of Massachusetts Amherst}
  \country{United States}}
\email{shenoy@cs.umass.edu}

\begin{document}

\begin{abstract}
Prolonged digital device use has made poor posture 
and musculoskeletal discomfort pervasive among knowledge 
workers. Existing ergonomic wearables rely solely on 
posture thresholds, frequently interrupting users during 
high-focus moments and leading to alert fatigue and 
abandonment. Yet posture and cognitive load are closely
coupled, and most systems remain cognitively unaware. We 
present \system, a head-worn ergonomic assistant that 
detects poor posture using IMU-based head tracking and 
estimates task-induced cognitive load using a 
consumer-grade EEG headband for continuous everyday use.
In a controlled lab study, \system achieves 81\% posture 
classification and 90.2\% task induced cognitive load estimation 
accuracy under leave-one-subject-out evaluation. In a 
preliminary real-time deployment, cognition-aware alerting 
reduces alert frequency by 81\%, improves perceived 
usability by 43\%, task performance by 25\%, and improves 
posture correction rate by 38\%, delivering fewer 
but better-timed interventions rather than merely 
suppressing alerts.
\end{abstract}



\setcopyright{cc}
\setcctype{by}
\acmJournal{PACMHCI}
\acmYear{2026}
\acmVolume{10}
\acmNumber{5}
\acmArticle{MHCI8780}
\acmMonth{8}
\acmDOI{10.1145/3821691}

\maketitle

\input{acmart-primary/intro_new}
\FloatBarrier

\input{acmart-primary/background_and_motivation}
\FloatBarrier

\input{acmart-primary/related-work-majorRevisions}
\FloatBarrier

\input{acmart-primary/ErgoAwareMHC}
\input{acmart-primary/UserStudyMHCI}
\FloatBarrier

\input{acmart-primary/Evaluation}
\input{acmart-primary/UserFeasibilityMHCI_REvision}

\input{acmart-primary/Discussions_new}

\FloatBarrier

\input{acmart-primary/Conclusion}
\FloatBarrier

\begin{acks}
We thank the anonymous reviewers for their suggestions for improving the paper. This research was supported in part by NSF grants 2211302, 2211888, 2213636, 2325956, and 2105494, and by U.S. Army contract W911NF-17-2-0196. Any opinions, findings, conclusions, or recommendations expressed in this paper are those of the authors and do not necessarily reflect the views of the funding agencies.
\end{acks}

\bibliographystyle{ACM-Reference-Format}
\bibliography{acmart-primary/ref,acmart-primary_ISWC/acmart}

\input{acmart-primary/Appendix}

\end{document}

%% file: acmart-primary/intro_new.tex
\section{Introduction}
Prolonged use of digital devices has made discomfort, musculoskeletal strain, eye strain, and fatigue pervasive among office workers and students \cite{borhany2018musculoskeletal,devi2023hazards}. In particular, \rev{forward head posture (FHP)} substantially increases load on the cervical spine, with even modest neck flexion leading to fatigue and long-term strain \cite{arooj2022forward,neupane2017text}.
In response, there has been a rich literature on sensing user posture using different modalities like wireless \cite{li2021sitsen}, acoustics \cite{qu2023sitpaa,bi2024smartsit}, and vision \cite{chen2019sitting}. A growing body of commercial and research systems leverage inertial measurement units (IMUs) to monitor posture and provide corrective feedback through wearables and workstation aids \cite{elliott2019upright,luo2023skin}.
Despite their promise, posture-aware ergonomic systems often fail to achieve sustained adoption, deployment, and social acceptance. 
The major challenges faced by these posture assistants are \textit{alert fatigue, increased anxiety or over-correction, and inaccurate feedback }\cite{smartwatch_posture_alerts_harmful,caixeiro2025effectiveness}. When interventions occur too frequently, users become desensitised, ignore alerts, or actively disable them, reducing overall effectiveness and increasing annoyance. Studies of alerting systems in healthcare and human-computer interaction have documented how excessive or poorly timed alerts significantly degrade user experience and compliance \cite{hellier2010review,adamczyk2004if}. 
Existing systems typically issue alerts whenever a posture threshold is violated, regardless of the user’s mental state or task context. As a result, users are frequently interrupted during moments of high concentration, leading to annoyance and eventual abandonment of the system \cite{baer2022posture}. These observations highlight a fundamental limitation of posture-only ergonomic assistants: they treat feedback as a purely physical correction problem, rather than an interruptibility-sensitive interaction \cite{adamczyk2004if,iqbal2007disruption}.
Prior work suggests that posture and cognition are tightly coupled. Poor posture can negatively impact cognitive performance \cite{jung2024effect}, while high cognitive load often exacerbates postural degradation and muscular strain \cite{le2024neck}. Importantly, moments of high cognitive demand are also when users are least receptive to interruptions \cite{obuchi2016investigating}. Yet, \textit{most existing ergonomic systems remain cognitively unaware}, issuing alerts when users are deeply focused, leading to annoyance.
This motivates a key question: \emph{How can ergonomic feedback be made more effective by being cognition-aware?} Rather than alerting users whenever posture deteriorates, an intelligent ergonomic assistant should reason about \emph{when} to intervene delivering fewer, better-timed corrections during moments of low cognitive focus or natural breaks in attention. \textit{We explore how we can sense holistic state of user (including both their posture and cognitive context) to generate human-centered adaptive interventions. }

Recent advances in wearable sensing make this possible. Consumer-grade head-worn EEG devices and ear-integrated sensors enable continuous, non-invasive estimation of \rev{task-induced cognitive load} in everyday settings \cite{FrenzBrainband,demirel2025beyond,kodikara2025fatiguesense,kosmyna2019attentivu}. When combined with posture sensing, these modalities offer complementary signals: IMUs capture physical strain, while EEG provides insight into \rev{task-induced cognitive load} and receptivity to feedback. \rev{We study how these physical and physiological signals vary over time and correlate with each other when a user is performing different tasks, and leverage these insights to design context-sensitive interventions. In this work, we present \system, a wearable-based cognition-aware ergonomic assistant that integrates posture sensing with cognitive-state-aware feedback scheduling.} Our system continuously monitors user's neck posture using an IMUs and estimates \rev{task-induced cognitive load} using wearable EEG. Instead of issuing posture alerts immediately, the system adapts feedback timing based on the user’s cognitive context, delaying or suppressing alerts during periods of high focus and prioritizing interventions during low-load moments or natural breakpoints \cite{obuchi2016investigating}. 

We evaluate our approach through a controlled user study (N=24) in which participants perform three tasks (Rest, Stroop, Numerical Calculation) under multiple postural conditions while wearing our head-mounted IMU and EEG band. We compare two feedback strategies: (1) posture-only alerts triggered by neck angle thresholds, and (2) cognition-aware alerts that fuse posture and \rev{task-induced cognitive load} to schedule feedback. Our results provide empirical evidence that posture-only alerts are often poorly timed, frequently interrupting users during periods of high cognitive focus. We show that cognition-aware alerts are significantly less frequent (81\% less), less disruptive, and more effective, leading to improved perceived usability (43\% more), higher posture correction rate (61.9\% to 100\%), and improved task performance (25\% more) compared to posture-only feedback.
Overall, this paper makes the following contributions:

\begin{enumerate}[leftmargin=15pt,itemsep=2pt,topsep=2pt]
    \item We design and implement a wearable ergonomic assistant that integrates IMU-based posture sensing with EEG-based \rev{task-induced cognitive load} estimation to continuously track the holistic state of the user. The system achieves \textbf{90.2\%} accuracy for task-induced cognitive load classification and \textbf{81.4\%} posture classification under LOSO evaluation.
    \item We analyze the interaction of physical (postural) and physiological (cognitive) signals during cognitively demanding tasks, and examine how cognitive state influences posture behavior.
    \item We predict perceived discomfort levels reported by users using multimodal sensor fusion and lightweight machine learning models, achieving \textbf{75\%} accuracy with a latency of \textbf{12.98 ms}.
    \item Through a \rev{preliminary} real-time within-subject study (N=13), we show that cognition-aware interventions reduce alert frequency by \textbf{81\%}, higher posture correction rates by \textbf{~38\%,} and improve perceived usability by\textbf{ 43\%} compared to posture-only feedback.

\end{enumerate}

%% file: acmart-primary/background_and_motivation.tex
\section{Background and Motivation}
In this section, we motivate the need for a cognition-aware ergonomic assistants in everyday work environments to promote good postural habits without affecting performance.
\subsection{Need for ergonomic assistant: A User Survey}
\begin{figure}[t]
  \centering
  \begin{subfigure}[t]{0.55\linewidth}
    \centering
    \includegraphics[width=\linewidth,height=4cm,keepaspectratio]{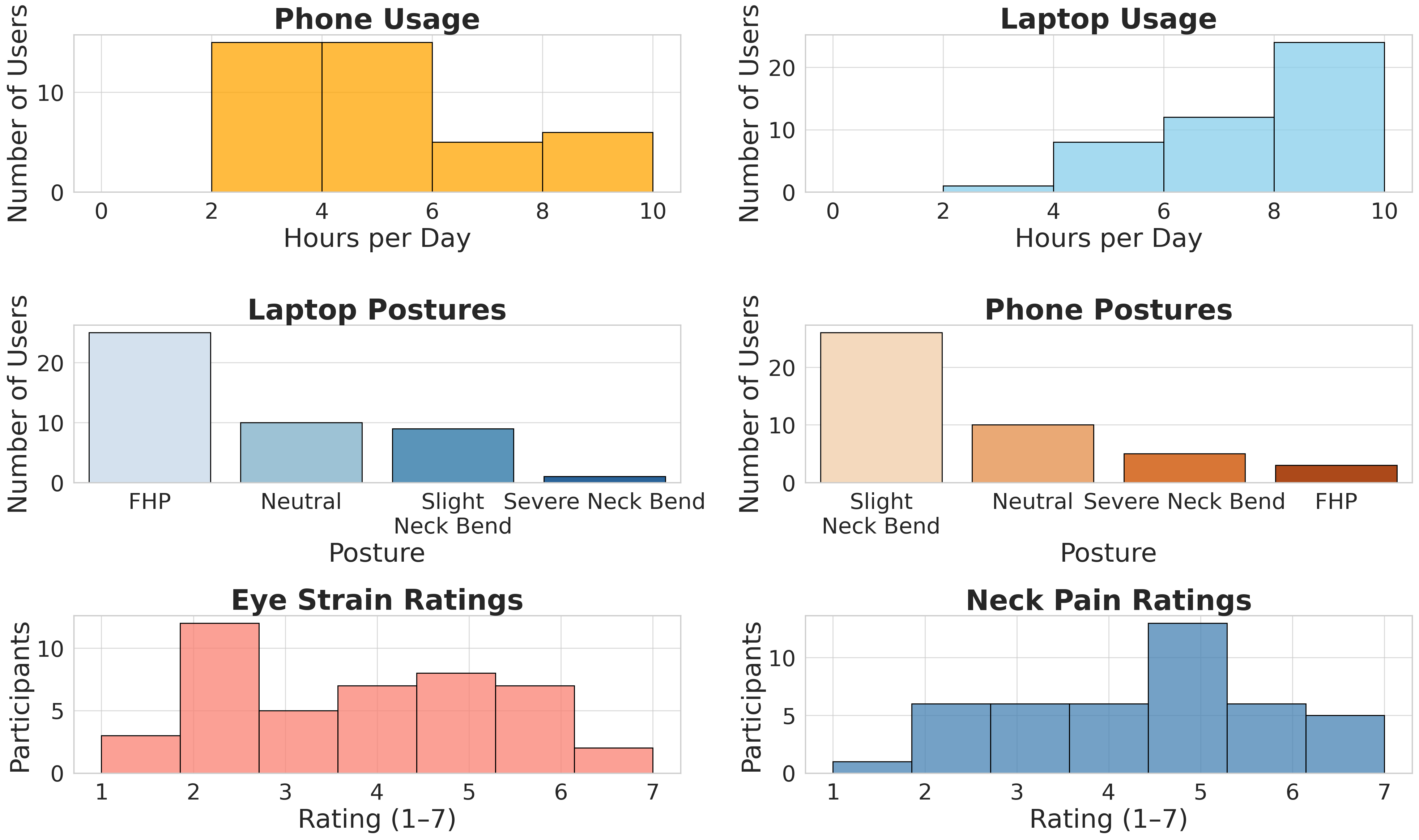}
    \caption{\textbf{User Pre-survey Results.}}
    \label{fig:survey_results}
  \end{subfigure}
  \hfill
  \begin{subfigure}[t]{0.40\linewidth}
    \centering
    \includegraphics[width=\linewidth,height=4cm,keepaspectratio]{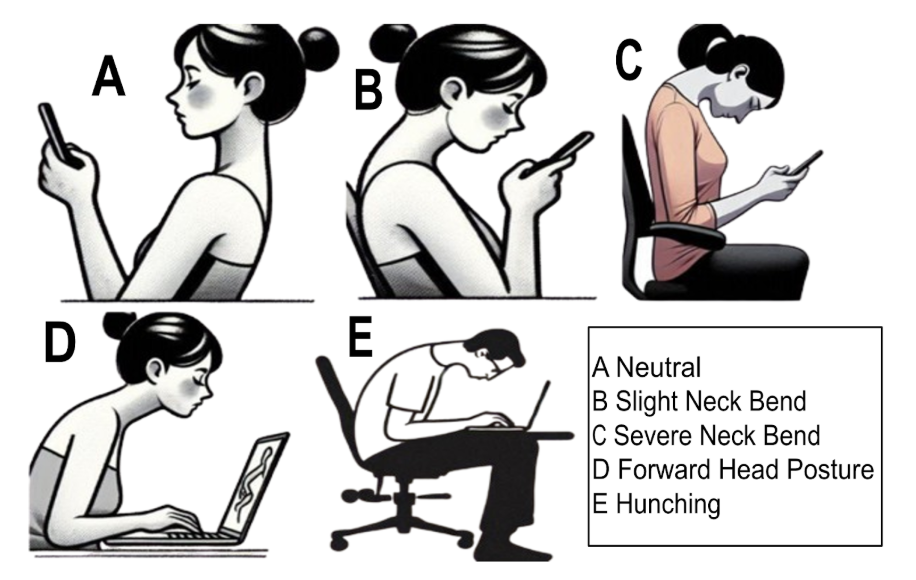}
    \caption{\textbf{Posture Categories ~\cite{chhaglani2024neckcare}.}}
    \label{fig:posture_types}
  \end{subfigure}
  \caption{User Pre- Survey context. (a) Results from the motivation survey and (b) Demo posture categories included in survey.}
  \label{fig:user_survey}
  \vspace{-0.3cm}
\end{figure}

Prolonged screen-based work is increasingly associated 
with musculoskeletal discomfort, eye strain, and 
fatigue~\cite{borhany2018musculoskeletal,devi2023hazards}. 
\rev{FHP} is particularly concerning as even modest neck 
flexion increases mechanical load on the cervical spine 
and accumulates into pain over 
time~\cite{arooj2022forward}. To understand real-world 
challenges, we conduct an IRB-approved pre-survey with 50 
participants aged 18--45. Participants report an average 
of 7.36 hours/day on laptops and 4.24 hours/day on phones, 
with 77.3\% and 75.0\% respectively using ergonomically 
poor postures. \rev{FHP} is the most common laptop posture, 
contributing to neck pain and cervical spondylosis 
\rev{(tech neck~\cite{atakla2023tech,chhaglani2024neckcare})}. 
Overall, 93.2\% of users report eye strain and 95.5\% 
report neck or upper back pain, yet none use any posture 
assistant (Figure~\ref{fig:survey_results}). Thus, there is a need for an ergonomic assistant that continuously monitors our state and provide timely alerts.
\subsection{\rev{Towards Cognition-Aware Ergonomic Assistance}}
\rev{A growing body of commercial and research systems use IMUs 
to detect slouching or excessive flexion and trigger 
corrective feedback~\cite{elliott2019upright,luo2023skin}. 
However, these systems often struggle with sustained 
adoption in real-world settings. A key reason is that 
alerts are frequently delivered at the wrong time. Posture 
thresholds are evaluated independent of the user's 
\rev{task-induced cognitive load}, task demands, or 
receptivity to interruption, leading to repeated 
notifications during high-focus moments that feel 
disruptive and contribute to alert 
fatigue~\cite{baer2022posture}. Interruption science 
further shows that timing matters, and alerts delivered 
during high-focus moments are more disruptive and less 
likely to be heeded than those aligned with natural 
breakpoints~\cite{adamczyk2004if,iqbal2007disruption,
dang2025co,obuchi2016investigating}.}
\rev{Recent advances in consumer-grade wearable EEG devices now 
enable practical non-invasive monitoring of 
\rev{task-induced cognitive load} in everyday 
contexts~\cite{FrenzBrainband,demirel2025beyond}, 
enabling real-time integration of cognitive state into 
systems that decide \emph{when} to deliver feedback. 
This is further supported by the strong bidirectional 
relationship between posture and cognition. Cognitive load 
can degrade posture and increase muscle 
strain~\cite{igarashi2016effect}, and poor posture can 
impair cognitive performance and alter neural 
patterns~\cite{le2024neck,jung2024effect}. Yet most 
existing posture feedback tools such as 
Upright~\cite{elliott2019upright} and 
SuperBreak~\cite{singh2019shape} rely solely on posture 
thresholds independent of cognitive context, risking 
over-intervention during precisely those moments when 
users are least receptive~\cite{baer2022posture,luo2018time}.}

\rev{Taken together, these observations motivate the need for 
cognition-aware ergonomic assistants that reason about a 
holistic user state encompassing both physical strain 
(posture) and psychological state (\rev{task-induced 
cognitive load} and receptivity). By fusing IMU-based 
posture sensing with EEG-based \rev{task-induced cognitive 
load} estimation, a wearable assistant can deliver 
better-timed interventions, suppressing or deferring 
alerts during high-focus periods and prioritizing feedback 
during natural breakpoints. Our proposed system aims to 
sense holistic user state and focuses on delivering 
\textit{human-centered adaptive interventions}.}

%% file: acmart-primary/related-work-majorRevisions.tex
\section{Related Work}
Wearable sensing has been widely adopted for monitoring either
physical ergonomics or cognitive states. However, few systems
have explored their integration to understand the interplay between
mental workload and physical posture.
\subsection{Posture Sensing for Ergonomic Risk}
\rev{Accurate sensing of head and neck posture has been widely explored to detect ergonomically harmful configurations such as forward head flexion and sustained non-neutral alignment. Vision-based systems~\cite{chen2019sitting,lee2024mocap,zago20203d,edriss2025commercial} leverage pose estimation and skeletal tracking to classify a wide range of postures with high spatial fidelity. However, they are computationally intensive, raise privacy concerns, and degrade under occlusion or low-light conditions. Wireless sensing approaches~\cite{li2021sitsen,qu2023sitpaa,feng2019you,su2025dynamic} infer posture through RF signal perturbations without requiring body-worn devices, but are highly sensitive to environmental dynamics such as furniture layout and user position, reducing robustness across settings. Pressure-based systems~\cite{tsai2023automated,mutlu2007robust} instrument seating interfaces to estimate body posture via contact distribution, yet remain constrained to fixed locations and cannot directly capture head or neck kinematics. Similarly, motion capture systems provide high-precision joint tracking and serve as a gold standard, but require specialized infrastructure and are impractical outside controlled laboratory environments~\cite{luo2023skin,wanying2024robot,wang2022comparison}. Wearable sensing provides a more practical alternative for continuous monitoring. IMU-based systems estimate orientation, angular velocity, and derived neck angles to classify posture in real time~\cite{elliott2019upright,petropoulos2017spomo,khurana2014neckgraffe,lee2014d06}. While accurate, these approaches often require rigid placement on the cervical spine or upper back, impacting comfort and long-term adherence. An increasingly practical direction is to leverage opportunistic sensing from existing head-worn devices, such as headphones or smart wearables, to estimate head kinematics \cite{chhaglani2024neckcare,kim2022validity,ferlini2019head,han2023test}. These systems utilize embedded inertial sensors to continuously capture orientation and motion dynamics with minimal user burden. Such approaches enable real-time, fine-grained tracking of head posture in everyday settings without requiring dedicated instrumentation, making them well-suited for scalable ergonomic monitoring.}
\subsection{Wearable EEG for Cognitive Load Monitoring}
\rev{Wearable brain sensing now supports inference of cognitive states such as mental workload, fatigue, and attentional engagement, with research moving from bulky EEG caps to lightweight headbands, earbuds, and EEG-equipped headphones~\cite{FrenzBrainband,knierim2025advancing,kwon2021realtimeeeg,emish2024remote,he2026wearable}. Knierim et al. showed that headphone-based EEG can classify cognitive load in lab and field settings~\cite{knierim2025advancing}, and Kwon et al. demonstrated continuous workload monitoring using wearable EEG bands~\cite{kwon2021realtimeeeg}. Most approaches represent EEG activity using band-limited spectral features, with frontal theta, alpha, and beta power systematically modulated by cognitive workload~\cite{chikhi2022eeg,puma2018using,raufi2022evaluation,brouwer2012estimating}. Frontal theta typically increases with task difficulty and resource demands, whereas alpha decreases under higher workload and is sensitive to fatigue and attentional load~\cite{puma2018using,stipacek2003sensitivity,hsu2022load}. Beta activity has been associated with visual attention and short-term memory 
processes in working-memory and vigilance tasks~\cite{lundqvist2024beta,wen2024beta,pershin2023vigilance}. Ratios such as theta--beta and alpha--beta, alongside spectral and sample entropy features, are widely used to discriminate cognitive load 
levels~\cite{raufi2022evaluation,chikhi2022eeg,liu2023fusion}. Recent work has also explored machine learning approaches on wearable EEG, ranging from classical models to deep learning architectures, enabling robust classification under real-world noise and motion artifacts \cite{yedukondalu2025cognitive,8972542,amin2015feature}. Despite this, real-world robustness remains limited. Dry electrodes, in-motion conditions, and uncontrolled environments are the three primary sources of signal degradation in wearable EEG \cite{mihajlovic2014wearable,casson2019wearable}, yet few studies address their combined effect in mobile settings. Real-world benchmarking shows that consumer-grade devices exhibit larger 
drops in spectral stability and weaker neurometric discrimination during 
multitasking and driving, with motion sensitivity and sparse electrode 
coverage as the main factors~\cite{vincenzo2026beyond}. Some existing systems also 
estimate cognitive state based on user behavior or physical 
context~\cite{gherman2025investigation,ding2020exploratory}, limiting their applicability 
for adaptive interactive systems. This creates opportunities to integrate cognitive state awareness into interactive systems that adapt in real time to both user state and context.}
\subsection{Multimodal Sensor Fusion for User State Modeling}
\rev{Prior studies have explored physical and physiological data streams and their combination for fatigue and ergonomics \cite{shao2024weaving,kalanadhabhatta2021fatigueset, wang2024ubiphysio,cen2022exploring}}. For example, Graña et al. demonstrated that combining an IMU’s motion features with EEG features yielded more accurate recognition of natural daily activities than IMU alone \cite{patel2025developing}. 
\rev{Nica et al.~\cite{nica2022fatigue} demonstrated that fusing EEG, ECG, and EMG data enables improved prediction of mental and physical fatigue, especially under multitasking conditions. Their work underscores the importance of multimodal sensing in understanding how cognitive and physical demands co-occur.} 
Demirel et al.~\cite{demirel2025beyond} address cybersickness in VR environments using head and physiological sensors.
\rev{The field of neuroergonomics also promotes this integration, emphasizing how mental workload and ergonomic performance are tightly coupled, particularly in safety-critical and prolonged-use scenarios~\cite{dehais2020neuroergonomics,mehta2013neuroergonomics}.}
\rev{This motivates the need for an assistant that continuously monitors both how you sit/stand and how you think/feel, enabling context-sensitive support.}
\rev{While prior work has explored combining physiological and physical signals for user state recognition, these approaches primarily focus on improving classification accuracy for tasks such as activity recognition or fatigue detection. They do not address how multimodal sensing can inform decision-making for real-time interventions. In particular, existing systems do not leverage multimodal signals to determine when feedback should be delivered in interactive settings, which is critical for minimizing disruption and improving user compliance.}

\rev{Overall, prior work has independently explored posture sensing, cognitive load estimation, multi-modal activity sensing and interruptibility-aware feedback. However, these domains remain largely disconnected. Posture-aware systems lack cognitive awareness and cognitive sensing systems do not inform intervention timing.
To this end, \textit{ErgoAssist} integrates IMU-based posture sensing and EEG-based cognitive load estimation to drive adaptive feedback scheduling. Rather than focusing solely on state recognition, our system addresses the key question of when to intervene, enabling fewer but better-timed ergonomic corrections.}

%% file: acmart-primary/ErgoAwareMHC.tex
\section{\system Architecture}
\label{sec:ergosystem}
\begin{figure}[t]
    \centering
\includegraphics[width=\linewidth,height=0.9\textheight,keepaspectratio]{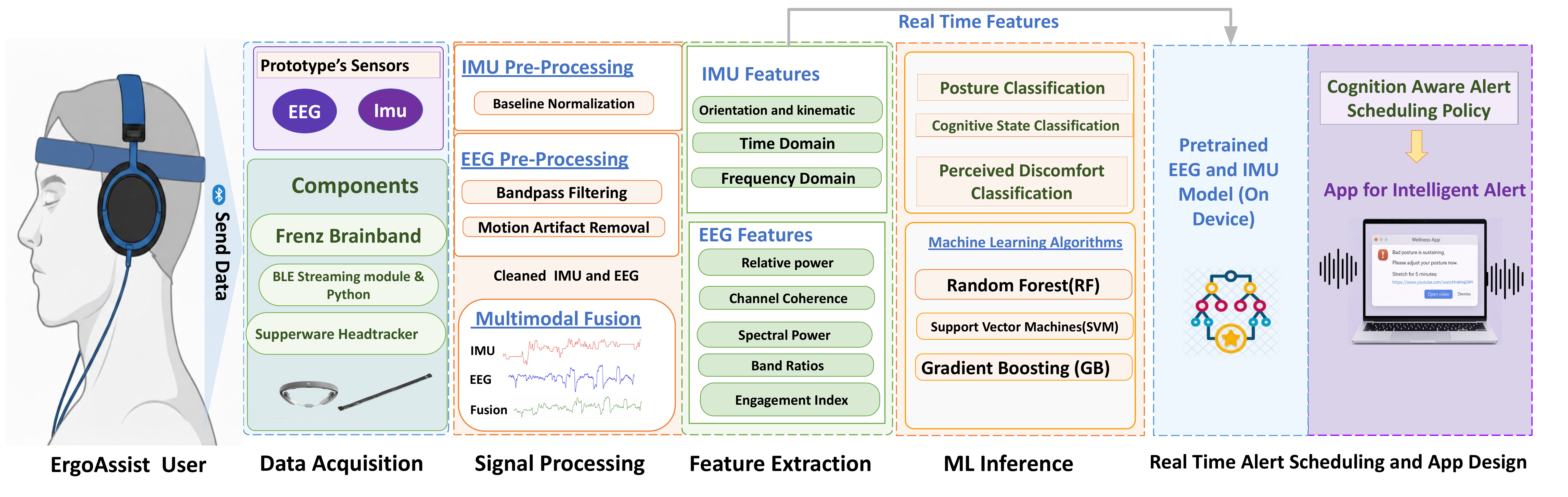}
    \caption{\system system pipeline. The prototype captures EEG and IMU signals and runs pretrained sensor-based machine learning models in real time to infer cognition-aware alerts.}
    \label{fig:pipeline}
\end{figure}

We design \textit{ErgoAssist}, a head-worn wearable system 
that provides cognition-aware posture assistance by sensing 
posture via a head tracker and cognitive state via an EEG 
headband. The system continuously tracks the holistic user 
state during everyday computer use and explores the 
relationship between physical and physiological signals 
during cognitively demanding tasks. The system pipeline is 
shown in Figure~\ref{fig:pipeline}. The core modules of 
\system are: (1) a head-worn multimodal sensing device, 
(2) a signal processing and machine learning pipeline for 
posture, discomfort, and \rev{task-induced cognitive load} 
inference, and (3) an interruptibility-aware feedback 
module to balance ergonomic benefit and productivity.
\subsection{Data Acquisition}
The headworn prototype consists of two components: a 
Supperware Head Tracker~\cite{SupperwareHeadTracker} and 
a Frenz Brainband~\cite{FrenzBrainband}. \rev{The head tracker includes a 6 axis IMU (3-axis 
accelerometer, gyroscope, and compass) providing pitch, 
roll, and yaw at 50 Hz with angular accuracy $<1^\circ$. We collect EEG using the Frenz Brainband, a 4-channel wearable device (LF-FpZ, OTE\_L-FpZ, RF-FpZ, OTE\_R-FpZ) 
sampled at 125 Hz. We use the \texttt{frenztoolkit} (v0.3.2) to stream 
timestamped EEG signals and spectral band powers over 
Bluetooth. From these, \rev{we extract \(\delta\) (1--4 Hz), 
\(\theta\) (4--7 Hz), \(\alpha\) (8--13 Hz), \(\beta\) 
(13--30 Hz), and \(\gamma\) (30--45 Hz) bands}. IMU data 
are collected asynchronously via a USB serial interface. EEG and IMU streams are aligned using system timestamps while retaining their native sampling rates (EEG: 125 Hz, 
IMU: 50 Hz).}
\subsection{Signal Processing }
ErgoAssist performs continuous, real-time signal processing on multimodal sensor streams to extract posture and cognition-related features. We perform calibration and noise removal to 
ensure robustness to inter-participant variability and everyday motion artifacts. IMU and EEG signals are processed independently and later fused at the model level.
\vspace{-8pt}
\subsubsection{\textbf{IMU Data Pre-processing}}
\label{sec:imudata}
\rev{We record IMU data at 50\,Hz while participants hold three head postures: D1 (neutral), D2 (mild forward flexion), and D3 (deep forward flexion). Among yaw, roll, and pitch, we use \textit{pitch} as our primary signal because it best captures sagittal head--neck flexion, consistent with biomechanical literature on \rev{FHP}~\cite{fercho2023kinematic,chhaglani2024neckcare}}. A key challenge in wearable-based posture monitoring is inter-participant variability in neutral head orientation. Raw pitch values during the same labeled posture can differ significantly across users due to anatomical differences and sensor placement variations~\cite{koca2025anatomy}. We mitigate this by normalizing pitch relative to each user's own neutral posture rather than absolute sensor coordinates.
\rev{To remove startup and ending artifacts, we discard the first and last 10 seconds of each IMU recording. This alignment matches the boundary trimming applied to EEG, ensuring that subsequent multimodal analysis is performed on temporally corresponding segments.} Then, for each user, we compute a neutral baseline \(P_{\text{base}}\) as the median pitch across all their D1 segments:
\[
P_{\text{base}} = \operatorname{median}\big(\text{pitch}_{\text{raw}}^{\text{D1}}\big).
\]
We define the baseline-centered flexion signal as
\[
\text{flex}(t) = \text{pitch}_{\text{raw}}(t) - P_{\text{base}}.
\]
This shifts each user’s neutral posture to approximately \(0^\circ\), so D2 and D3 are expressed as deviations from their personal neutral position.
\rev{As shown in Figure~\ref{fig:baseline_correction}, baseline normalization improves the interpretability of posture-dependent pitch patterns by reducing participant-specific offset in the raw signal. In the unnormalized representation, the pitch distributions exhibit substantial inter-subject variability. After personalized baseline calibration, the median relative pitch decreases monotonically from D1 to D3, showing clearer flexion ordering across posture conditions. 
 \begin{figure}[t]
    \centering
    \includegraphics[width=0.7\columnwidth]{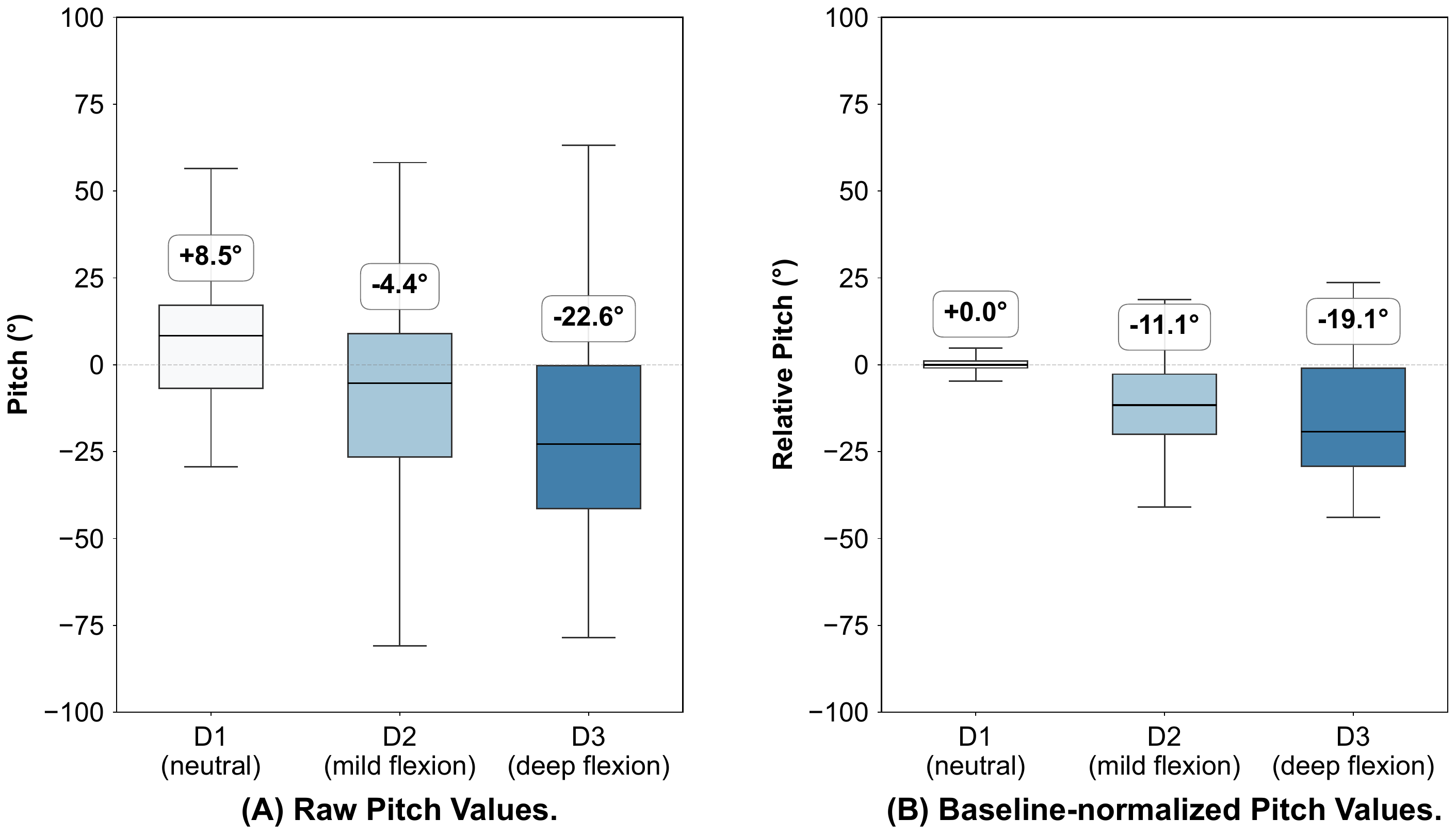}
    \caption{\rev{Distribution of raw and baseline-normalized pitch across posture conditions. Personalized baseline normalization reduces participant-specific offset and makes the progression from D1 to D3 more interpretable.}}
    \label{fig:baseline_correction}
\end{figure}
Although the normalized distributions still partially overlap, they exhibit clearer ordering across D1, D2, and D3, with separation that is sufficient for binary and three-class posture classification.}
\subsubsection{\textbf{EEG Data Pre processing}}
\begin{figure*}[t]
    \centering
    \begin{subfigure}[b]{0.235\textwidth}
        \centering
        \includegraphics[width=\linewidth,height=3.8cm,keepaspectratio]{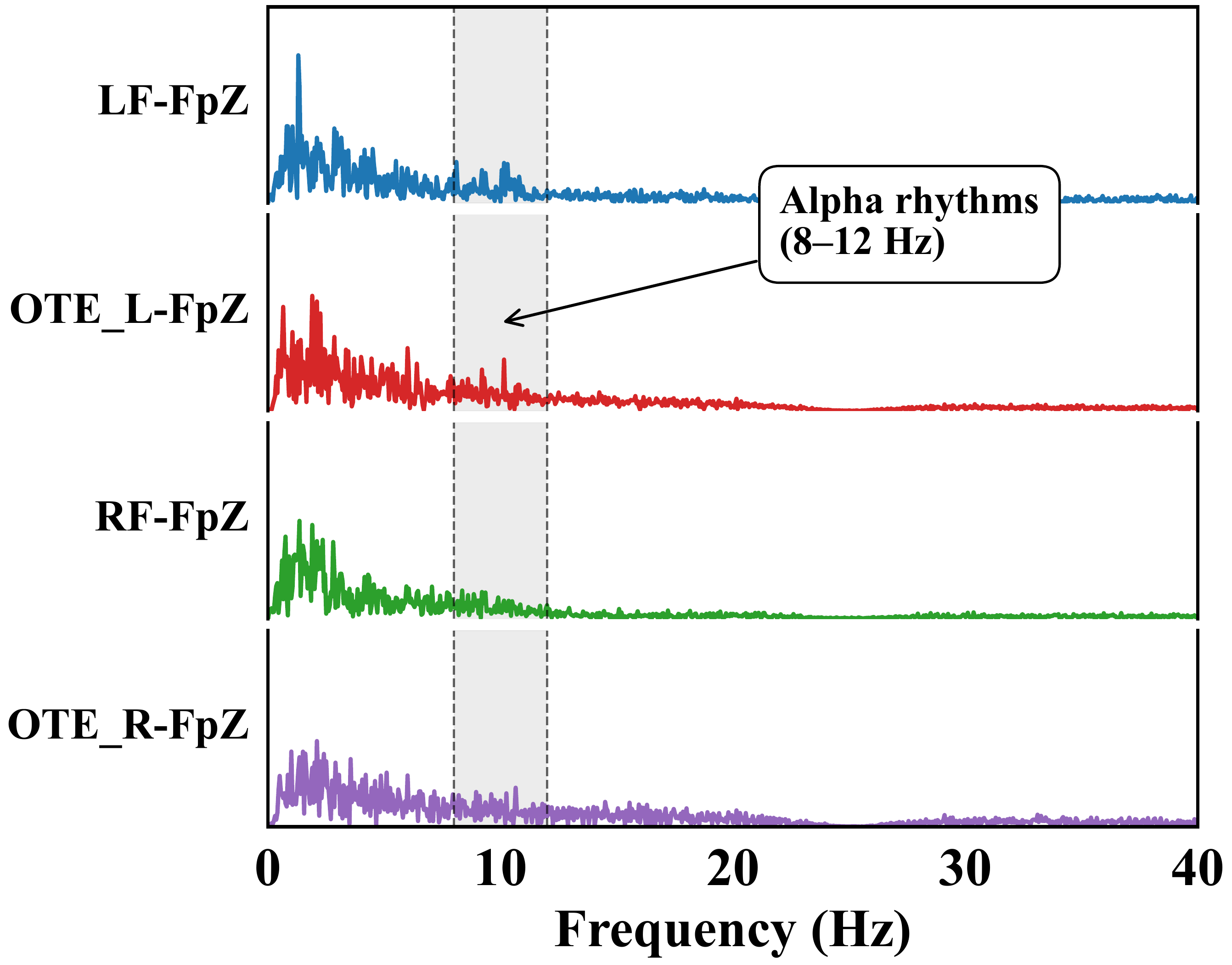}
        \caption{\textbf{FFT of EEG signals}}
        \label{fig:eeg_fft}
    \end{subfigure}\hspace{0.004\textwidth}%
    %
    \begin{subfigure}[b]{0.23\textwidth}
        \centering
        \includegraphics[width=\linewidth,height=2.9cm,keepaspectratio]{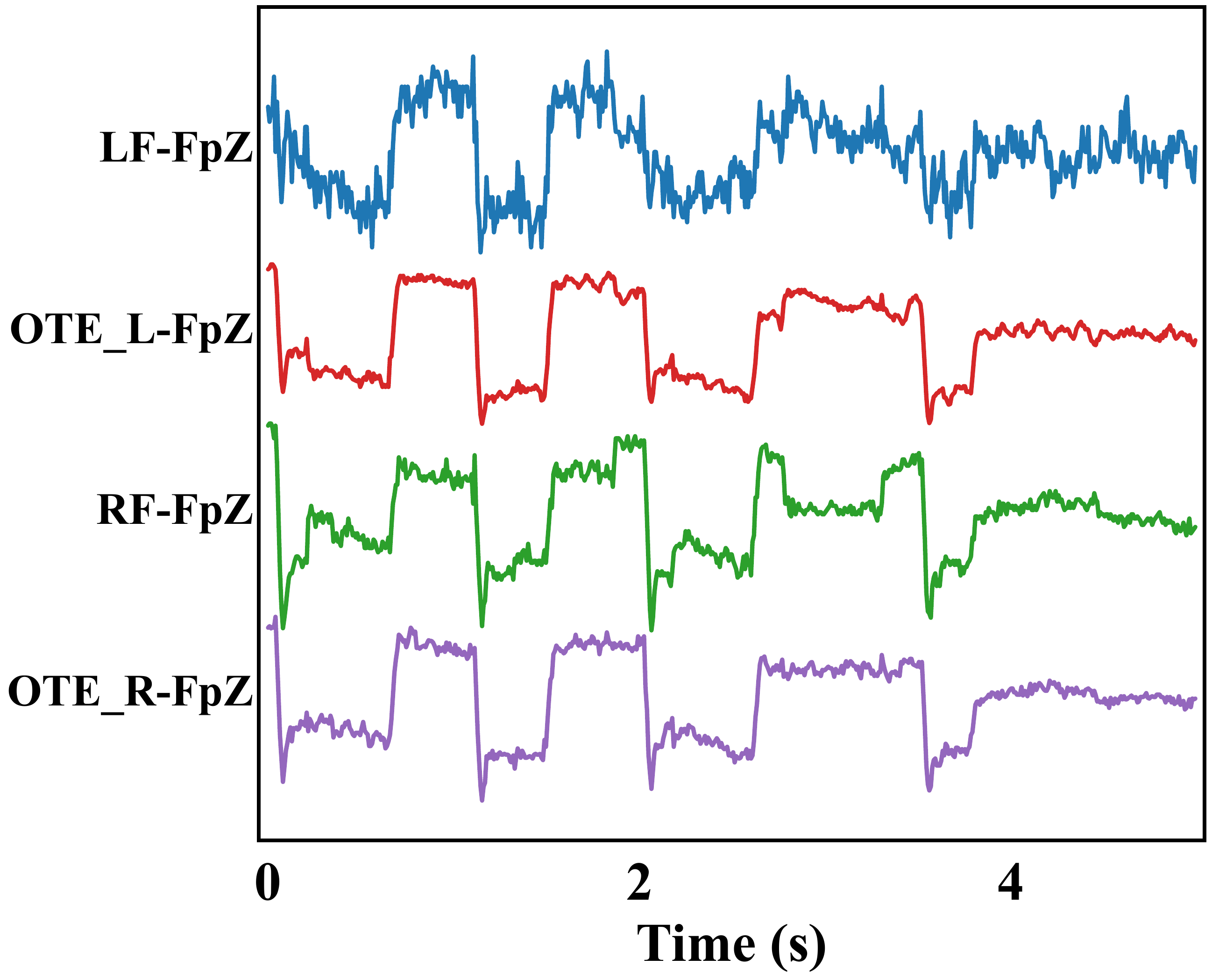}
        \caption{\textbf{EEG comparison}}
        \label{fig:eeg_time}
    \end{subfigure}\hspace{0.004\textwidth}%
    %
    \begin{subfigure}[b]{0.235\textwidth}
        \centering
        \includegraphics[width=\linewidth,height=2.6cm,keepaspectratio]{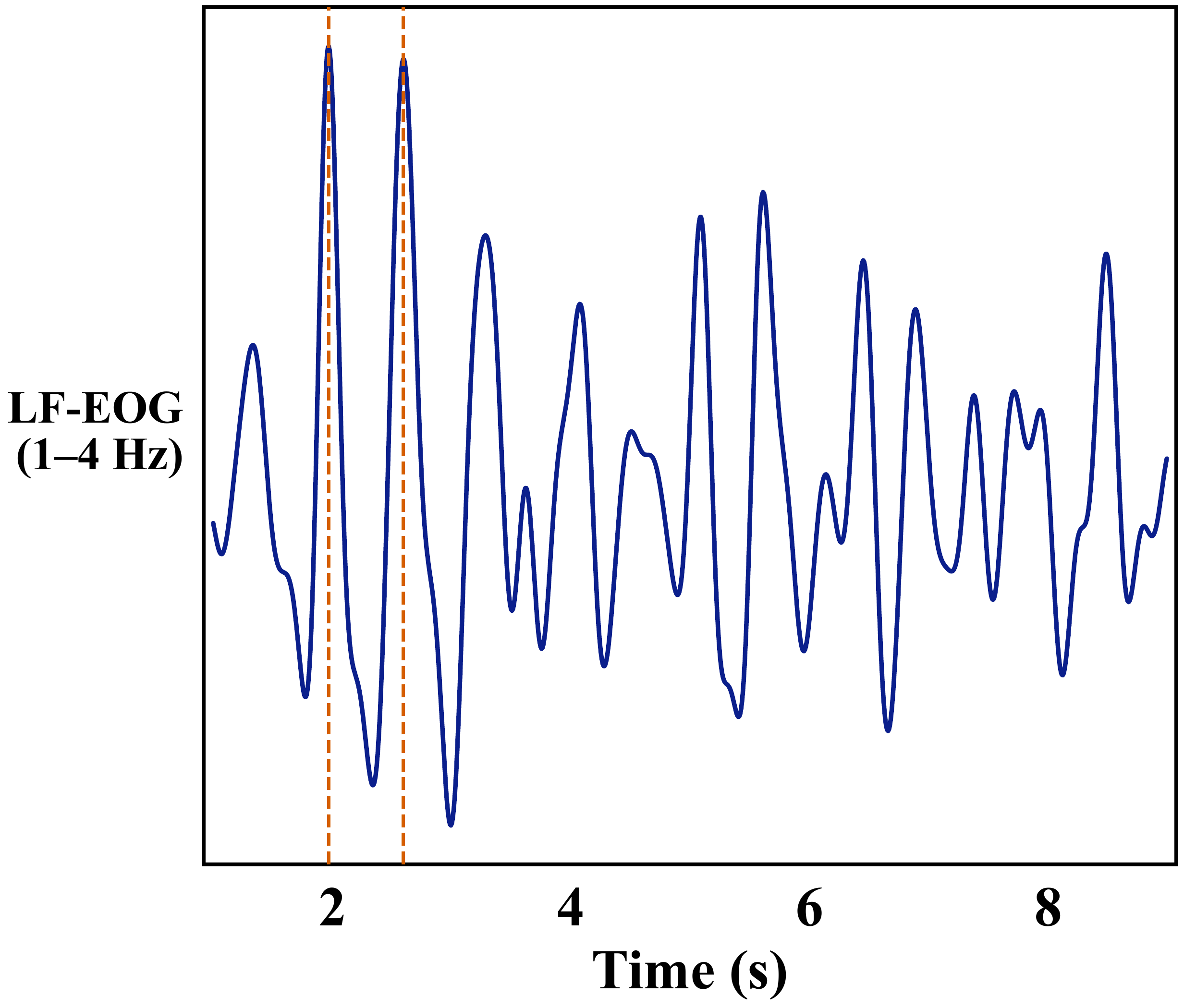}
        \caption{\textbf{EOG signals}}
        \label{fig:eog}
    \end{subfigure}\hspace{0.004\textwidth}%
    %
    \begin{subfigure}[b]{0.23\textwidth}
        \centering
        \includegraphics[width=\linewidth,height=3.1cm,keepaspectratio]{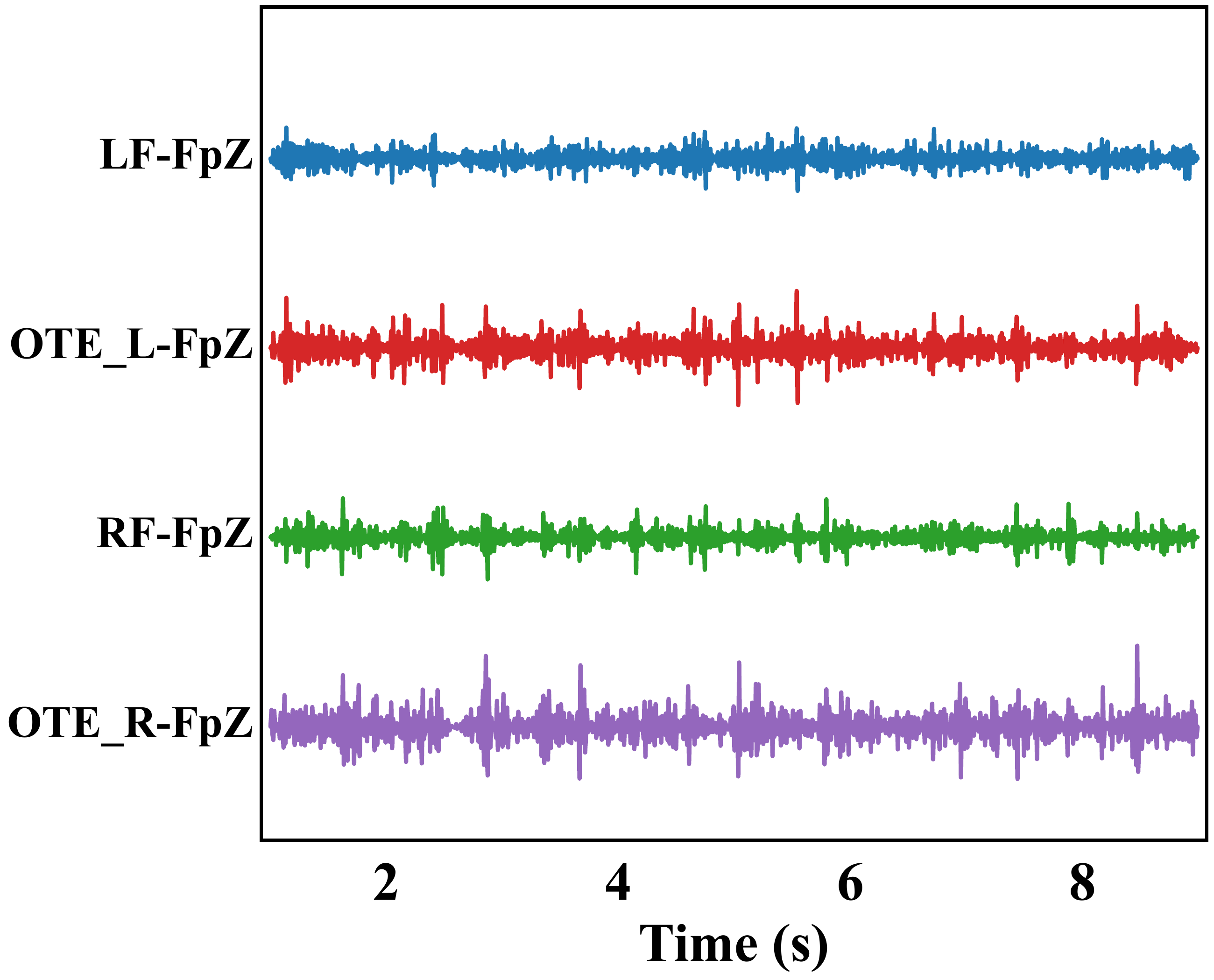}
        \caption{\textbf{EMG comparison}}
        \label{fig:emg}
    \end{subfigure}
    \caption{EEG, EOG, and EMG signals acquired with the FRENZ Brainband.
        The figure illustrates frequency characteristics and time-domain of four EEG
        channels, eye blinks captured as EOG activity, and posture related muscle
        activity captured as EMG.
    }
    \label{fig:Frenz_comparison}
\end{figure*}

We collect raw EEG data at a sampling rate of 125\,Hz using the FRENZ Brainband. \rev{We discard the first and last 10 seconds of each EEG recording to remove boundary transients caused by device settling, participant adjustment, and task onset and offset. This step is not due to synchronization error. EEG and IMU streams are aligned using their timestamps, and the same boundary trimming is applied to IMU recordings so that both modalities remain temporally aligned for downstream multimodal analysis.} We validate physiological sensitivity using a standard 
eyes-open versus eyes-closed benchmark, where eyes-closed 
recordings show increased alpha-band power (8--12\,Hz) 
consistent with canonical neurophysiological activity 
(Figure~\ref{fig:eeg_fft}). Per-channel Z-normalization is 
applied before visualization (Figure~\ref{fig:eeg_time}), 
and the device additionally captures blink-related EOG 
(Figure~\ref{fig:eog}) and posture-related EMG activity 
(Figure~\ref{fig:emg}).EEG recordings often exhibit a low signal-to-noise ratio due to contamination from ocular activity, muscle contractions, and environmental interference~\cite{6609968}. These artifacts overlap with typical EEG frequency bands and introduce bias in downstream features~\cite{amin2023normal}.
\noindent{\textbf{Band-pass Filtering}} We first band-pass filter (0.5--40Hz) EEG signal using a zero-phase fourth-order Butterworth filter \cite{widmann2015digital}. The high-pass cutoff reduces slow drifts caused by sweat, electrode polarization, and skin--electrode interface instabilities and the low-pass cutoff attenuates high-frequency muscle activity and other non-neural noise outside typical EEG rhythms. Zero-phase filtering (forward and reverse application) preserves the temporal alignment of neural events.

\subsubsection{\textbf{IMU-based Motion Artifact Removal}}
\label{para:eeg_imu_cleaning}
\begin{figure*}[t]
    \centering

    \begin{subfigure}[t]{0.495\textwidth}
        \centering
        \includegraphics[width=\linewidth]{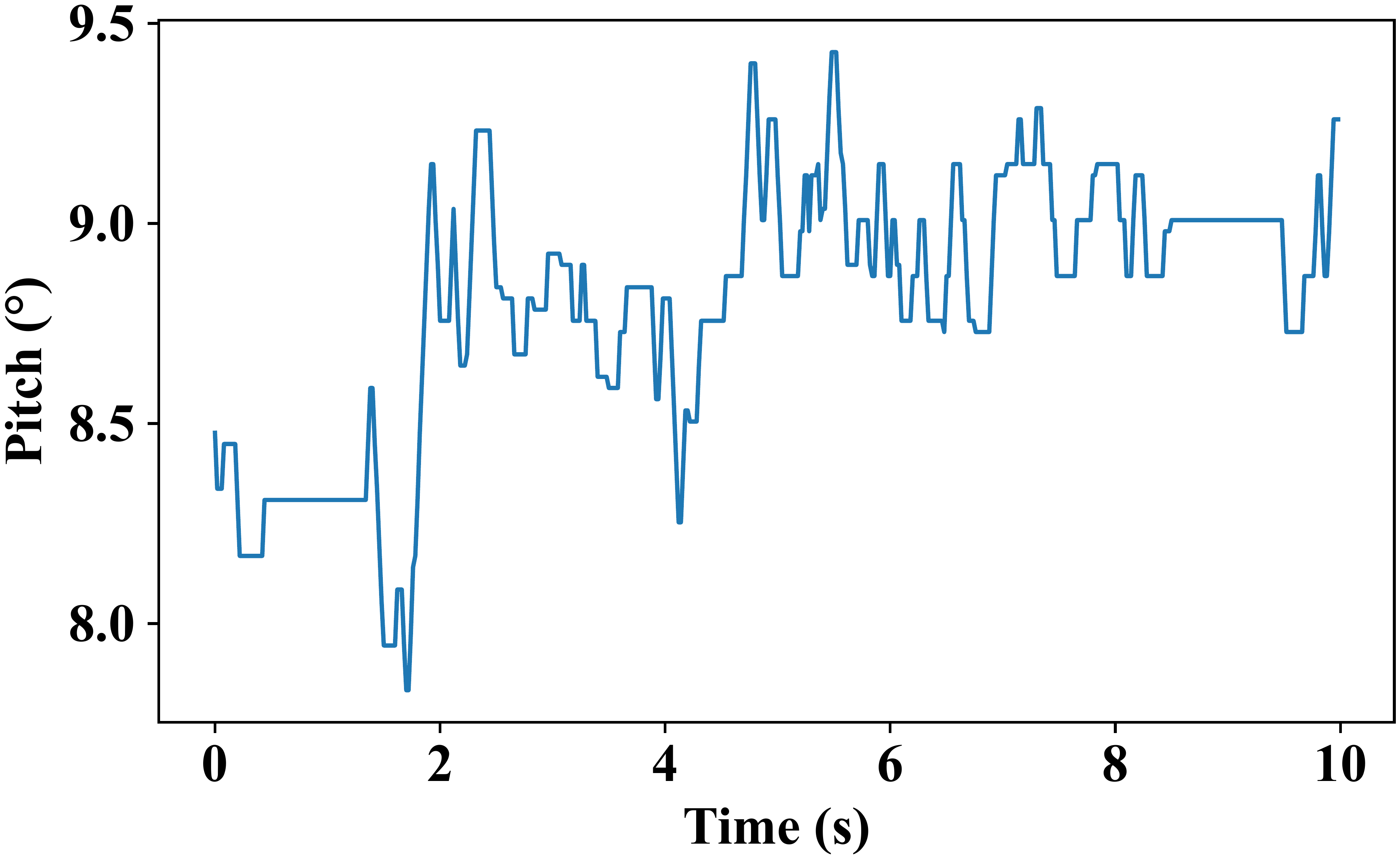}
        \caption{\textbf{IMU head orientation}}
        \label{fig:imu_pitch}
    \end{subfigure}\hfill
    \begin{subfigure}[t]{0.495\textwidth}
        \centering
        \includegraphics[width=\linewidth]{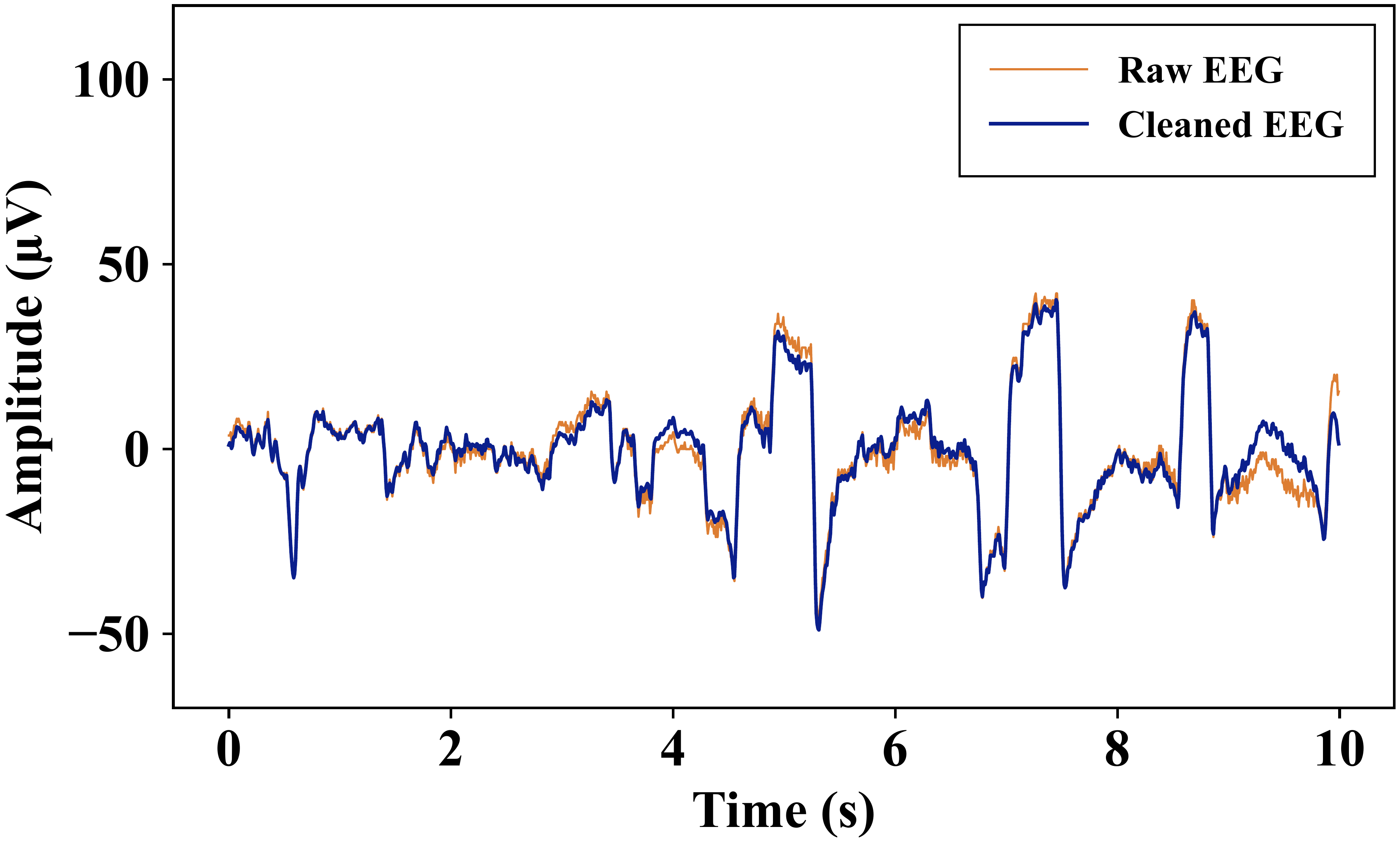}
        \caption{\textbf{Raw vs.\ Cleaned EEG}}
        \label{fig:eeg_comparison}
    \end{subfigure}

    \vspace{2pt}

    \begin{subfigure}[t]{0.495\textwidth}
        \centering
        \includegraphics[width=\linewidth]{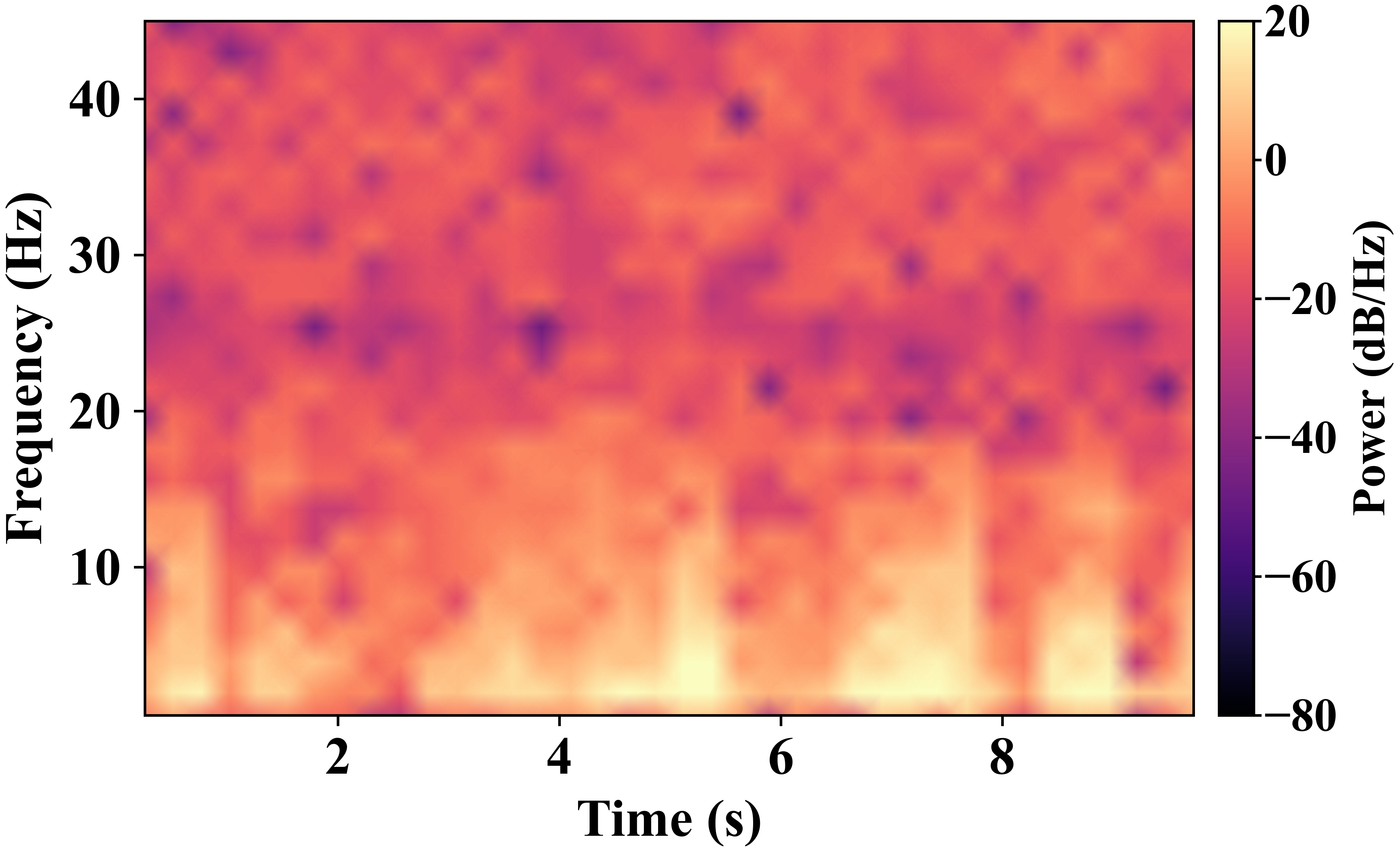}
        \caption{\textbf{Raw EEG Spectrogram}}
        \label{fig:raw_spectrogram}
    \end{subfigure}\hfill
    \begin{subfigure}[t]{0.495\textwidth}
        \centering
        \includegraphics[width=\linewidth]{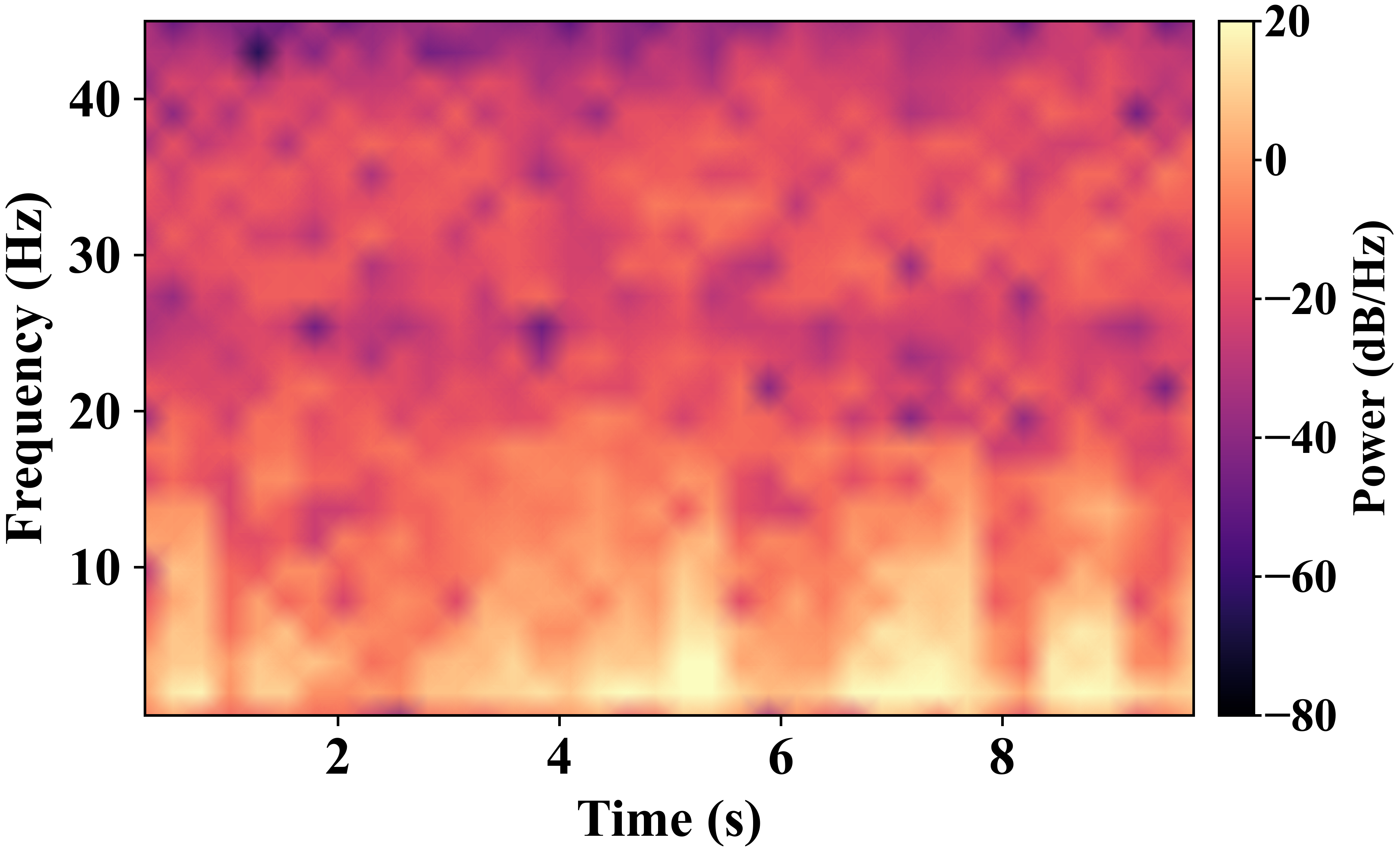}
        \caption{\textbf{Clean EEG Spectrogram}}
        \label{fig:clean_spectrogram}
    \end{subfigure}

    \caption{EEG preprocessing pipeline for motion artifact removal.The figure demonstrates the multimodal preprocessing approach:
    (a) IMU data (Pitch) capturing head movements,
    (b) temporal comparison of raw EEG and cleaned EEG signal,
    (c) spectrogram of raw EEG, and
    (d) spectrogram of cleaned EEG.}
    
    \label{fig:preprocessing_pipeline}
    \vspace{-10pt}
\end{figure*}

For multimodal data processing, we suppress motion-induced artifacts by leveraging synchronized IMU data to identify contaminated EEG segments. We compute angular velocity from IMU orientation signals and estimate RMS velocity using a \rev{1 s} sliding window. For each recording, we define an adaptive threshold based on the RMS velocity distribution, capturing subject- and trial-specific movement variability. Segments exceeding this threshold are labeled as motion-contaminated~\cite{o2012automatic}.
\rev{We then convert these labels into a binary IMU-derived motion mask at the native 50 Hz IMU rate and map this mask onto the 125 Hz EEG sampling grid using nearest-neighbor assignment, so that each EEG sample inherits the motion label of the temporally closest IMU sample. Because this step transfers binary artifact labels rather than interpolating continuous signal values, it does not introduce synthetic IMU measurements. We then band-pass filter the EEG (0.5 to 40 Hz), segment it into 5 s windows with 50\% overlap, and discard any window in which more than 40\% of samples are flagged as motion-contaminated. As shown in Figure \ref{fig:preprocessing_pipeline}, periods of increased head motion captured by the IMU align with large-amplitude, broadband distortions in the raw EEG signal. After IMU-guided artifact rejection, these motion-related components are substantially attenuated in both the time-domain waveform and the corresponding spectrogram.}
\subsection{Feature Extraction}
\paragraph{\textbf{IMU Feature Extraction}}
\label{para:imu_feature}
After IMU data processing (Section~\ref{sec:imudata}), we segment the pitch signal into 5s windows with 50\% overlap. \rev{From each window,
we extract time-domain, derivative-based, and frequency-domain features to capture both posture magnitude and posture dynamics. We focus on extracting all features from pitch as pitch better reflects \rev{FHP} than yaw and roll \cite{chhaglani2024neckcare}. Time domain features such as Mean, median, and root mean square summarize the sustained level of neck flexion \cite{jimenez2022machine} . Standard deviation, range, interquartile range, skewness, and kurtosis characterize within-window variability and waveform shape~\cite{zhu2017feature,6365160}. Zero-crossing rate and line length capture short-term fluctuations and movement irregularity~\cite{pan2024automated}. We also compute velocity and acceleration based features from the first and second temporal derivatives of pitch to describe postural adjustment dynamics \cite{jimenez2022machine}. Derivative signals of this kind, analogous to jerk in accelerometer-based HAR, have been shown to improve discrimination between posture classes~\cite{zhu2017feature}. In addition, we extract Hjorth activity, mobility, and complexity as compact descriptors of signal power and spectral shape~\cite{oppel2021analysis}. In the frequency domain, dominant frequency, spectral entropy, and low- and high-frequency power ratios characterize the rate and regularity of postural change~\cite{6365160,zhu2017feature}. IMU-based kinematic features of this kind have been validated for posture recognition across multiple body placements and sensor configurations~\cite{tang2021upper,zmitri2019human}. To improve interpretability, we report model-specific feature importance scores with a linear-kernel SVM weight analysis for the same IMU feature set in Section \ref{para:imu_based_posture}.} We standardize the resulting feature vectors and use them to train machine learning models for both multi-class and binary posture classification.
\paragraph{\textbf{EEG Feature Extraction}}
\label{para:eegfeature}
We focus on EEG band-power ratios that serve as established cognitive workload
(CWL) biomarkers, including $\theta/\beta$, $\theta/\alpha$, and $\alpha/\beta$
to capture changes in mental effort and attentional engagement \cite{raufi2022evaluation}.
Accordingly, we focus on the $\theta$, $\alpha$, and $\beta$ frequency bands.
We do not extract features from the $\delta$ band or higher $\gamma$ frequencies,
as these ranges are more susceptible to motion, muscle, and environmental noise
in wearable EEG recordings.\rev{We extract absolute band power because it captures the strength of oscillatory activity within each band \cite{fernandez1993test, sandre2026power}. We also compute relative band power to normalize each band by total spectral power and reduce sensitivity to subject-specific amplitude differences \cite{wang2016relative}. Inter-band power ratios and the engagement index are included because they more directly reflect workload-related shifts between slower and faster rhythms than band powers alone \cite{wang2023improved, nuamah2018support}. We further compute magnitude-squared coherence for all channel pairs averaged within each band to quantify cross-channel functional coupling \cite{la2014human, roach2008event}. Finally, we estimate periodic alpha power by fitting a spectral parameterization model over 1 to 30 Hz and integrating the periodic component after removing the aperiodic background. This isolates oscillatory alpha activity from broadband spectral slope effects and improves physiological specificity \cite{virtue2025task}. To improve feature interpretability, Section \ref{para:eegbasedactivity} reports model-specific feature rankings together with a linear-kernel SVM baseline on the same EEG feature set.} 
\paragraph{\textbf{Multimodal Feature Extraction}}
\label{para:multimodalFeature}
We temporally align EEG and IMU features using matched 5 s windows with 50\% overlap and retain only windows that pass IMU-guided artifact rejection (described in section \ref{para:eeg_imu_cleaning}). For each trial, we concatenate the modality-specific feature vectors into a single multimodal representation. We then aggregate window-level features to train machine learning models for our downstream classification tasks.
\subsection{\textbf{ML Inference}}
\rev{We select classifier settings to balance predictive performance and computational efficiency for lightweight wearable inference. To validate these choices, we perform a post hoc nested LOSO grid search over the spaces in Table~\ref{tab:search_spaces}, with best configurations reported in Table~\ref{tab:all_grid_search}. For SVM, we search $C \in \{0.1, 1, 10\}$ and $\gamma \in \{\texttt{scale}, \texttt{auto}, .01\}$ to cover standard regularization and kernel ranges~\cite{hsu2003practical}.  Results show that selected configurations remain competitive for binary IMU posture and \rev{task-induced cognitive load} tasks, with small F1 differences. Larger gaps appear in multiclass settings, indicating higher sensitivity to hyperparameters. For example, in IMU binary classification, grid-selected SVM (\(C{=}10\), \(\gamma{=}.01\)) achieves F1\(=.82\), compared to F1\(=.78\) for our selected setting (\(C{=}1.0\), \(\gamma{=}\texttt{scale}\)). We apply the same validation to unimodal discomfort classification. For multimodal discomfort classification, hyperparameters are selected via nested cross-validation with inner grid search due to the higher-dimensional feature space and limited data (Table~\ref{tab:hyperparams}). For class imbalance, we apply SMOTE (\(k{=}5\)) only on training folds, with class weighting used when the minority class has fewer than two samples. Test partitions are never oversampled to prevent leakage.}
\begin{table}[t]
\footnotesize
\centering
\caption{\rev{Hyperparameter search spaces. The same 
spaces apply to IMU posture, task-induced cognitive load, and unimodal 
discomfort classifiers.}}
\label{tab:search_spaces}
\setlength{\tabcolsep}{10pt}
\renewcommand{\arraystretch}{1.25}
\begin{tabular}{ll} 
\toprule
\rowcolor[HTML]{D9ECE8}
\textbf{Model} & \textbf{Search Space} \\
\midrule
\rev{RF} & \rev{$n_{\text{est}} \in \{100,200,300,500\}$, depth $\in \{5,8,10,\texttt{None}\}$, min\_leaf $\in \{1,3,5,8\}$} \\
\rev{GB} & \rev{$n_{\text{est}} \in \{100,200,300\}$, lr $\in \{0.01,0.05,0.1,0.2\}$, depth $\in \{2,3,5\}$} \\
\rev{SVM} & \rev{$C \in \{0.1,1,10\}$, $\gamma \in \{\texttt{scale},\texttt{auto},0.01\}$} \\
\rev{Linear SVM} & \rev{$C \in \{0.1,1,10\}$} \\
\bottomrule
\end{tabular}
\end{table}
\begin{table*}[t]
\footnotesize
\centering
\caption{\rev{Best hyperparameter configurations and 
LOSO performance identified by nested grid search. Search 
spaces are given in Table \ref{tab:search_spaces}. All metrics are 
macro-averaged.}}
\label{tab:all_grid_search}
\setlength{\tabcolsep}{6pt}
\renewcommand{\arraystretch}{1.18}
\begin{tabular}{l l l p{0.28\textwidth} c c}
\toprule
\rowcolor[HTML]{D9ECE8}
\centering\arraybackslash \textbf{Task} &
\centering\arraybackslash \textbf{Sub-task} &
\centering\arraybackslash \textbf{Model} &
\centering\arraybackslash \textbf{Best configuration} &
\centering\arraybackslash \textbf{Acc} &
\centering\arraybackslash \textbf{F1} \\

\midrule

\multirow{8}{*}{\rev{\shortstack[l]{IMU posture\\classification}}}
& \multirow{4}{*}{\rev{Binary}}
  & \rev{RF}         & \rev{$n_{\text{est}}{=}200$, depth${=}5$, min\_leaf${=}1$} & \rev{.796} & \rev{.786} \\
& & \rev{GB}         & \rev{$n_{\text{est}}{=}200$, lr${=}0.2$, depth${=}3$}      & \rev{.783} & \rev{.762} \\
& & \rev{SVM}        & \rev{$C{=}10$, $\gamma{=}0.01$}                            & \rev{\textbf{.822}} & \rev{\textbf{.818}} \\
& & \rev{Linear SVM} & \rev{$C{=}0.1$}                                            & \rev{.796} & \rev{.793} \\

\cmidrule{2-6}

& \multirow{4}{*}{\rev{Multiclass}}
  & \rev{RF}         & \rev{$n_{\text{est}}{=}200$, depth${=}8$, min\_leaf${=}5$} & \rev{.658} & \rev{.643} \\
& & \rev{GB}         & \rev{$n_{\text{est}}{=}300$, lr${=}0.01$, depth${=}5$}     & \rev{\textbf{.658}} & \rev{\textbf{.649}} \\
& & \rev{SVM}        & \rev{$C{=}10$, $\gamma{=}\texttt{scale}$}                  & \rev{.612} & \rev{.598} \\
& & \rev{Linear SVM} & \rev{$C{=}1$}                                              & \rev{.658} & \rev{.641} \\

\midrule

\multirow{8}{*}{\rev{\shortstack[l]{EEG-based task-induced cognitive load\\classification}}}
& \multirow{4}{*}{\rev{Binary}}
  & \rev{RF}         & \rev{$n_{\text{est}}{=}200$, depth${=}5$, min\_leaf${=}3$} & \rev{.902} & \rev{.902} \\
& & \rev{GB}         & \rev{$n_{\text{est}}{=}300$, lr${=}0.2$, depth${=}3$}      & \rev{\textbf{.916}} & \rev{\textbf{.916}} \\
& & \rev{SVM}        & \rev{$C{=}1$, $\gamma{=}0.01$}                             & \rev{.832} & \rev{.832} \\
& & \rev{Linear SVM} & \rev{$C{=}10$}                                             & \rev{.867} & \rev{.867} \\

\cmidrule{2-6}

& \multirow{4}{*}{\rev{Multiclass}}
  & \rev{RF}         & \rev{$n_{\text{est}}{=}100$, depth${=}8$, min\_leaf${=}8$} & \rev{\textbf{.685}} & \rev{\textbf{.615}} \\
& & \rev{GB}         & \rev{$n_{\text{est}}{=}100$, lr${=}0.05$, depth${=}5$}     & \rev{.643} & \rev{.562} \\
& & \rev{SVM}        & \rev{$C{=}10$, $\gamma{=}\texttt{scale}$}                  & \rev{.636} & \rev{.548} \\
& & \rev{Linear SVM} & \rev{$C{=}0.1$}                                            & \rev{.643} & \rev{.603} \\

\midrule

\multirow{4}{*}{\rev{\shortstack[l]{Unimodal\\discomfort\\(IMU only)}}}
& \multirow{4}{*}{\rev{Binary}}
  & \rev{RF}         & \rev{$n_{\text{est}}{=}100$, depth${=}8$, min\_leaf${=}5$} & \rev{.580} & \rev{.577} \\
& & \rev{GB}         & \rev{$n_{\text{est}}{=}100$, lr${=}0.1$, depth${=}2$}      & \rev{.565} & \rev{.554} \\
& & \rev{SVM}        & \rev{$C{=}0.1$, $\gamma{=}\texttt{scale}$}                 & \rev{\textbf{.623}} & \rev{\textbf{.622}} \\
& & \rev{Linear SVM} & \rev{$C{=}0.1$}                                            & \rev{.551} & \rev{.545} \\

\midrule

\multirow{4}{*}{\rev{\shortstack[l]{Unimodal\\discomfort\\(EEG only)}}}
& \multirow{4}{*}{\rev{Binary}}
  & \rev{RF}         & \rev{$n_{\text{est}}{=}100$, depth${=}8$, min\_leaf${=}5$} & \rev{.522} & \rev{.511} \\
& & \rev{GB}         & \rev{$n_{\text{est}}{=}200$, lr${=}0.1$, depth${=}3$}      & \rev{\textbf{.580}} & \rev{\textbf{.579}} \\
& & \rev{SVM}        & \rev{$C{=}0.1$, $\gamma{=}0.01$}                           & \rev{.565} & \rev{.549} \\
& & \rev{Linear SVM} & \rev{$C{=}0.1$}                                            & \rev{.522} & \rev{.515} \\

\bottomrule

\multicolumn{6}{l}{%
\rev{\footnotesize $^\dagger$ Multimodal discomfort uses 
prospective nested cross-validation; search space in 
Table \ref{tab:hyperparams}.}} \\

\end{tabular}

\vspace{-8pt}
\end{table*}
\normalsize

\subsubsection{\textbf{IMU-Based Posture Classification}}
\label{subse:imu basedclssification}
To support on-device inference, we use four traditional 
machine learning models for IMU-based posture 
classification: Radial basis function Support Vector Machine (SVM), Linear SVM, Random Forest (RF), 
and Gradient Boosting (GB). These models provide a good 
tradeoff between accuracy, interpretability, and 
computational efficiency for real-time use. \rev{For the 
main reported experiments, model parameters were chosen 
to balance predictive performance and computational 
efficiency for on-device deployment. Linear SVM uses a 
linear kernel with $C{=}1.0$ and balanced class weighting; 
$C{=}1.0$ provides moderate regularization that is 
well-suited to normalised inertial features without 
requiring heavy tuning. SVM uses $C{=}1.0$ and 
$\gamma{=}\texttt{scale}$, where \texttt{scale} 
automatically adapts the kernel width to the feature 
variance, avoiding manual bandwidth selection. RF uses 
$n_{\text{est}}{=}300$, max\_depth${=}10$, and 
min\_leaf${=}5$ as the moderate depth and leaf size prevent 
overfitting on the limited per-subject windows while 
retaining sufficient model capacity. GB uses 
$n_{\text{est}}{=}200$, lr${=}0.1$, and max\_depth${=}3$; 
shallow trees with a moderate learning rate are standard 
for structured sensor ovefitting issue ~\cite{friedman2001greedy}.} We evaluate these 
classifiers on a multiclass task with labels $D1$, $D2$, 
and $D3$, and on a binary task separating neutral posture 
($D1$) from combined slouched postures ($D2{+}D3$). We 
use two validation strategies: leave-one-subject-out 
cross-validation and a within-subject 80/20 split. In 
each fold, \texttt{StandardScaler} is fit only on the 
training partition to prevent data leakage.


\subsubsection{\textbf{EEG-Based \rev{Task-induced Cognitive Load} Classification}}
\label{sec:eegact}
We use the same four machine learning models for EEG-based 
activity recognition, applied to the EEG features described 
in Section~\ref{para:eegfeature}. \rev{For the main 
reported EEG experiments, RF uses $n_{\text{est}}{=}300$ 
with balanced class weighting in the binary setting: 300 
trees provide stable importance estimates for the 
high-dimensional EEG feature set. GB uses 300 boosting 
iterations with lr${=}0.1$ and max\_depth${=}3$ because EEG spectral features are noisy and 
high-dimensional as deeper ensembles risk overfitting 
across subjects~\cite{friedman2001greedy}. SVM uses 
$C{=}1.0$ and $\gamma{=}\texttt{scale}$ with balanced 
class weighting to capture non-linear decision boundaries. 
Linear SVM uses a linear kernel with $C{=}1.0$ and 
balanced class weighting as it serves as an interpretable 
baseline whose feature importance can be examined through 
held-out permutation importance.} We evaluate on both 
binary (Rest vs.\ Active) and multiclass (Rest, Stroop, 
Numerical) tasks using an 80/20 stratified split and LOSO 
cross-validation. For the binary task, we apply SMOTE 
with $k{=}5$ nearest neighbours only to the training 
partition when the minority class has at least two samples. 
Class weighting is used instead when the minority class is 
too small to support synthetic generation, and test 
partitions are never oversampled. For the multiclass task, 
we do not apply SMOTE or class weighting. Normalization 
follows the same training-only procedure as IMU 
classification.

\subsubsection{\textbf{Perceived Discomfort Classification}}
We perform unimodal and multimodal discomfort 
classification on the labelled dataset 
(Section~\ref{para:discomfortla}) using 80/20 stratified 
split and LOSO evaluation. We select the top $K{=}10$ 
features per training partition using mutual information, 
reducing $K$ when features or trials are limited. Features 
are standardized with \texttt{StandardScaler} fit on 
training data only to prevent leakage. \rev{The RF classifier uses 100 decision 
trees with square-root feature sampling at each split, a 
minimum leaf size of 8, and balanced class weighting; the 
larger leaf size regularizes the model given the limited 
labelled trials (57). The GB classifier uses 100 boosting 
iterations with lr${=}0.1$ and max\_depth${=}3$ as shallow 
trees with a moderate learning rate consistently prevent 
overfitting ~\cite{friedman2001greedy}. The SVM uses 
$C{=}1.0$, $\gamma{=}\texttt{scale}$, and balanced class 
weighting. The Linear SVM uses a linear kernel with 
$C{=}1.0$ and balanced class weighting; $C{=}1.0$ provides 
standard regularization appropriate for the small 
normalised feature set.} To address class imbalance, we 
apply random undersampling on the training partition to 
balance class counts; test partitions are never resampled.\\
For multimodal discomfort classification, we fuse EEG and 
IMU features (Section~\ref{para:multimodalFeature}) into a 
single trial-level representation. Before training, we 
apply median imputation to fill missing values, remove 
near-zero-variance features (threshold${=}.01$), and drop 
highly correlated features ($r{>}.9$). We then select the 
top $K{=}10$ features and standardise using training data 
only. RF, SVM, and GB hyperparameters are selected 
via nested cross-validation using a 3-fold stratified 
inner loop and \rev{Linear SVM uses fixed $C{=}1.0$ as the 
regularization strength is stable across the small 
labelled set}. The search space for RF, SVM, and GB is 
reported in Table~\ref{tab:hyperparams}.

\rev{Overall for all evaluations, we use standard implementations from scikit-learn and imbalanced-learn. Unless explicitly stated otherwise, all unspecified hyperparameters follow library defaults.}
\begin{table}[ht]

\vspace{-6pt}

\footnotesize
\centering

\caption{Hyperparameter search space for multimodal discomfort classification.}

\label{tab:hyperparams}

\setlength{\tabcolsep}{5pt}
\renewcommand{\arraystretch}{1.15}

\begin{tabular}{>{\centering\arraybackslash}p{0.18\columnwidth} p{0.72\columnwidth}}

\toprule

\rowcolor[HTML]{D9ECE8}
\centering\arraybackslash \textbf{Model} &
\centering\arraybackslash \textbf{Hyperparameter search space} \\

\midrule

RF &
$n_{\text{estimators}} \in \{100, 500\}$; 
$max\_depth \in \{8, 12, \texttt{None}\}$; 
$min\_samples\_leaf \in \{2, 3\}$; 
$min\_samples\_split \in \{4, 6\}$; 
$max\_features = \texttt{sqrt}$; 
$class\_weight = \texttt{balanced}$ \\

\midrule

SVM &
Kernel = RBF; 
$C \in \{.1, 1, 10\}$; 
$\gamma \in \{\texttt{scale}, \texttt{auto}, .001, .01\}$; 
$class\_weight = \texttt{balanced}$ \\

\midrule

GB &
$n_{\text{estimators}} \in \{100, 300\}$; 
learning rate $\in \{.01, .1, .2\}$; 
$max\_depth \in \{3, 5, 7\}$; 
$min\_samples\_leaf \in \{2, 3\}$; 
subsample $\in \{.8, 1.0\}$; 
balanced sample weights \\

\bottomrule

\end{tabular}

\vspace{-12pt}

\end{table}

\normalsize
\subsection{\textbf{Real-Time Alert Scheduling Algorithm}}
\label{sec:algorithmposturecognition}
\rev{
\noindent\textbf{Model Selection.}
{We deploy Random Forest as the single classifier in our alert policies for both posture and task-induced cognitive load inference, based on its stable performance in both 80–20 and LOSO evaluations reported in Section \ref{sec:systemperformance}. We use SVM, GB, and Linear SVM only as comparative baselines to contextualize RF performance. We do not use ensemble combinations at inference time, as combining multiple models introduces computational overhead incompatible with the latency and memory constraints of continuous on-device wearable deployment \cite{ahmed2024tiny,daghero2023dynamic}.}}
We implement two real-time alert intervention policies. The posture-only policy (P-only) triggers alerts solely based on sustained non-neutral head posture. The cognition-aware policy (P+C) applies the same posture correction logic but modulates alert timing using an estimated \rev{task-induced cognitive load}, delaying or suppressing alerts during periods of high mental engagement and releasing them during low-focus breakpoints. \\
\textbf{Posture-Only Alert (P-only):}  In this alert system, an alert is generated when the user’s head posture deviates by more than $10^\circ$ from their personalized neutral posture (described in section \ref{sec:imudata}), and remains in same posture for at least 3 seconds \cite{dang2025co}. The posture-only alert is summarized in Algorithm~\ref{alg:posture_only}.\\
\noindent\textbf{Cognition-Aware Alert Policy (P+C):} We trigger alerts based on the user's real-time 
\rev{task-induced cognitive load} rather than posture alone. \rev{At each inference step, posture is classified using $f_{\text{IMU}}(X_{\text{imu}})$, where \texttt{isBad} denotes the current prediction. \texttt{badCount} tracks consecutive bad-posture ticks to determine \texttt{sustainedBad}, and \texttt{badDur} tracks continuous duration for the 30-second override. Both reset when posture returns to neutral.} If bad posture sustains beyond 30 seconds, the system overrides the cognitive load estimate and triggers an alert regardless of the interruptibility state. The cognition-aware alert policy is summarized in Algorithm~\ref{alg:pc_alert}.
\begin{figure*}[t]
\centering
\begin{minipage}[t]{0.48\textwidth}
\scriptsize\raggedright
\begin{algorithm}[H]
\SetAlgoSkip{empty}
\caption{\textbf{Posture-only Alert Scheduling (P-only)}}
\label{alg:posture_only}
\KwInput{Sampling rate $f_s = 50$ Hz;
baseline duration $T_b$; alert threshold $\theta$;
persistence window $T_p$; alert interval $T_i$}
\KwOutput{Set of alert events $\mathcal{E}$}
\BlankLine
\textbf{Stage 1: Personalized Posture Calibration}\\
Collect pitch samples for $T_b$ seconds while the user maintains neutral posture\;
$P_{\text{base}} \leftarrow \mathrm{median}(\text{baseline pitch samples})$\;
$N_p \leftarrow \lceil T_p \cdot f_s \rceil$\;
$\mathcal{E} \leftarrow \emptyset$, $lastAlert \leftarrow -\infty$, $badCount \leftarrow 0$\;
\BlankLine
\textbf{Stage 2: Real-time Monitoring}\\
\While{new pitch sample $(t, pitch)$ is available}{
    $d \leftarrow pitch - P_{\text{base}}$\;
    $a \leftarrow |d|$\;
    \eIf{$a > \theta$}{
        $badCount \leftarrow badCount + 1$\;
    }{
        $badCount \leftarrow 0$\;
    }
    $sustainedBad \leftarrow (badCount \ge N_p)$\;
    $inInterval \leftarrow (t - lastAlert < T_i)$\;
    \If{$sustainedBad \land \neg inInterval$}{
        trigger alert with deviation $d$\;
        $\mathcal{E} \leftarrow \mathcal{E} \cup \{t\}$\;
        $lastAlert \leftarrow t$\;
        $badCount \leftarrow 0$\;
    }
}
\Return{$\mathcal{E}$}
\end{algorithm}
\end{minipage}
\hfill
\begin{minipage}[t]{0.48\textwidth}
\scriptsize\raggedright
\begin{algorithm}[H]
\SetAlgoSkip{empty}
\caption{\textbf{Cognition-aware Alert Scheduling with IMU + EEG (P+C)}}
\label{alg:pc_alert}
\KwInput{IMU stream sampling rate $f_s = 50$ Hz;
baseline duration $T_{\text{base}}$;
IMU window length $T_{\text{imu}}$;
posture persistence window $T_p$;
continuous bad-posture window $T_{\text{cont}} = 30$ s;
alert interval $T_i$;
EEG window length $T_e$;
pretrained IMU posture classifier $f_{\text{IMU}}(\cdot)$;
EEG classifier $f_{\text{EEG}}(\cdot)$}
\KwOutput{Set of alert events $\mathcal{E}$}
\textbf{Initialization: }
$\Delta t \leftarrow 1/f_s$,
$N_p \leftarrow \lceil T_p \cdot f_s \rceil$,
$badCount \leftarrow 0$, $badDur \leftarrow 0$,
$lastAlert \leftarrow -\infty$,
$eegState \leftarrow \texttt{Unknown}$, $lastEEG \leftarrow -\infty$,
$\mathcal{E} \leftarrow \emptyset$,
$baselineReady \leftarrow \texttt{false}$,
$t_0 \leftarrow$ session start time\;
\textbf{Real-time Monitoring: }
\While{session is running}{
read next IMU sample at time $t$\;
\If{$\neg baselineReady$}{
accumulate IMU samples for baseline\;
\If{$t - t_0 \ge T_{\text{base}}$}{
compute baseline normalization parameters\;
$baselineReady \leftarrow \texttt{true}$\;}
\textbf{continue}\;}
\If{not enough IMU samples for a $T_{\text{imu}}$ window}{
\textbf{continue}\;}
$X_{\text{imu}} \leftarrow$ most recent IMU window of length $T_{\text{imu}}$\;
$X_{\text{imu}} \leftarrow$ baseline-normalize $(X_{\text{imu}})$\;
$\hat{y}_{\text{imu}} \leftarrow f_{\text{IMU}}(X_{\text{imu}})$\;
$isBad \leftarrow (\hat{y}_{\text{imu}} = \texttt{Bad})$\;
\eIf{$isBad$}{
$badCount \leftarrow badCount + 1$\;
$badDur \leftarrow badDur + \Delta t$\;}
{
$badCount \leftarrow 0$\;
$badDur \leftarrow 0$\;}
$sustainedBad \leftarrow (badCount \ge N_p)$\;
$continuousBad \leftarrow (badDur \ge T_{\text{cont}})$\;
\If{$t - lastEEG \ge T_e$}{
$X_e \leftarrow$ most recent EEG window of length $T_e$\;
$eegState \leftarrow f_{\text{EEG}}(X_e)$\;
$lastEEG \leftarrow t$\;}
$interruptible \leftarrow (eegState = \texttt{Rest})$\;
$inInterval \leftarrow (t - lastAlert < T_i)$\;
$trigger \leftarrow \texttt{false}$\;
\If{$\neg inInterval$}{
\If{$continuousBad$}{
$trigger \leftarrow \texttt{true}$\;}
\ElseIf{$sustainedBad \land interruptible$}{
$trigger \leftarrow \texttt{true}$\;}}
\If{$trigger$}{
trigger alert at time $t$\;
$\mathcal{E} \leftarrow \mathcal{E} \cup \{t\}$\;
$lastAlert \leftarrow t$\;
$badCount \leftarrow 0$\;
$badDur \leftarrow 0$\;}}
\Return{$\mathcal{E}$}
\end{algorithm}
\end{minipage}
\vspace{-0.5em}
\Description{Two pseudocode algorithms shown side by side. Left: the posture-only (P-only) policy triggers an alert when the head pitch deviation exceeds a threshold and is sustained for a persistence window, subject to a minimum inter-alert interval. Right: the cognition-aware (P+C) policy applies the same posture criterion but also reads an EEG-based interruptibility state, deferring alerts during high cognitive load and forcing an alert through a 30-second continuous bad-posture override.}
\label{fig:alert_algorithms}
\end{figure*}
\normalsize

%% file: acmart-primary/UserStudyMHCI.tex
\section{Experimental Setup and Data Collection}
\label{para:ergouserstudy}
\begin{figure}[t]
    \centering
    \includegraphics[width=0.7\columnwidth]{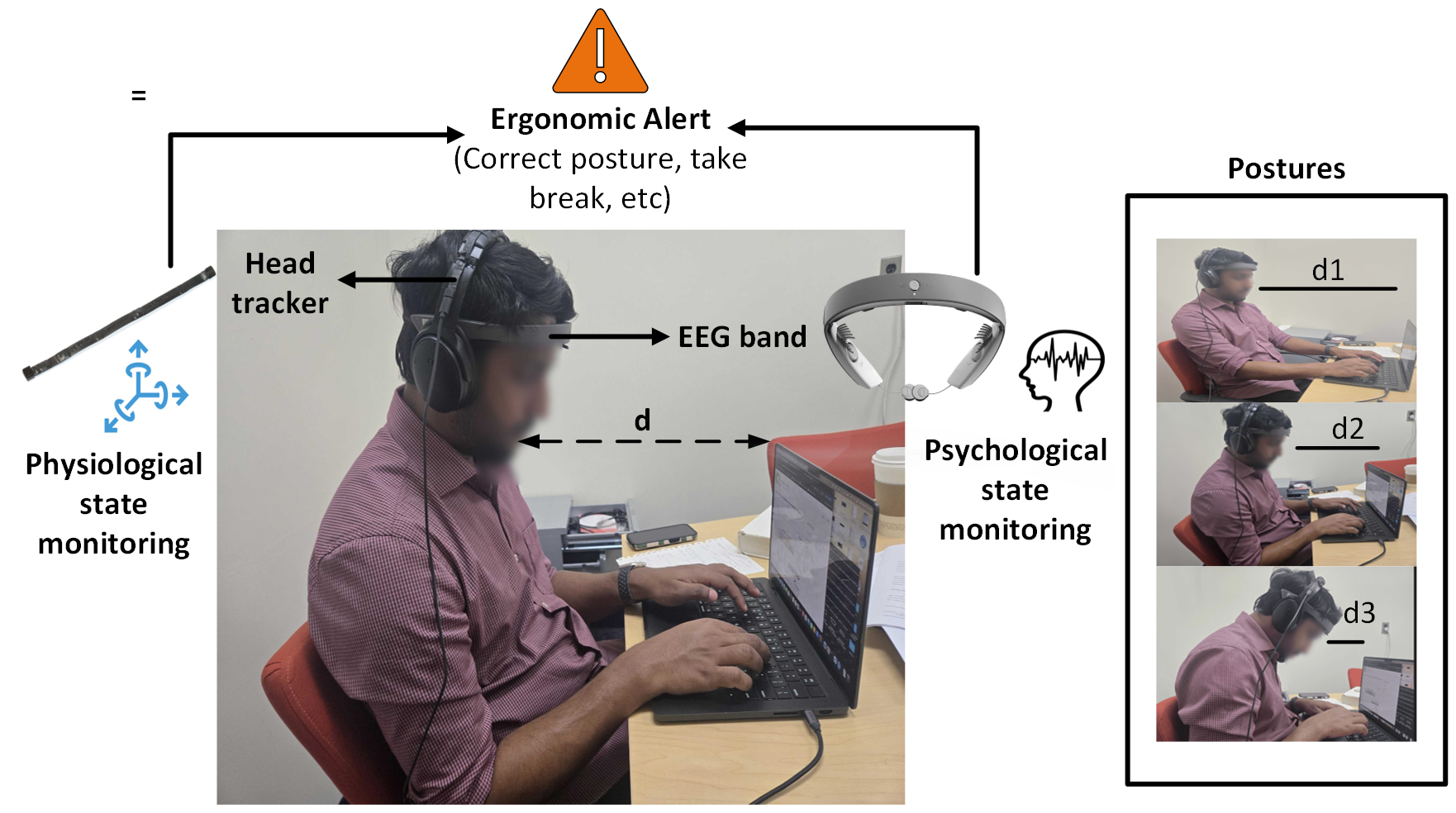}
    \caption{Data collection setup. Participant performing activities at three screen distance conditions during \system data collection.}
    \label{fig:setup}
\end{figure}

In this experiment, we conduct a user study to determine the system design and evaluate the feasibility of low cost wearable sensors for intelligent intervention and discomfort analysis. This study was approved by our Institutional Review Board. The details of the main experiment are summarized below.
\paragraph{\textbf{Participants}}
\label{sec:participant1details}We recruit 24 participants aged 18 years or above for this 
IRB-approved study (mean age = 29.46 years, SD = 4.91). 
All participants report daily use of electronic devices and 
complete the study after providing informed consent. Each 
session lasts approximately 20 to 25 minutes, and 
participants receive a \$10 Amazon gift card as 
compensation. Detailed participant demographics are 
provided in Appendix~\ref{app:participants}.
\rev{\paragraph{\textbf{Data Collection Procedure}} The user study was conducted in a controlled laboratory setting. After providing written consent, participants completed a brief questionnaire regarding their typical electronic device usage, posture habits, and experiences with muscle or eye fatigue during prolonged device use. \rev{The physical setup included three posture conditions defined by screen distance: D1 (70 cm, neutral), D2 (50 cm, mild forward flexion), and D3 (30 cm, deep forward flexion), as shown in Figure~\ref{fig:setup}}. We employed a mixed design in which each participant experienced all three posture conditions. The order of these conditions was randomized to reduce order effects and bias. At each distance, every participant perform:
\begin{enumerate}
\item \textbf{Rest Activity:} Participants sit quietly with eyes closed for approximately 1 minute, establishing a baseline for neurophysiological measurements.
\item \textbf{Active Activity:} Participants either complete a Stroop task (with both congruent and incongruent stimuli) or perform numerical calculation (addition, subtraction, multiplication, division) in a dynamic setting. We select the Stroop task and numerical calculation as established paradigms to induce task-induced cognitive load~\cite{nirabi2025cognitive, xiong2020pattern}. The Stroop task captures selective attention under interference~\cite{dos2007computerized,kingma1996stroop}, and numerical calculation engages sustained mental effort~\cite{cirett2012eeg, spuler2016eeg}. Together, they provide complementary cognitive load patterns. Participants are randomly assigned to one task to control for task-specific bias.
\end{enumerate}}
\rev{We use a three-minute active task window to elicit measurable task-induced cognitive load within a controlled protocol. Prior work shows that EEG-based task-induced cognitive load estimation is feasible from short windows ranging from 10--20 seconds to one minute~\cite{nirabi2025cognitive,xiong2020pattern}, and that sustained task engagement induces measurable changes in seated posture and neck muscle co-activation within comparable timeframes~\cite{igarashi2016effect}. Both tasks (Stroop and numerical calculation) are established proxies for mental effort at comparable difficulty levels~\cite{nirabi2025cognitive,ghosh2022sam}, and task order was counterbalanced across participants to control for ordering effects. Throughout the experiment, each participant wore both the Supperware head tracker~\cite{SupperwareHeadTracker} and the FRENZ Brainband~\cite{FrenzBrainband} for continuous data collection.}
\rev{\paragraph{\textbf{Instructions}:} We do not enforce fixed angular targets for forward flexion due to inter-participant variability in cervical anatomy and natural head posture~\cite{christensen1998natural,stenneberg2023variability}. Instead, we define posture conditions using controlled viewing distances \(D_1\), \(D_2\), and \(D_3\). We define neutral posture as a natural upright sitting position at the farthest viewing distance, \(D_1 = 70\) cm. Viewing distance is measured using a measuring tape from the laptop screen to the participant’s neck position. Participants are instructed to sit upright with relaxed shoulders and look at the screen comfortably, without intentional forward leaning. We do not impose a fixed angular definition of neutral posture, as natural head alignment varies across individuals~\cite{christensen1998natural,stenneberg2023variability}. Instead, we illustrate representative posture angles in Figure~\ref{fig:posture_types} to contextualize the degree of neck flexion. At reduced viewing distances \(D_2\) and \(D_3\), participants naturally adopt increasing forward flexion driven by proximity to the screen, without explicit instructions to match a specific angle. This design maintains ecological validity while ensuring consistent distance across conditions. We validate posture differences using IMU-derived pitch measurements with a clear separation across conditions (Section~\ref{sec:imudata}, Figure~\ref{fig:baseline_correction}).}
\paragraph{\textbf{Perceived Discomfort Level Questionnaire}}
\label{para:discomfortla}
To assess subjective \rev{task-induced cognitive load} and 
musculoskeletal strain, we administer the NASA Task Load 
Index (NASA-TLX)~\cite{hart1988development} and the Cornell 
Musculoskeletal Discomfort Questionnaire 
(CMDQ)~\cite{hedge1999cornell} after each posture condition. 
Both instruments are rated on a 7-point Likert scale and 
provide complementary insights into perceived mental demand 
and physical discomfort. The full questionnaire items are 
provided in Appendix~\ref{app:questionnaire}.

%% file: acmart-primary/Evaluation.tex
\section{\system Performance Evaluation}
\label{sec:systemperformance}
In this section, we present the effectiveness of the \system for posture, \rev{task-induced cognitive load}, and discomfort prediction.
\subsection{IMU-based Posture Classification}
\label{para:imu_based_posture}
For binary posture classification, we group mild and severe 
neck flexion into a single slouched class ($D2 + D3$) 
against neutral posture ($D1$), achieving consistent 
performance across all models. Under 80/20 split, accuracy 
ranges from .84 to .88, and under LOSO evaluation from 
.79 to .81, with RF performing best in both settings. 
\rev{A linear kernel SVM evaluated on the same standardized 
IMU feature set achieves .89 under 80/20 split and 
.78 under LOSO evaluation, confirming that binary 
posture separation is largely captured by a linear decision 
boundary.}

For multiclass classification ($D1$ vs.\ $D2$ vs.\ $D3$), 
models reach .72 to .74 accuracy under 80/20 split, with 
$D1$ the most reliably identified class and $D2$ 
consistently showing the lowest scores. Under LOSO 
evaluation, accuracy drops to .57 to .61, indicating 
stronger subject-specific variability in neck flexion 
dynamics. \rev{The linear kernel SVM achieves .77 under 
80/20 split and .63 under LOSO evaluation. Across all 
models, the dominant features are consistently pitch RMS, 
mean, and median, together with derivative-based 
descriptors such as acceleration and velocity variability, 
indicating that IMU posture classification is driven 
primarily by sustained flexion magnitude rather than 
model-specific artifacts.} Overall performance is 
summarized in Table~\ref{tab:imu_posture_performance}, and 
feature rankings are shown in 
Table~\ref{tab:imu_feature_importance}.
\begin{table}[t]
\footnotesize
\centering
\caption{IMU-based posture classification performance for binary (D1 vs.\ D2 + D3) and multiclass (D1/D2/D3) tasks. Metrics include accuracy (Acc), macro-averaged precision, recall, and F1 under 80/20 and leave-one-subject-out (LOSO) evaluation.}
\label{tab:imu_posture_performance}

\setlength{\tabcolsep}{5pt} 
\renewcommand{\arraystretch}{1.2} 
\begin{tabular}{llcccccccc}
\toprule

\rowcolor[HTML]{D9ECE8}
& & \multicolumn{4}{c}{\textbf{80/20}} & \multicolumn{4}{c}{\textbf{LOSO validation}} \\

\cmidrule(lr){3-6} \cmidrule(lr){7-10}

\rowcolor[HTML]{D9ECE8}
\textbf{Task} & \textbf{Model} & \textbf{Acc} & \textbf{Prec.} & \textbf{Rec.} & \textbf{F1} & \textbf{Acc} & \textbf{Prec.} & \textbf{Rec.} & \textbf{F1} \\

\midrule

Binary & RF & .88 & .88 & .87 & .87 & \textbf{.81} & \textbf{.80} & \textbf{.83} & \textbf{.80} \\
       & SVM & .86 & .85 & .84 & .85 & .79 & .79 & .82 & .78 \\
       & GB & .84 & .83 & .83 & .83 & .80 & .78 & .80 & .78 \\
       & \rev{LinearSVM} & \rev{\textbf{.89}} & \rev{.88} & \rev{.88} & \rev{.88} & \rev{.78} & \rev{.78} & \rev{.81} & \rev{.77} \\

\midrule

Multiclass & RF & .74 & .73 & .73 & .73 & .61 & .60 & .61 & .60 \\
           & SVM & .72 & .72 & .71 & .71 & .61 & .60 & .61 & .59 \\
           & GB & .72 & .71 & .71 & .70 & .57 & .55 & .56 & .55 \\
           & \rev{LinearSVM} & \rev{\textbf{.77}} & \rev{.75} & \rev{.75} & \rev{.75} & \rev{\textbf{.63}} & \rev{\textbf{.63}} & \rev{\textbf{.63}} & \rev{\textbf{.62}} \\

\bottomrule
\end{tabular}
\end{table}

\begin{table*}[t]
\centering
\footnotesize

\caption{Top five IMU features for posture classification. Values denote model-specific feature importance scores. For RF, GB, and XGB, these scores are derived from tree-based importance measures. For LinearSVM, they are derived from normalized absolute coefficient magnitudes.}

\label{tab:imu_feature_importance}

\setlength{\tabcolsep}{12pt}
\renewcommand{\arraystretch}{1.25}

\begin{tabular}{llll}

\toprule

\rowcolor[HTML]{D9ECE8}
\textbf{Task} & \textbf{Eval} & \textbf{Model} & \textbf{Top features (importance)} \\

\midrule

Binary & 80/20 & RF & rms (.19), median (.18), mean (.18), acc\_std (.07), vel\_max (.05) \\
       &       & GB & rms (.39), median (.29), mean (.13), acc\_std (.09), acc\_max (.04) \\
       &       & XGB & mean (.20), median (.17), vel\_max (.12), acc\_std (.11), rms (.09) \\
       &       & \rev{LinearSVM} & \rev{rms (.89), mean (.80), median (.76), acc\_std (.46), vel\_max (.29)} \\

\midrule

Binary & LOSO  & RF & median (.28), rms (.26), mean (.26), vel\_max (.02), acc\_std (.02) \\
       &       & GB & rms (.59), median (.13), mean (.09), vel\_max (.02), acc\_std (.02) \\
       &       & XGB & rms (.17), median (.13), vel\_mean (.10), mean (.09), ll (.04) \\
       &       & \rev{LinearSVM} & \rev{rms (1.00), vel\_std (.27), mean (.13), acc\_std (.13), range (.11)} \\

\midrule

Multiclass & 80/20 & RF & median (.20), mean (.19), rms (.19), vel\_max (.06), acc\_std (.06) \\
           &       & GB & rms (.33), median (.27), mean (.14), acc\_std (.09), vel\_max (.08) \\
           &       & XGB & mean (.22), median (.16), vel\_max (.11), rms (.10), acc\_std (.08) \\
           &       & \rev{LinearSVM} & \rev{rms (.92), mean (.89), median (.87), acc\_std (.44), vel\_max (.32)} \\

\midrule

Multiclass & LOSO  & RF & rms (.22), median (.21), mean (.21), vel\_max (.05), acc\_std (.03) \\
           &       & GB & rms (.40), mean (.16), median (.16), vel\_max (.06), acc\_std (.04) \\
           &       & XGB & rms (.12), median (.11), mean (.10), vel\_mean (.07), vel\_max (.05) \\
           &       & \rev{LinearSVM} & \rev{rms (1.00), vel\_std (.29), mean (.21), acc\_std (.20), std (.13)} \\

\bottomrule

\end{tabular}
\end{table*}

\noindent\textbf{Key Takeaway}: Head-mounted IMU sensing enables reliable detection of \rev{FHP} with high accuracy after personalized baseline normalization. Binary posture classification (neutral vs. slouched) is robust across users, achieving up to 88\% accuracy (RF) and maintaining strong LOSO performance. 

\normalsize
\subsection{EEG-based \rev{Task-induced Cognitive Load} Classification}
\label{para:eegbasedactivity}
\rev{For the binary task, RF and GB achieve the highest 
LOSO accuracies of .902 and .895 respectively, indicating 
that Rest versus Active discrimination is robust across 
subjects and that a substantial part of this separation 
is captured by a linear decision boundary. The multiclass 
task is more challenging as the model must separate two 
active conditions (\textit{Stroop} and \textit{Numerical}) 
in addition to \textit{Rest}. Under 80/20 split, RF 
achieves the highest accuracy of .828, and under LOSO 
evaluation performance drops for all models with RF 
reaching .664. This generalization gap indicates that 
models reliably separate Rest from Active but 
distinguishing the two active tasks is more sensitive to 
subject-specific variation. Overall results are summarized 
in Table~\ref{tab:eeg_performance}.}
\begin{table}[ht]
\centering
\caption{EEG-based \rev{task-induced cognitive load} classification performance for binary and multiclass tasks. Metrics include accuracy (Acc), macro-averaged precision, recall, and F1 under 80/20 and leave-one-subject-out (LOSO) evaluation.}
\label{tab:eeg_performance}

\footnotesize
\setlength{\tabcolsep}{4pt} 
\renewcommand{\arraystretch}{1.2}

\begin{tabular}{llcccccccc} 
\toprule

\rowcolor[HTML]{D9ECE8}
& & \multicolumn{4}{c}{\textbf{80/20}} & \multicolumn{4}{c}{\textbf{LOSO validation}} \\

\cmidrule(lr){3-6} \cmidrule(lr){7-10}

\rowcolor[HTML]{D9ECE8}
\textbf{Task} & \textbf{Model} & \textbf{Acc} & \textbf{Prec.} & \textbf{Rec.} & \textbf{F1} & \textbf{Acc} & \textbf{Prec.} & \textbf{Rec.} & \textbf{F1} \\

\midrule

Binary     & RF  & .759 & .760 & .757 & .757 & \textbf{.902} & \textbf{.902} & \textbf{.902} & \textbf{.902} \\
           & SVM & .793 & .799 & .790 & .791 & .818 & .818 & .818 & .818 \\
           & GB  & \textbf{.828} & \textbf{.829} & \textbf{.829} & \textbf{.828} & .895 & .895 & .895 & .895 \\
           & \rev{Linear SVM} & \rev{.793} & \rev{.799} & \rev{.790} & \rev{.791} & \rev{.804} & \rev{.804} & \rev{.804} & \rev{.804} \\

\midrule

Multiclass & RF  & \textbf{.828} & \textbf{.800} & \textbf{.787} & \textbf{.779} & \textbf{.664} & .564 & .569 & .564 \\
           & SVM & .690 & .612 & .622 & .613 & .587 & .471 & .477 & .460 \\
           & GB  & .793 & .750 & .765 & .754 & .636 & .542 & .546 & .541 \\
           & \rev{Linear SVM} & \rev{.724} & \rev{.711} & \rev{.746} & \rev{.714} & \rev{.601} & \rev{\textbf{.571}} & \rev{\textbf{.573}} & \rev{\textbf{.567}} \\

\bottomrule
\end{tabular}
\end{table}
\normalsize
\rev{The feature importance analysis is summarized in 
Table~\ref{tab:eeg_features_8020_loso}. Across models, 
the most informative EEG features are consistently 
\(\theta/\beta\), TBR, relative theta, and relative alpha 
power. These patterns remain stable under LOSO evaluation, 
and for the multiclass task feature importance becomes more 
distributed but the same ratios remain dominant. The linear 
SVM additionally highlights periodic alpha features, 
suggesting that both spectral balance and oscillatory alpha 
structure contribute to \rev{task-induced cognitive load} 
discrimination.}

\begin{table*}[t]
\centering
\footnotesize
\caption{Top five EEG features for \rev{task-induced cognitive load} classification under 80/20 and LOSO evaluation. For RF, GB, and XGB, values denote model-based feature importance scores. \rev{For Linear SVM, values denote normalized absolute coefficient-based importance scores.}}
\label{tab:eeg_features_8020_loso}

\setlength{\tabcolsep}{8pt}
\renewcommand{\arraystretch}{1.25}

\resizebox{\textwidth}{!}{%
\begin{tabular}{llll} 
\toprule

\rowcolor[HTML]{D9ECE8}
\textbf{Task} & \textbf{Eval} & \textbf{Model} & \textbf{Top features (importance)} \\

\midrule
Binary & 80/20 & RF & $\theta/\beta$\_Ch4 (.09), TBR\_Ch4 (.08), TBR\_mean (.07), $\theta_{rel}$\_Ch4 (.06), $\beta_{rel}$\_Ch4 (.05) \\
       &       & GB & $\theta_{rel}$\_Ch4 (.47), BandAlpha\_Ch1 (.07), TBR\_Ch4 (.06), $\theta/\beta$\_Ch4 (.05), AlphaPower\_Ch1 (.05) \\
       &       & XGB & BandAlpha\_Ch3 (.09), $\alpha_{rel}$\_Ch2 (.07), TBR\_mean (.07), $\theta/\beta$\_Ch4 (.07), $\theta_{rel}$\_Ch2 (.06) \\
       &       & \rev{Linear SVM} & \rev{EngagementIndex\_std (1.00), $\alpha_{rel}$\_Ch1 (.99), $\theta/\alpha$\_Ch3 (.69), $\alpha_{rel}$\_Ch3 (.61), $\theta_{rel}$\_Ch4 (.60)} \\

\midrule
Binary & LOSO  & RF & $\theta/\beta$\_Ch4 (.07), TBR\_Ch4 (.06), TBR\_mean (.06), $\alpha_{rel}$\_Ch3 (.05), $\theta_{rel}$\_Ch4 (.05) \\
       &       & GB & $\theta_{rel}$\_Ch4 (.21), $\alpha_{rel}$\_Ch3 (.21), TBR\_Ch4 (.15), $\theta/\beta$\_Ch4 (.09), TBR\_mean (.04) \\
       &       & XGB & TBR\_mean (.10), $\alpha_{rel}$\_Ch3 (.06), $\theta/\beta$\_Ch4 (.06), TBR\_Ch4 (.05), $\theta_{rel}$\_Ch4 (.04) \\
       &       & \rev{Linear SVM} & \rev{$\alpha_{rel}$\_Ch1 (.95), $\theta/\alpha$\_Ch3 (.85), $\alpha_{rel}$\_Ch3 (.84), EngagementIndex\_std (.82), $\alpha_{rel}$\_Ch4 (.71)} \\

\midrule
Multiclass & 80/20 & RF & $\theta/\beta$\_Ch4 (.05), TBR\_Ch4 (.04), TBR\_mean (.03), $\alpha_{rel}$\_Ch3 (.03), $\theta_{rel}$\_Ch4 (.03) \\
           &       & GB & $\theta/\beta$\_Ch4 (.18), $\alpha_{rel}$\_Ch3 (.10), $\theta_{rel}$\_Ch4 (.06), TBR\_Ch4 (.06), $\alpha/\beta$\_Ch3 (.05) \\
           &       & XGB & BandBeta\_Ch2 (.11), $\theta/\beta$\_Ch4 (.05), TBR\_Ch4 (.05), TBR\_Ch3 (.04), $\theta_{rel}$\_Ch4 (.04) \\
           &       & \rev{Linear SVM} & \rev{$\alpha_{rel}$\_Ch1 (1.00), $\alpha_{rel}$\_Ch3 (.93), Coh\_C1C2\_Alpha (.88), $\alpha/\beta$\_Ch3 (.74), $\alpha_{rel}$\_Ch4 (.71)} \\

\midrule
Multiclass & LOSO  & RF & $\theta/\beta$\_Ch4 (.04), TBR\_mean (.04), TBR\_Ch4 (.04), $\alpha_{rel}$\_Ch3 (.03), $\theta_{rel}$\_Ch4 (.03) \\
           &       & GB & $\alpha_{rel}$\_Ch3 (.11), $\theta_{rel}$\_Ch4 (.10), $\theta/\beta$\_Ch4 (.09), $\theta/\alpha$\_Ch4 (.05), TBR\_Ch4 (.05) \\
           &       & XGB & TBR\_mean (.05), $\theta_{rel}$\_Ch4 (.05), TBR\_Ch3 (.04), $\alpha_{rel}$\_Ch3 (.03), TBR\_Ch4 (.03) \\
           &       & \rev{Linear SVM} & \rev{$\alpha_{rel}$\_Ch1 (1.00), $\alpha_{rel}$\_Ch3 (.78), $\alpha/\beta$\_Ch3 (.70), $\theta_{rel}$\_Ch3 (.65), $\alpha_{rel}$\_Ch4 (.65)} \\

\bottomrule
\end{tabular}
}

\end{table*}

\rev{\textbf{Window overlap sensitivity. }}\rev{ We conduct a 
window-overlap ablation by varying feature-window overlap 
from 0\% to 50\% while keeping the Random Forest pipeline 
fixed. Table~\ref{tab:eeg_overlap_ablation} reports macro 
F1 under both stratified 80/20 and LOSO evaluation. Under 
LOSO, binary macro F1 ranges from .88 to .91 and 
multiclass macro F1 ranges from .56 to .61, indicating 
broadly stable performance across the full overlap range. 
We retain 50\% overlap as the default because it produces 
more feature windows per recording, improving feature 
stability by reducing sensitivity to any single window. 
Features from all valid windows are averaged into a single 
fixed-length representation per recording before any 
train-test split, so overlapping windows are never treated 
as separate training examples and the overlap does not 
introduce train-test information leakage ~\cite{brookshire2024data}.}
\begin{table}[t]
\centering
\caption{\rev{Effect of feature-window overlap on EEG classification (macro F1, Random Forest). We report both stratified 80/20 and LOSO evaluation. LOSO is the primary cross-subject estimate.}}
\label{tab:eeg_overlap_ablation}
\footnotesize

\setlength{\tabcolsep}{8pt} 
\renewcommand{\arraystretch}{1.2}

\begin{tabular}{lcccc} 
\toprule

\rowcolor[HTML]{D9ECE8}
\rev{\textbf{Window Overlap}} & \rev{\textbf{Binary 80/20}} & \rev{\textbf{Binary LOSO}} & \rev{\textbf{Multiclass 80/20}} & \rev{\textbf{Multiclass LOSO}} \\

\midrule
\rev{0\%}            & \rev{.828} & \rev{.881} & \rev{.834} & \rev{.561} \\
\rev{10\%}           & \rev{.793} & \rev{.888} & \rev{.694} & \rev{.561} \\
\rev{20\%}           & \rev{.828} & \rev{.909} & \rev{.782} & \rev{.585} \\
\rev{30\%}           & \rev{.757} & \rev{.902} & \rev{.726} & \rev{.607} \\
\rev{40\%}           & \rev{.757} & \rev{.909} & \rev{.694} & \rev{.600} \\
\rev{50\%$^\dagger$} & \rev{.757} & \rev{.902} & \rev{.779} & \rev{.564} \\

\bottomrule
\multicolumn{5}{l}{\rev{\scriptsize $^\dagger$ Setting used in our main experiments.}}
\end{tabular}
\end{table}

\noindent\textbf{Key Takeaway}: Wearable EEG can reliably distinguish \rev{task-induced cognitive load} such as Rest vs. Active , achieving ~90\% LOSO accuracy, even with lightweight models and consumer-grade hardware.
\subsection{Feature Group Ablation Analysis}
\rev{We conduct a group-wise feature ablation study to assess the contribution of each IMU and EEG feature group to posture and cognitive load classification respectively.}

\paragraph{\textbf{\rev{IMU Feature Group Ablation}}}
\rev{We conduct a group-wise ablation study using Random 
Forest to evaluate each feature group for posture 
classification. Table~\ref{tab:imuablation} shows that 
time-domain features are the dominant features. Under 80/20 
evaluation, the accuracy is .87 for binary and .74 for 
multiclass classification. The full feature set generalizes 
best under LOSO with .81 binary accuracy and .61 multiclass 
accuracy. Other feature groups perform poorly in isolation 
with LOSO binary accuracies between .52 and .56. This 
indicates they capture complementary dynamics rather than 
primary discriminative signals. For multiclass LOSO and 
80/20, the results show the same trend. This suggests that 
derivative, Hjorth, and frequency-domain features improve 
within-subject discrimination but not cross-subject 
generalization. We conclude that time-domain features are 
the primary signal source but combining all feature groups 
remains necessary for robust binary cross-subject 
performance.}

\begin{table*}[t]

\centering
\footnotesize

\caption{\rev{IMU Feature Group Ablation Study using Random Forest. All metrics are macro-averaged. Top features are ranked by RF importance score (in parentheses)}}

\label{tab:imuablation}

\renewcommand{\arraystretch}{1.18}

\resizebox{\textwidth}{!}{%
\begin{tabular}{lllccccp{0.25\linewidth}p{0.25\linewidth}} 

\toprule

\rowcolor[HTML]{D9ECE8}
& & &
\multicolumn{2}{c}{\rev{\textbf{80/20}}} &
\multicolumn{2}{c}{\rev{\textbf{LOSO}}} &
& \\

\cmidrule(lr){4-5}
\cmidrule(lr){6-7}

\rowcolor[HTML]{D9ECE8}
\rev{\textbf{Task}} &
\rev{\textbf{Feature Group}}
& &
\rev{\textbf{Acc}} &
\rev{\textbf{F1}}
& \rev{\textbf{Acc}} &
\rev{\textbf{F1}}
& \centering\arraybackslash \rev{\textbf{Top-5 Features (80/20)}}
& \centering\arraybackslash \rev{\textbf{Top-5 Features (LOSO)}} \\

\midrule

\multirow{5}{*}{\rev{Binary}}

& \rev{Time-domain} & &
\rev{.87} & \rev{.87} & \rev{.81} & \rev{.80}
& \rev{rms (.25), median (.25), mean (.22), ll (.07), mav (.07)}
& \rev{median (.30), mean (.27), rms (.27), std (.03), ll (.03)} \\

& \rev{Derivative} & &
\rev{.67} & \rev{.64} & \rev{.52} & \rev{.50}
& \rev{acc\_std (.25), vel\_max (.20), vel\_std (.16), acc\_mean (.15), acc\_max (.12)}
& \rev{acc\_std (.22), vel\_mean (.21), vel\_std (.18), acc\_mean (.17), vel\_max (.13)} \\

& \rev{Hjorth} & &
\rev{.60} & \rev{.57} & \rev{.56} & \rev{.51}
& \rev{hj\_activity (.51), hj\_mobility (.24), hj\_complexity (.24)}
& \rev{hj\_complexity (.35), hj\_activity (.34), hj\_mobility (.31)} \\

& \rev{Frequency-domain} & &
\rev{.58} & \rev{.55} & \rev{.55} & \rev{.51}
& \rev{high\_freq\_power (.39), low\_freq\_power (.25), spec\_entropy (.11), high\_freq\_ratio (.10), low\_freq\_ratio (.10)}
& \rev{high\_freq\_power (.21), low\_freq\_power (.19), spec\_entropy (.19), high\_freq\_ratio (.17), low\_freq\_ratio (.17)} \\

& \textbf{\rev{All Features}} & &
\rev{\textbf{.88}} & \rev{\textbf{.87}} & \rev{\textbf{.81}} & \rev{\textbf{.80}}
& \rev{rms (.19), median (.18), mean (.18), acc\_std (.07), vel\_max (.05)}
& \rev{median (.28), rms (.26), mean (.26), vel\_max (.02), acc\_std (.02)} \\

\midrule

\multirow{5}{*}{\rev{Multiclass}}

& \rev{Time-domain} & &
\rev{.74} & \rev{.73} & \rev{.61} & \rev{.59}
& \rev{median (.27), rms (.26), mean (.23), ll (.06), mav (.06)}
& \rev{rms (.25), median (.25), mean (.23), std (.05), ll (.04)} \\

& \rev{Derivative} & &
\rev{.46} & \rev{.46} & \rev{.35} & \rev{.35}
& \rev{acc\_std (.24), vel\_max (.21), vel\_std (.16), acc\_mean (.15), acc\_max (.13)}
& \rev{acc\_std (.21), vel\_mean (.18), vel\_std (.17), vel\_max (.16), acc\_mean (.16)} \\

& \rev{Hjorth} & &
\rev{.37} & \rev{.37} & \rev{.36} & \rev{.36}
& \rev{hj\_activity (.50), hj\_mobility (.25), hj\_complexity (.25)}
& \rev{hj\_activity (.39), hj\_complexity (.32), hj\_mobility (.30)} \\

& \rev{Frequency-domain} & &
\rev{.37} & \rev{.37} & \rev{.35} & \rev{.35}
& \rev{high\_freq\_power (.36), low\_freq\_power (.26), spec\_entropy (.12), low\_freq\_ratio (.11), high\_freq\_ratio (.11)}
& \rev{high\_freq\_power (.23), low\_freq\_power (.20), spec\_entropy (.18), high\_freq\_ratio (.16), low\_freq\_ratio (.16)} \\

& \textbf{\rev{All Features}} & &
\rev{\textbf{.74}} & \rev{\textbf{.73}} & \rev{\textbf{.61}} & \rev{\textbf{.60}}
& \rev{median (.20), mean (.19), rms (.19), vel\_max (.06), acc\_std (.06)}
& \rev{rms (.22), median (.21), mean (.21), vel\_max (.05), acc\_std (.03)} \\

\bottomrule

\end{tabular}
}

\end{table*}

\normalsize

\paragraph{\textbf{\rev{EEG Feature Group Ablation}}}
\rev{We perform a feature-level ablation study with a Random Forest classifier to assess each EEG feature group for task-induced cognitive load classification (Table~\ref{tab:eegablationfeature}). Coherence features reach perfect performance under the 80/20 split (Acc: 1.00) but drop to Acc: .93 under LOSO. This suggests that connectivity patterns are partly subject-specific and can inflate performance when subjects overlap between train and test. Spectral ratio features show the most stable performance across settings (80/20: Acc: .72; LOSO: Acc: .92), indicating strong generalization of relative band dynamics. In contrast, absolute power (LOSO: Acc: .76) and periodic alpha features (LOSO: Acc: .59) provide weaker discrimination. For multiclass classification, spectral and coherence features obtain similar performance (LOSO: Acc: .69), and other groups perform lower. The combined feature set consistently performs better for all classification tasks (binary LOSO: Acc: .90; multiclass LOSO: Acc: .66), indicating complementary information across feature groups.}

\begin{table*}[t]

\centering
\footnotesize

\caption{\rev{EEG Feature Group Ablation Study using Random Forest. All metrics are macro-averaged. Top features are ranked by RF importance score (in parentheses).}}

\label{tab:eegablationfeature}

\renewcommand{\arraystretch}{1.18}

\resizebox{\textwidth}{!}{%
\begin{tabular}{lllccccp{0.25\linewidth}p{0.25\linewidth}}

\toprule

\rowcolor[HTML]{D9ECE8}
& & &
\multicolumn{2}{c}{\rev{\textbf{80/20}}} &
\multicolumn{2}{c}{\rev{\textbf{LOSO}}} &
& \\

\cmidrule(lr){4-5}
\cmidrule(lr){6-7}

\rowcolor[HTML]{D9ECE8}
\rev{\textbf{Task}} &
\rev{\textbf{Feature Group}}
& &
\rev{\textbf{Acc}} &
\rev{\textbf{F1}}
& \rev{\textbf{Acc}} &
\rev{\textbf{F1}}
& \centering\arraybackslash \rev{\textbf{Top-5 Features (80/20)}}
& \centering\arraybackslash \rev{\textbf{Top-5 Features (LOSO)}} \\

\midrule

\multirow{6}{*}{\parbox[c]{1cm}{\centering \rev{Binary}}}

& \rev{Spectral Ratios} & &
\rev{.72} & \rev{.72} & \rev{\textbf{.92}} & \rev{.92}
& \rev{Ch4 $\theta_{rel}$ (.17), Ch4 $\theta/\beta$ (.12), Ch4 $\beta_{rel}$ (.10), Ch3 $\alpha_{rel}$ (.07), Ch2 $\theta_{rel}$ (.06)}
& \rev{Ch4 $\theta_{rel}$ (.12), Ch3 $\alpha_{rel}$ (.10), Ch4 $\theta/\beta$ (.10), Ch4 $\beta_{rel}$ (.07), Ch1 $\alpha_{rel}$ (.06)} \\

& \rev{Absolute Power} & &
\rev{.69} & \rev{.69} & \rev{.76} & \rev{.76}
& \rev{Ch3 $\theta$ (.15), Ch1 $\alpha$ (.15), Ch3 $\alpha$ (.12), Ch3 $\beta$ (.12), Ch1 $\theta$ (.10)}
& \rev{Ch1 $\alpha$ (.15), Ch1 $\theta$ (.14), Ch3 $\theta$ (.13), Ch3 $\beta$ (.11), Ch3 $\alpha$ (.10)} \\

& \rev{TBR+EI} & &
\rev{.72} & \rev{.72} & \rev{.84} & \rev{.84}
& \rev{TBR\_ch4 (.21), TBR\_mean (.18), EI\_ch4 (.15), TBR\_std (.12), EI\_mean (.06)}
& \rev{TBR\_ch4 (.18), TBR\_mean (.17), TBR\_std (.11), EI\_ch4 (.10), TBR\_ch2 (.08)} \\

& \rev{\textbf{Coherence}} & &
\rev{\textbf{1.00}} & \rev{\textbf{1.00}} & \textbf{\rev{.93}} & \textbf{\rev{.93}}
& \textbf{\rev{C1C3 $\alpha$ (.12), C2C4 $\theta$ (.10), C3C4 $\theta$ (.10), C2C3 $\alpha$ (.09), C1C2 $\theta$ (.09)}}
& \textbf{\rev{C1C3 $\alpha$ (.12), C2C4 $\theta$ (.11), C1C2 $\alpha$ (.10), C3C4 $\theta$ (.09), C2C3 $\theta$ (.09)}} \\

& \rev{Periodic Alpha} & &
\rev{.59} & \rev{.59} & \rev{.59} & \rev{.59}
& \rev{log\_PA\_ch3 (.13), PA\_ch3 (.13), log\_PA\_mean (.10), log\_PA\_ch1 (.10), PA\_ch1 (.10)}
& \rev{PA\_ch3 (.13), log\_PA\_ch3 (.13), log\_PA\_mean (.10), PA\_ch1 (.10), log\_PA\_ch1 (.10)} \\

& \rev{All Features} & &
\rev{.76} & \rev{.76} & \rev{.90} & \rev{.90}
& \rev{$\theta/\beta$\_Ch4 (.09), TBR\_Ch4 (.08), TBR\_mean (.07), $\theta_{rel}$\_Ch4 (.06), $\beta_{rel}$\_Ch4 (.05)}
& \rev{$\theta/\beta$\_Ch4 (.07), TBR\_Ch4 (.06), TBR\_mean (.06), $\alpha_{rel}$\_Ch3 (.05), $\theta_{rel}$\_Ch4 (.05)} \\

\midrule

\multirow{6}{*}{\parbox[c]{1cm}{\centering \rev{Multiclass}}}

& \rev{Spectral Ratios} & &
\rev{.72} & \rev{.66} & \rev{\textbf{.69}} & \rev{.60}
& \rev{Ch4 $\theta/\beta$ (.08), Ch4 $\theta_{rel}$ (.08), Ch3 $\alpha_{rel}$ (.06), Ch1 $\alpha_{rel}$ (.06), Ch3 $\theta_{rel}$ (.05)}
& \rev{Ch4 $\theta_{rel}$ (.08), Ch4 $\theta/\beta$ (.07), Ch1 $\alpha_{rel}$ (.07), Ch3 $\alpha_{rel}$ (.07), Ch4 $\beta_{rel}$ (.05)} \\

& \rev{Absolute Power} & &
\rev{.76} & \rev{.74} & \rev{.54} & \rev{.46}
& \rev{Ch3 $\theta$ (.14), Ch3 $\beta$ (.11), Ch1 $\theta$ (.10), Ch3 $\alpha$ (.09), Ch1 $\beta$ (.09)}
& \rev{Ch3 $\theta$ (.12), Ch1 $\alpha$ (.11), Ch1 $\theta$ (.10), Ch3 $\beta$ (.10), Ch3 $\alpha$ (.10)} \\

& \rev{TBR+EI} & &
\rev{.76} & \rev{.70} & \rev{.63} & \rev{.55}
& \rev{TBR\_ch4 (.15), TBR\_mean (.11), TBR\_std (.10), EI\_ch4 (.09), TBR\_ch3 (.09)}
& \rev{TBR\_ch4 (.14), TBR\_mean (.12), TBR\_ch3 (.10), EI\_ch3 (.10), TBR\_std (.09)} \\

& \rev{Coherence} & &
\rev{.76} & \rev{.67} & \rev{\textbf{.69}} & \rev{\textbf{.61}}
& \rev{C1C2 $\alpha$ (.11), C1C3 $\alpha$ (.10), C2C4 $\theta$ (.10), C2C3 $\alpha$ (.08), C1C4 $\alpha$ (.08)}
& \rev{C1C3 $\alpha$ (.10), C2C4 $\theta$ (.09), C1C2 $\alpha$ (.09), C2C3 $\theta$ (.09), C2C3 $\alpha$ (.09)} \\

& \rev{Periodic Alpha} & &
\rev{.45} & \rev{.33} & \rev{.50} & \rev{.41}
& \rev{log\_PA\_ch3 (.12), log\_PA\_mean (.11), PA\_ch3 (.11), PA\_ch1 (.10), log\_PA\_ch4 (.10)}
& \rev{PA\_ch3 (.13), log\_PA\_ch3 (.12), log\_PA\_ch1 (.10), PA\_ch1 (.10), log\_PA\_ch2 (.10)} \\

& \textbf{\rev{All Features}} & &
\rev{\textbf{.83}} & \rev{\textbf{.78}} & \rev{\textbf{.66}} & \rev{\textbf{.56}}
& \rev{\textbf{$\theta/\beta$\_Ch4 (.05), TBR\_Ch4 (.04), TBR\_mean (.03), $\alpha_{rel}$\_Ch3 (.03), $\theta_{rel}$\_Ch4 (.03)}}
& \rev{\textbf{$\theta/\beta$\_Ch4 (.04), TBR\_mean (.04), TBR\_Ch4 (.04), $\alpha_{rel}$\_Ch3 (.03), $\theta_{rel}$\_Ch4 (.03)}} \\

\bottomrule
\end{tabular}
}

\end{table*}

\normalsize
\subsection{Correlation between posture and \rev{task-induced cognitive load}}

\begin{figure*}[htbp]
\centering
\begin{subfigure}[b]{0.30\textwidth}
    \includegraphics[width=\textwidth]{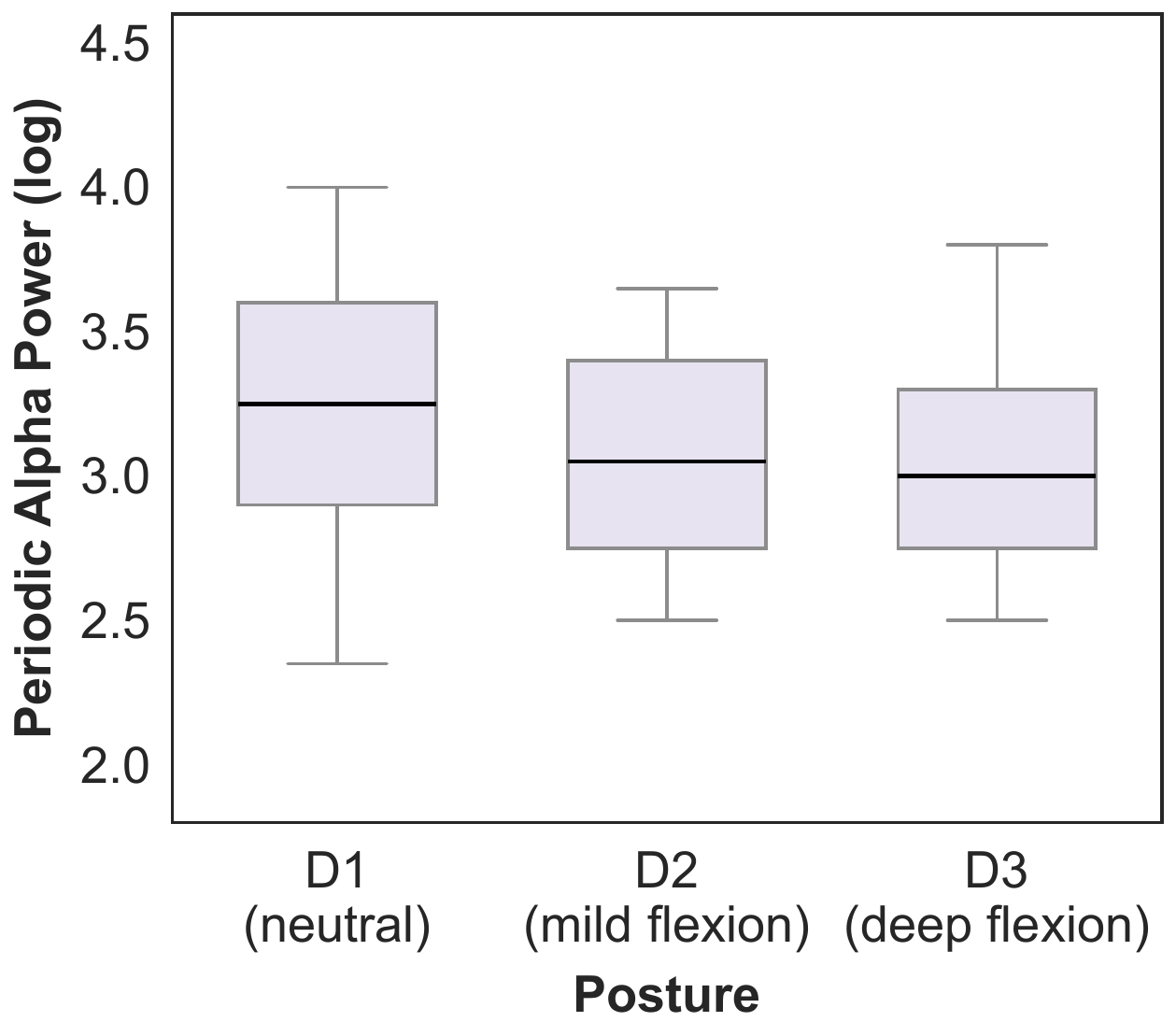}
    \caption{\textbf{Rest}}
    \label{fig:pa_rest}
\end{subfigure}
\hfill
\begin{subfigure}[b]{0.30\textwidth}
    \includegraphics[width=\textwidth]{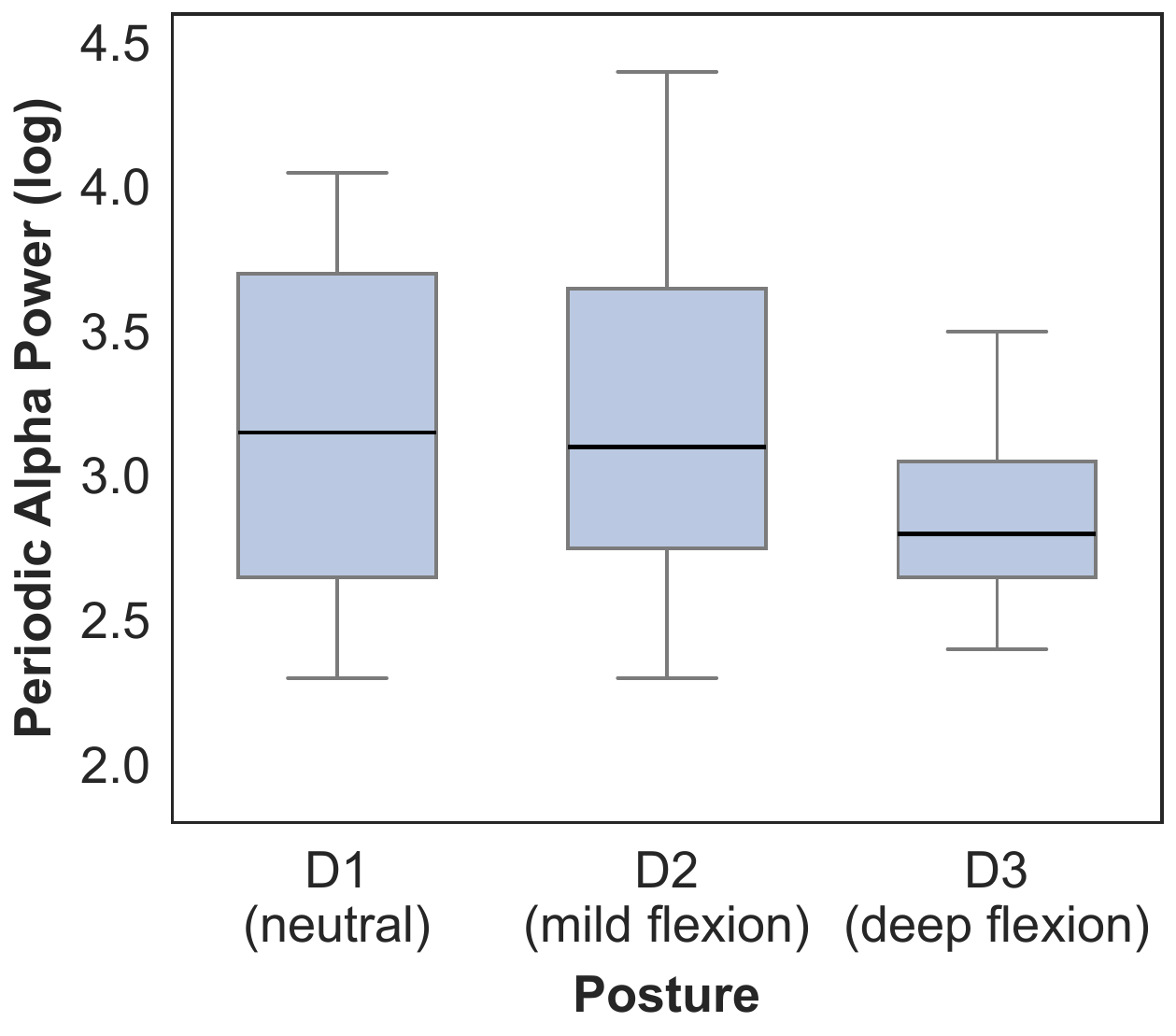}
    \caption{\textbf{Stroop}}
    \label{fig:pa_stroop}
\end{subfigure}
\hfill
\begin{subfigure}[b]{0.30\textwidth}
    \includegraphics[width=\textwidth]{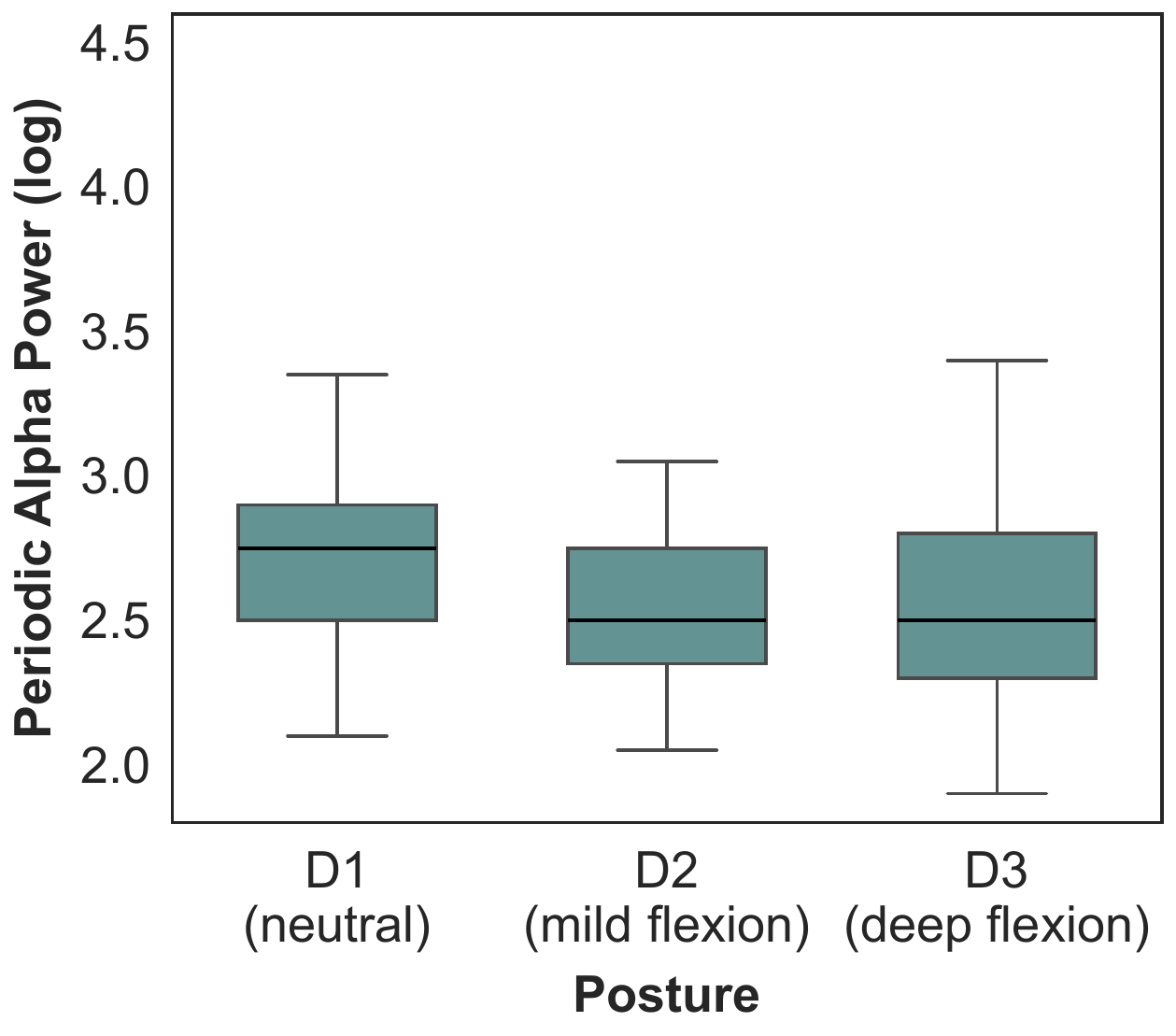}
    \caption{\textbf{Numerical Calculation}}
    \label{fig:pa_numerical}
\end{subfigure}
\begin{subfigure}[b]{0.30\textwidth}
    \includegraphics[width=\textwidth]{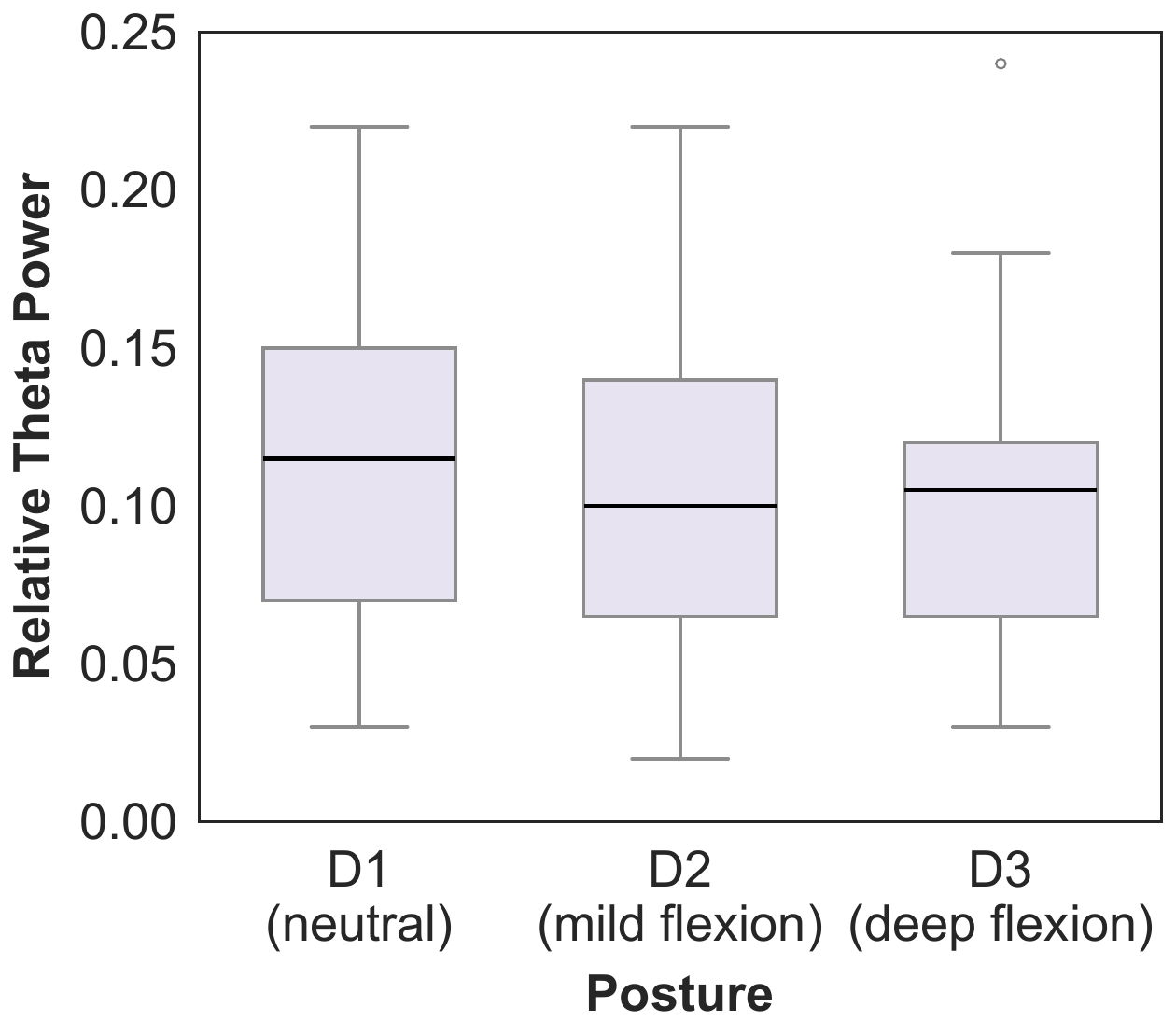}
    \caption{\textbf{Rest}}
    \label{fig:theta_rest}
\end{subfigure}
\hfill
\begin{subfigure}[b]{0.30\textwidth}
    \includegraphics[width=\textwidth]{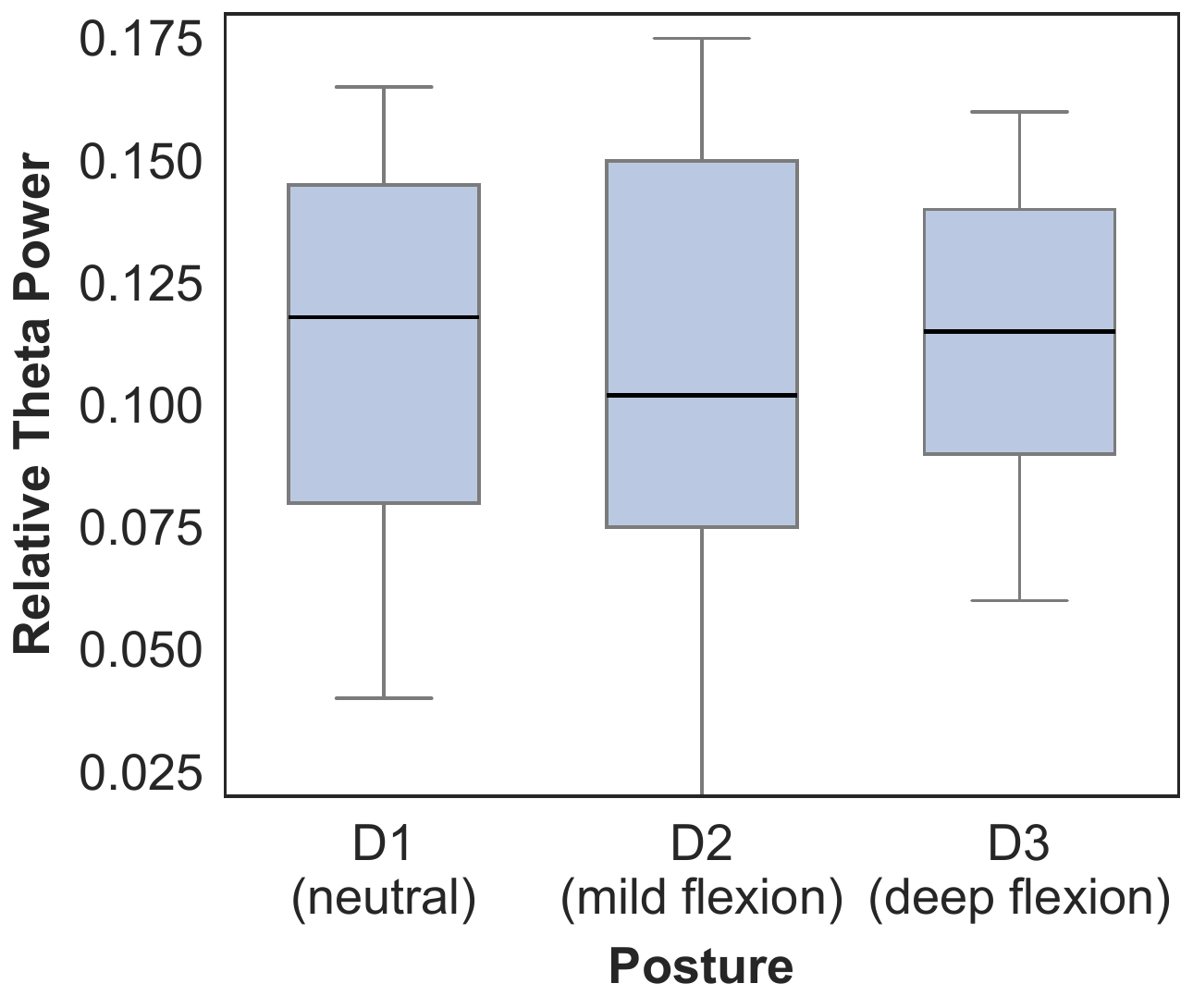}
    \caption{\textbf{Stroop}}
    \label{fig:theta_stroop}
\end{subfigure}
\hfill
\begin{subfigure}[b]{0.30\textwidth}
    \includegraphics[width=\textwidth]{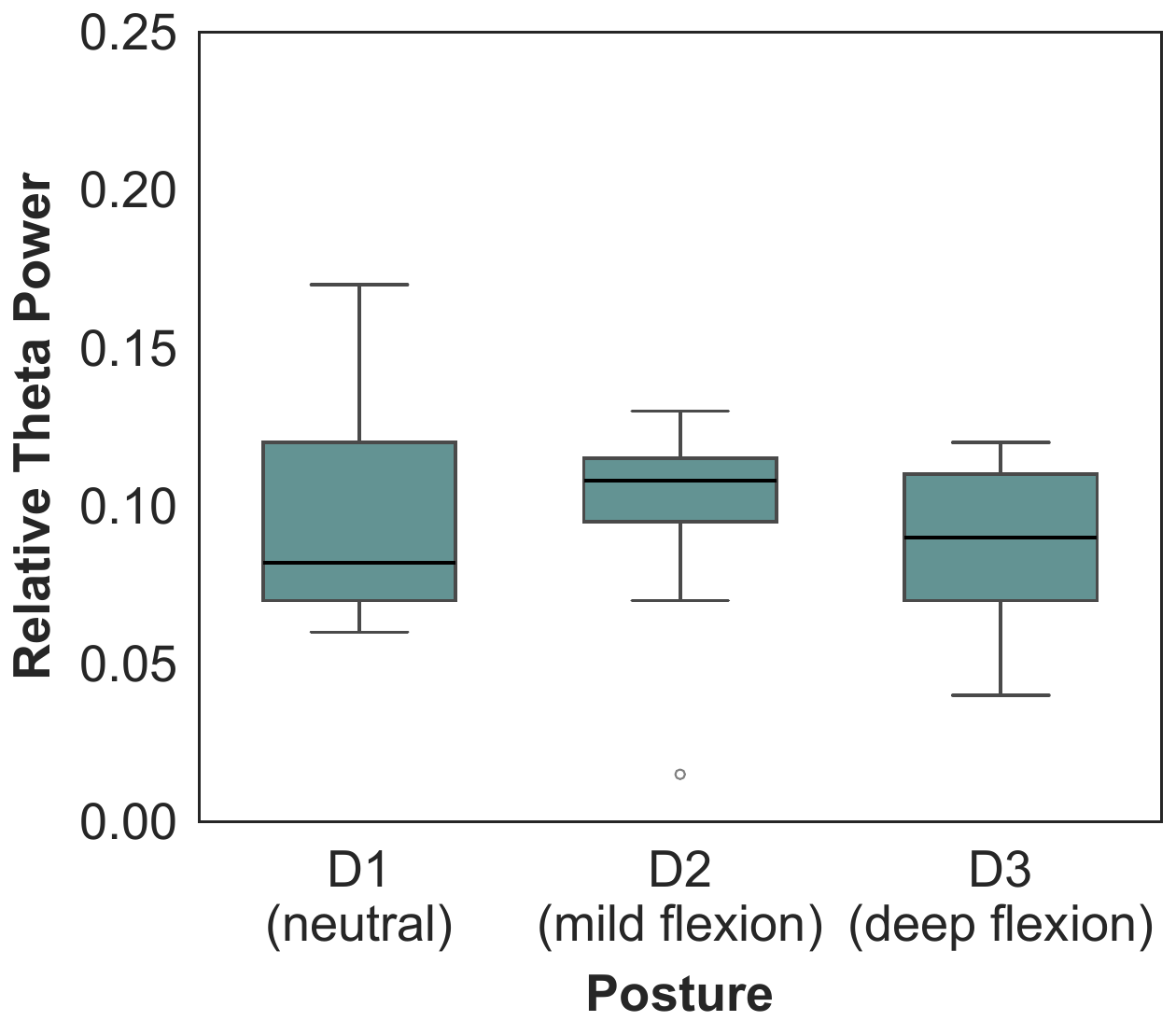}
    \caption{\textbf{Numerical Calculation}}
    \label{fig:theta_numerical}
\end{subfigure}
\begin{subfigure}[b]{0.30\textwidth}
    \includegraphics[width=\textwidth]{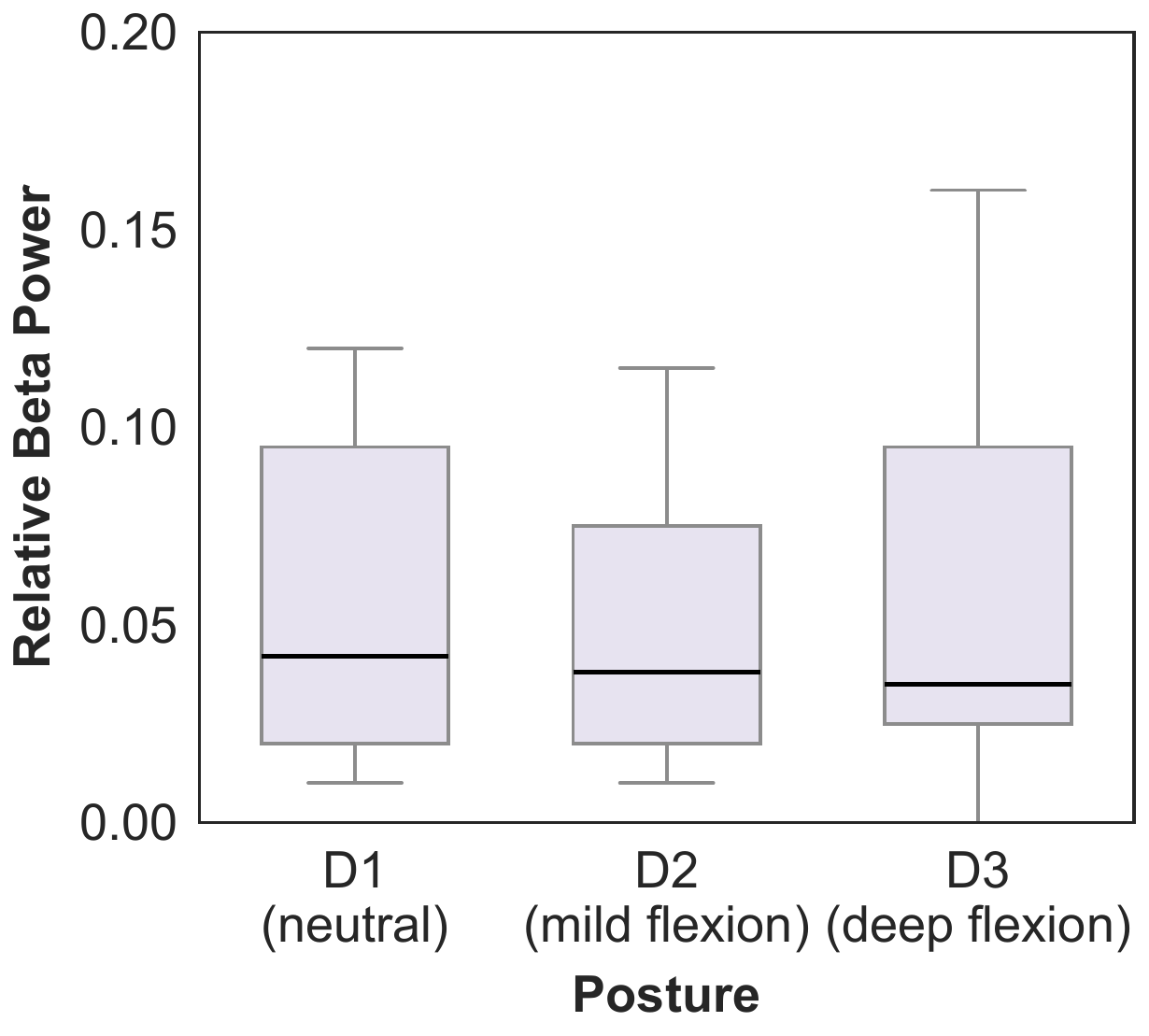}
    \caption{\textbf{Rest}}
    \label{fig:beta_rest}
\end{subfigure}
\hfill
\begin{subfigure}[b]{0.30\textwidth}
    \includegraphics[width=\textwidth]{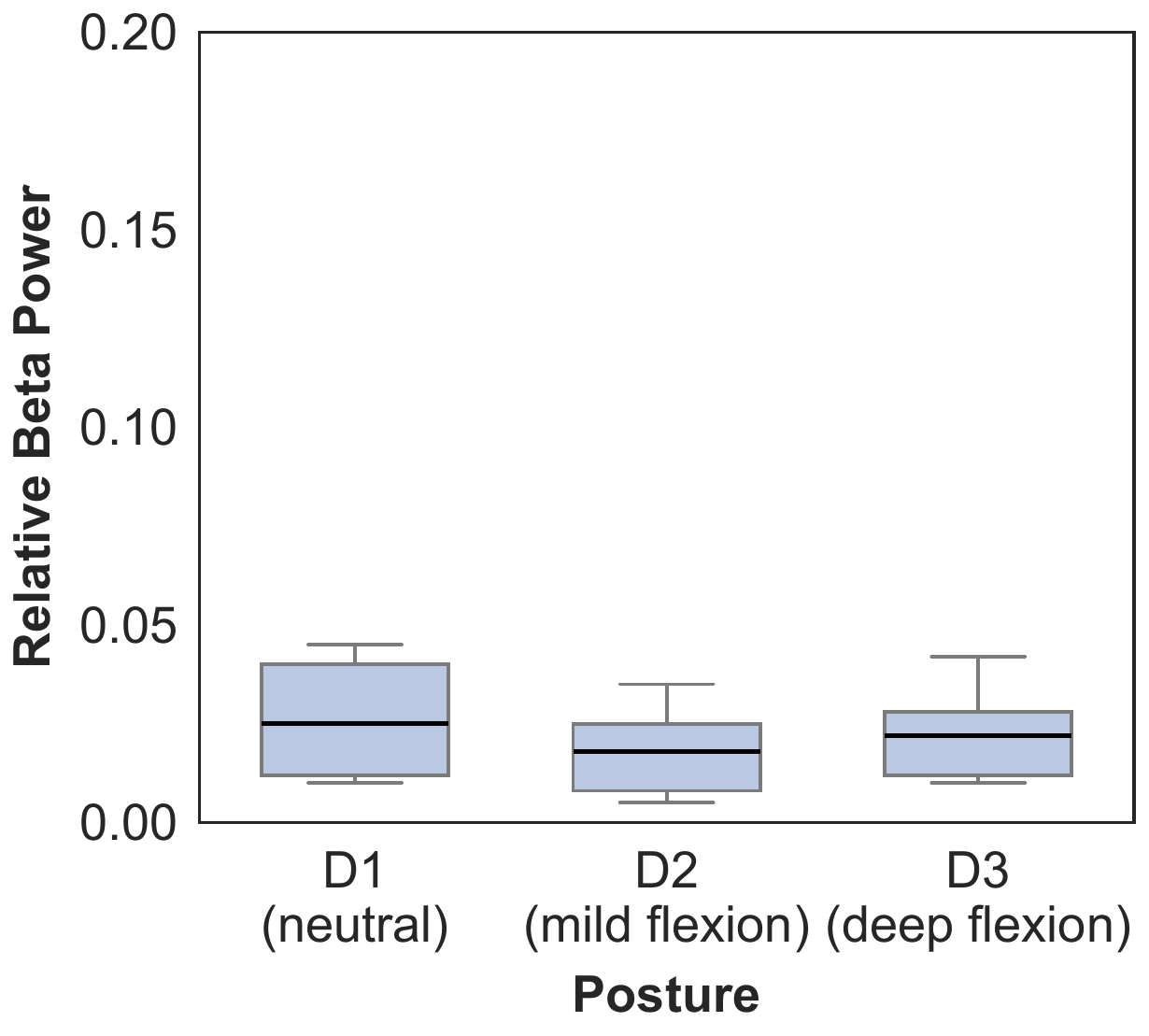}
    \caption{\textbf{Stroop}}
    \label{fig:beta_stroop}
\end{subfigure}
\hfill
\begin{subfigure}[b]{0.30\textwidth}
    \includegraphics[width=\textwidth]{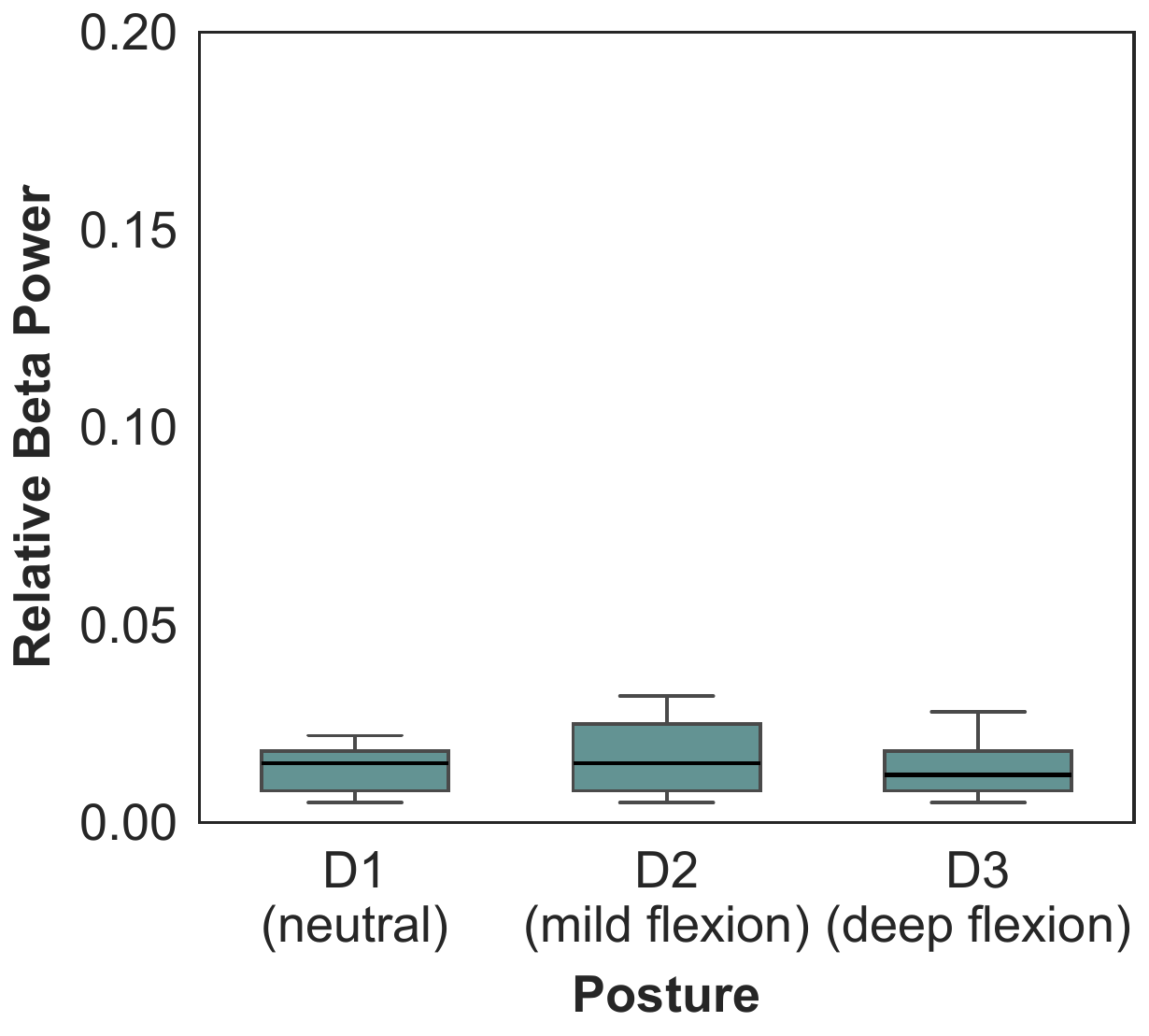}
    \caption{\textbf{Numerical Calculation}}
    \label{fig:beta_numerical}
\end{subfigure}
\caption{EEG spectral features across diverse tasks and postures.}
\label{fig:eeg_all_metrics}
\end{figure*}

\begin{table}[ht]

\centering

\caption{Linear mixed-effects model results for periodic and relative alpha power. All coefficients are fixed-effect estimates relative to the resting, upright ($D1$) baseline.}

\label{tab:lmm_alpha}

\footnotesize

\renewcommand{\arraystretch}{1.15}

\begin{tabular}{lllccc}

\toprule

\rowcolor[HTML]{D9ECE8}
\textbf{Metric} &
\textbf{Factor} &
\textbf{Contrast} &
$\boldsymbol{\beta}$ &
\textbf{Std. Error} &
$\boldsymbol{p}$ \\

\midrule

\multirow{5}{*}{Periodic Alpha}

& Baseline & Resting \& Neutral ($D1$) & 3.209 & .090 & $<.001^{*}$ \\

& Cognitive Activity & Numerical vs. Rest & -0.455 & .130 & $<.001^{*}$ \\

& Cognitive Activity & Stroop vs. Rest & -0.084 & .131 & .521 \\

& Posture & Mild flexion ($D2$) & -0.139 & .104 & .179 \\

& Posture & Deep flexion ($D3$) & -0.138 & .104 & .184 \\

\midrule

\multirow{2}{*}{Relative Alpha}

& Interaction & Stroop $\times$ Mild flexion ($D2$) & -0.031 & .013 & $.018^{*}$ \\

& Interaction & Stroop $\times$ Deep flexion ($D3$) & -0.020 & .013 & .122 \\

\bottomrule

\multicolumn{6}{l}{\small \textit{$^{*}p < .05$. Negative coefficients indicate reduced alpha power.}}

\end{tabular}

\end{table}
\normalsize
We analyze the relationship between \rev{task-induced 
cognitive load} and posture on EEG biomarkers using linear 
mixed-effects models~\cite{galecki2012linear} in 
Table~\ref{tab:lmm_alpha}. The resting neutral posture 
($D1$) serves as the baseline ($\beta = 3.209$, $SE = .090$, 
$p < .001$). Periodic alpha power shows a significant 
decrease during the numerical task ($\beta = -0.455$, 
$p < .001$), and relative alpha power shows a significant 
reduction during the Stroop task under mild flexion ($D2$) 
($\beta = -.031$, $p = .018$). No other significant 
interactions between \rev{task-induced cognitive load} and 
posture are observed. As shown in 
Figure~\ref{fig:eeg_all_metrics}, slouched postures exhibit 
reduced alpha band power relative to neutral, indicating 
lower relaxation levels~\cite{sharma2015assessing}, and 
increased beta and theta activity in $D2$ and $D3$ suggests 
elevated cognitive effort and mental 
strain~\cite{raufi2022evaluation,lansbergen2011increase}.

\noindent\textbf{Key Takeaway}: \system’s EEG biomarkers are driven mainly by \rev{task-induced cognitive load} but bad posture also plays a secondary role to increase cognitive strain in high demanding task.
\subsection{Perceived Discomfort Classification}
\begin{figure*}[t]
\centering
\setlength{\fboxsep}{0pt}
\setlength{\fboxrule}{0.5pt}
\captionsetup[subfigure]{justification=centering}

\begin{subfigure}[t]{0.32\linewidth}
    \centering
    \vspace{0pt}
    \fbox{\includegraphics[width=\linewidth,height=4cm,keepaspectratio=false]{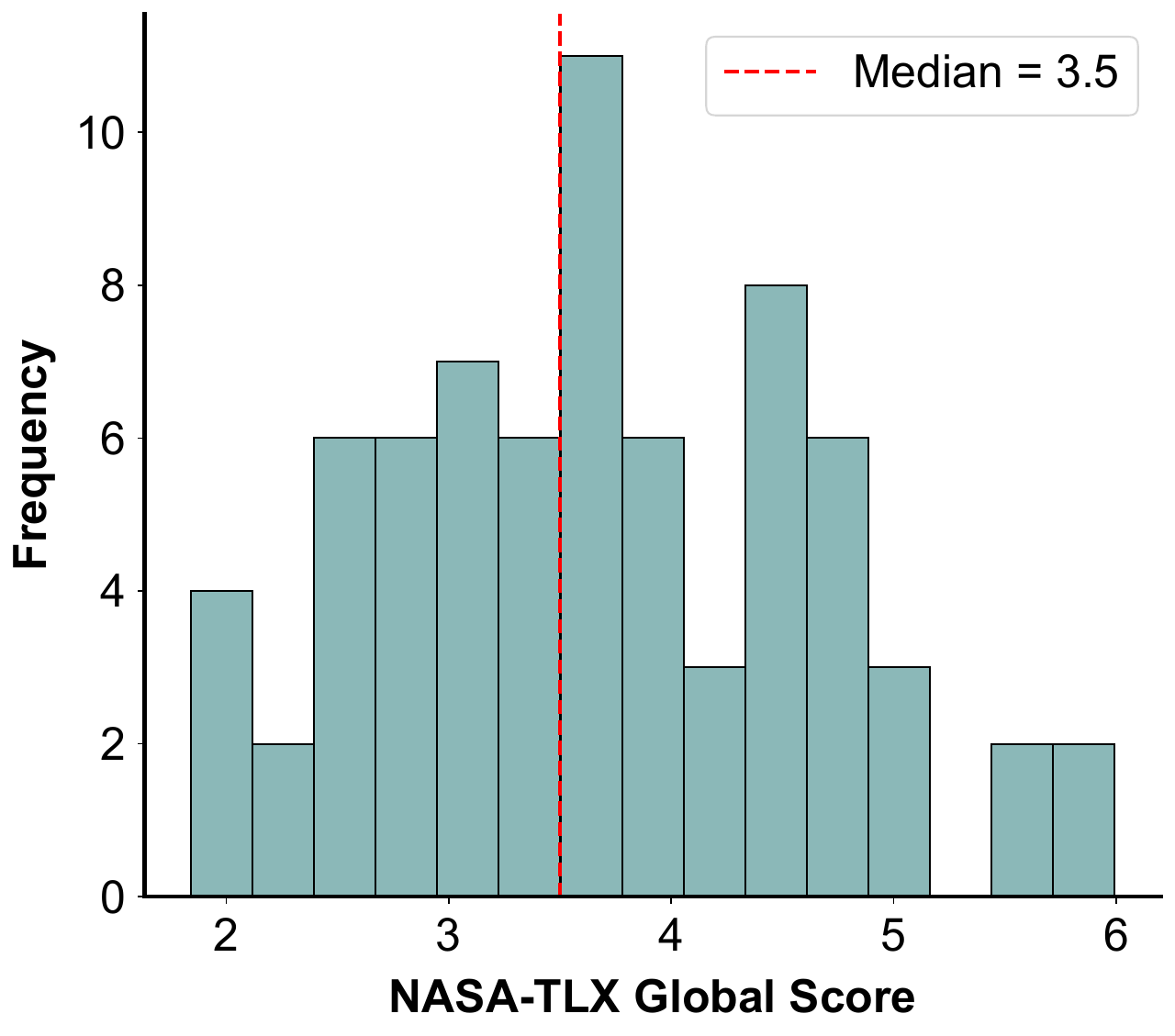}}
    \caption{\textbf{NASA-TLX distribution.}}
    \label{fig:tlx_histogram}
\end{subfigure}
\hfill
\begin{subfigure}[t]{0.32\linewidth}
    \centering
    \vspace{0pt}
    \fbox{\includegraphics[width=\linewidth,height=4cm,keepaspectratio=false]{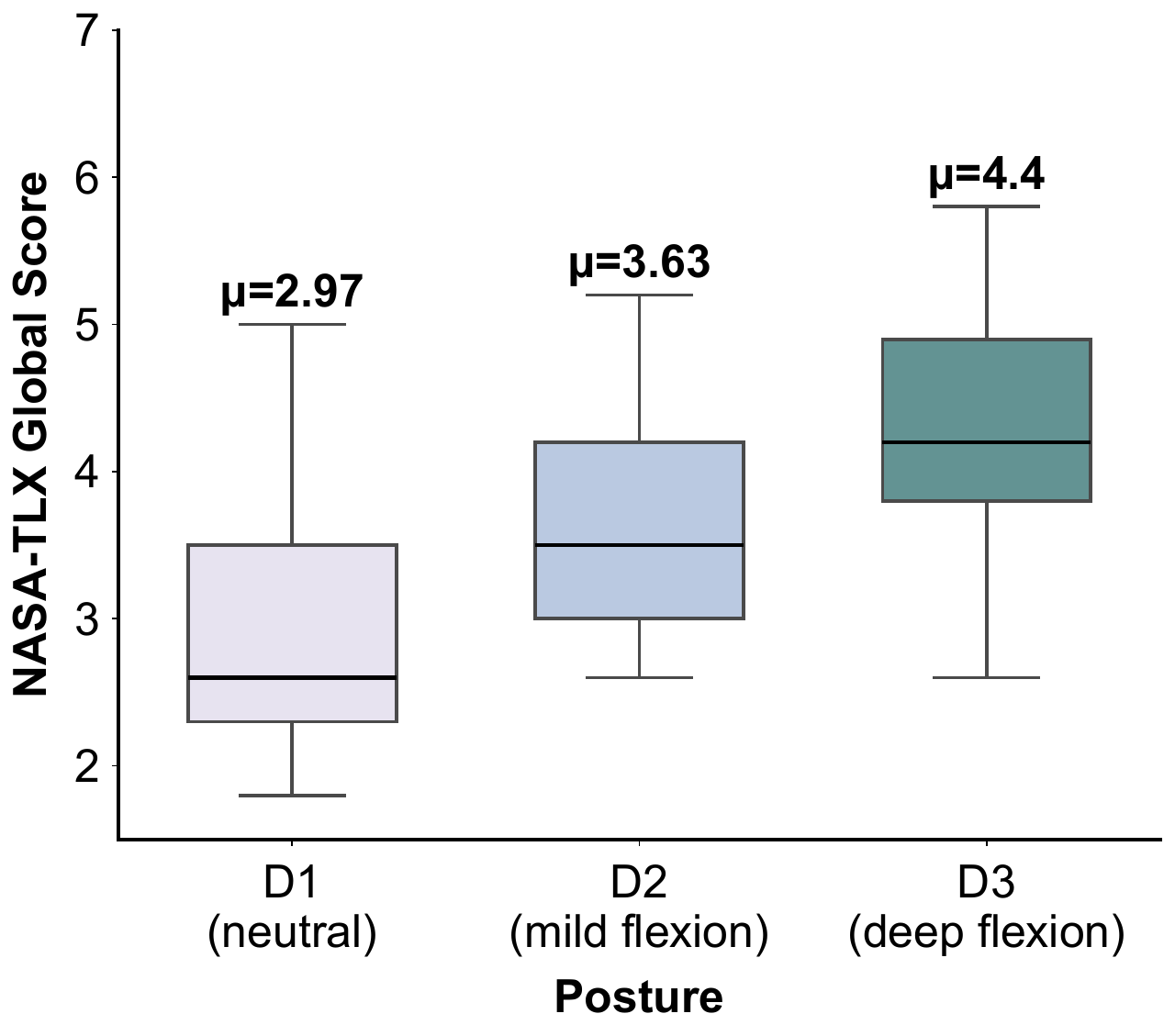}}
    \caption{\textbf{NASA-TLX by posture.}}
    \label{fig:tlx_posture_boxplot}
\end{subfigure}
\hfill
\begin{subfigure}[t]{0.32\linewidth}
    \centering
    \vspace{0pt}
    \fbox{\includegraphics[width=\linewidth,height=4cm,keepaspectratio=false]{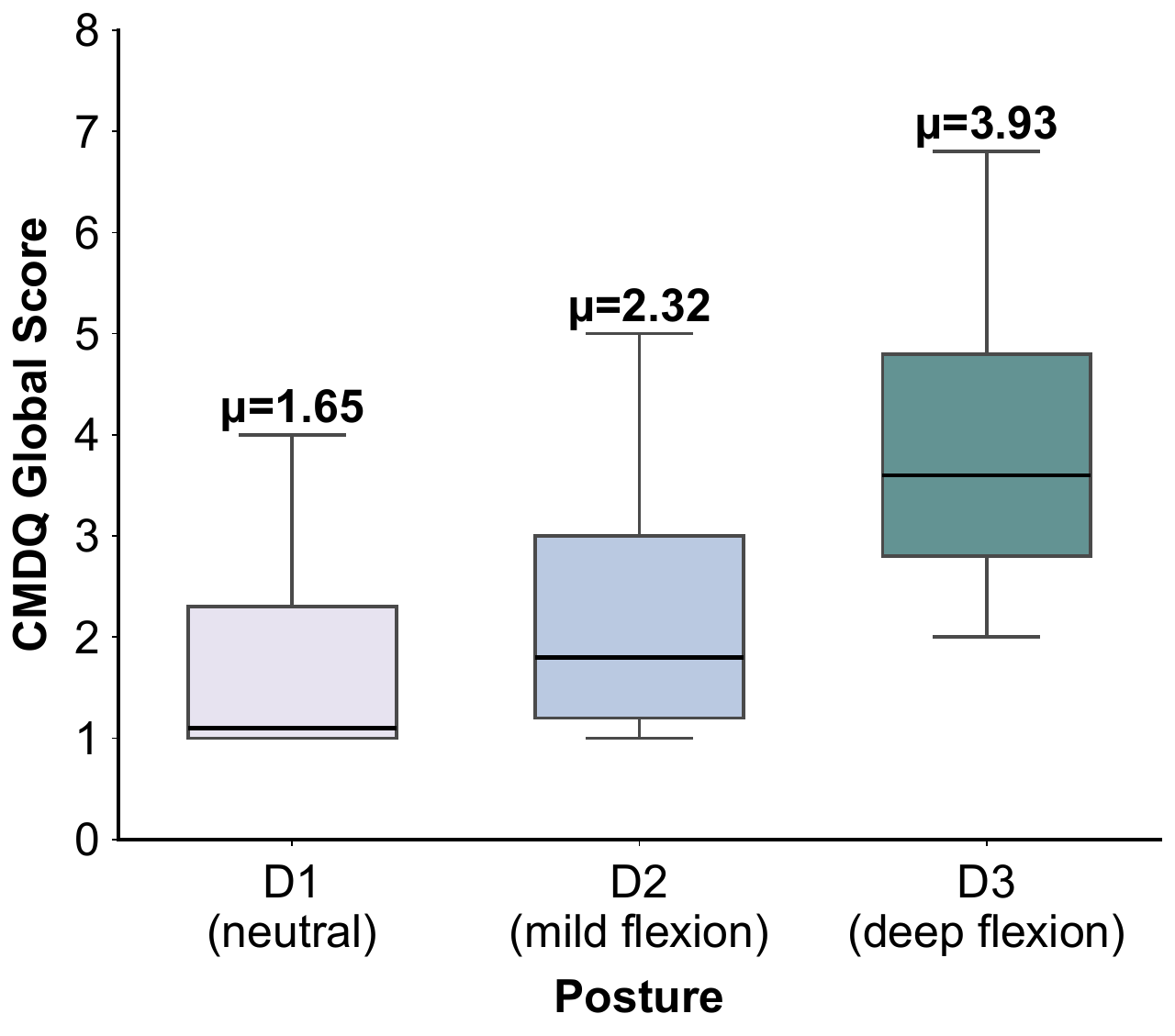}}
    \caption{\textbf{CMDQ by posture.}}
    \label{fig:cmdq_posture_boxplot}
\end{subfigure}

\caption{Subjective workload and discomfort measures across postures.
(a) Global NASA-TLX score distribution. 
(b) NASA-TLX global scores by posture. 
(c) CMDQ scores indicating increased physical discomfort during deeper neck flexion states.}
\label{fig:subjective_measures}
\end{figure*}

We measure perceived cognitive workload using NASA-TLX and 
physical discomfort using CMDQ on 7-point scales, both 
showing good internal consistency (TLX: $\alpha = .76$; 
CMDQ: $\alpha = .89$). TLX scores indicate moderate 
workload ($M = 3.67$, $SD = 1.01$) and CMDQ scores reflect 
mild to moderate physical discomfort ($M = 2.63$, 
$SD = 1.54$). We split TLX scores at the median (3.50) 
into low- and high-discomfort trials 
(Figure~\ref{fig:tlx_histogram}). A Kruskal--Wallis test 
confirms that posture significantly affects TLX 
($H = 23.76$, $p < .001$), with deeper neck flexion ($D3$) 
associated with higher perceived strain (Figures~\ref{fig:tlx_posture_boxplot},~\ref{fig:cmdq_posture_boxplot}).

\paragraph{\textbf{Unimodal Perceived Discomfort 
Classification}} For IMU-based models, RF achieves the highest within-subject 
accuracy (.79) under 80/20 split, and GB provides the most 
robust cross-participant accuracy (.65) under LOSO 
evaluation. \rev{The linear kernel SVM achieves .57 under 
80/20 and .57 under LOSO evaluation.} For EEG-based 
classification, RF achieves the strongest performance (.71 
under 80/20 and .64 under LOSO evaluation), with SVM and 
GB performing worse. \rev{The linear kernel SVM achieves 
.43 under 80/20 and .55 under LOSO evaluation.}

\paragraph{\textbf{Multimodal Perceived Discomfort 
Classification}}
We combine EEG and IMU features in a multimodal setting 
(Section~\ref{para:multimodalFeature}). Under 80/20 
evaluation, RF and GB both achieve the strongest 
within-subject accuracy (.75), and under LOSO evaluation 
RF provides the best cross-participant accuracy (.58), 
followed by GB (.56) and SVM (.53). \rev{The multimodal 
linear kernel SVM achieves .67 under 80/20 evaluation but 
drops to .42 under LOSO evaluation.} All results are 
summarized in Table~\ref{tab:discomfort_models_full}.

\paragraph{\textbf{Perceived Discomfort Classification 
Features}}
\rev{Feature rankings are summarized in 
Tables~\ref{tab:unimodal_feature_importance} 
and~\ref{tab:multimodal_feature_importance}. For IMU-only 
recognition, the most influential features are pitch mean 
and dominant frequency, followed by skewness, spectral 
entropy, and kurtosis, suggesting that perceived discomfort 
is associated with sustained head flexion magnitude and 
postural adjustment dynamics. For EEG-only recognition, 
dominant features include relative beta and alpha power, 
theta band coherence, Hjorth complexity, and higher-order 
statistical descriptors. Under LOSO evaluation, the EEG 
feature set shifts toward waveform shape statistics, Hjorth 
measures, and band relative power terms. In the multimodal 
setting, the strongest predictors combine frontal EEG 
descriptors such as relative theta power, TBR, kurtosis, 
Hjorth mobility, and \texttt{pitch\_mean\_corr}, consistent 
across both RF and Linear SVM.}

\noindent\textbf{Key Takeaway}: Both posture and \rev{task-induced cognitive load} contribute to user-perceived discomfort, as reflected in NASA-TLX and CMDQ scores.Multimodal (IMU + EEG) models outperform unimodal approaches, achieving better accuracy and generalization across users.

\rev{All ML evaluations establish the technical validity of ErgoAssist components and alert policies (Sec.~\ref{sec:ergosystem}). Specifically, posture classification verifies the reliability of detecting non-neutral head configurations, while EEG-based inference confirms the feasibility of estimating task-induced cognitive load. Further, the observed correlation between posture deviation and cognitive state provides empirical support for our core design assumption. Together, these results justify the integration of posture sensing and cognitive awareness in enabling context-aware alert scheduling.}

\begin{table}[t]

\centering

\caption{Discomfort classification performance across modalities and classifiers.}

\label{tab:discomfort_models_full}

\footnotesize

\setlength{\tabcolsep}{6pt}
\renewcommand{\arraystretch}{1.15}

\begin{tabular}{@{}llccccc@{}}

\toprule

\rowcolor[HTML]{D9ECE8}
\textbf{Modality} &
\textbf{Classifier} &
\textbf{Method} &
\textbf{Accuracy} &
\textbf{Precision} &
\textbf{Recall} &
\textbf{F1-score} \\

\midrule

\multirow{8}{*}{IMU only}

& RF  & 80/20 & \textbf{.786} & \textbf{.857} & \textbf{.750} & \textbf{.800} \\

& RF  & LOSO  & .652 & .743 & .634 & .684 \\

& SVM & 80/20 & .571 & .667 & .500 & .571 \\

& SVM & LOSO  & .551 & .647 & .537 & .587 \\

& GB  & 80/20 & .643 & .800 & .500 & .615 \\

& GB  & LOSO  & \textbf{.652} & \textbf{.730} & \textbf{.659} & \textbf{.692} \\

& \rev{Linear SVM} & \rev{80/20} & \rev{.571} & \rev{.625} & \rev{.604} & \rev{.562} \\

& \rev{Linear SVM} & \rev{LOSO}  & \rev{.565} & \rev{.582} & \rev{.583} & \rev{.565} \\

\midrule

\multirow{8}{*}{EEG only}

& RF  & 80/20 & \textbf{.714} & \textbf{.750} & \textbf{.750} & \textbf{.750} \\

& RF  & LOSO  & \textbf{.638} & \textbf{.750} & \textbf{.585} & \textbf{.658} \\

& SVM & 80/20 & .357 & .429 & .375 & .400 \\

& SVM & LOSO  & .551 & .647 & .537 & .587 \\

& GB  & 80/20 & .429 & .500 & .500 & .500 \\

& GB  & LOSO  & .536 & .655 & .463 & .543 \\

& \rev{Linear SVM} & \rev{80/20} & \rev{.429} & \rev{.438} & \rev{.438} & \rev{.429} \\

& \rev{Linear SVM} & \rev{LOSO}  & \rev{.551} & \rev{.552} & \rev{.554} & \rev{.547} \\

\midrule

\multirow{8}{*}{Multimodal}

& RF         & 80/20 & \textbf{.750} & \textbf{.824} & \textbf{.824} & \textbf{.824} \\

& RF         & LOSO  & .579          & .625          & .625          & .625          \\

& SVM        & 80/20 & .583          & .737          & .737          & .737          \\

& SVM        & LOSO  & .526          & .658          & .658          & .658          \\

& GB         & 80/20 & \textbf{.750} & \textbf{.824} & \textbf{.824} & \textbf{.824} \\

& GB         & LOSO  & .561          & .648          & .648          & .648          \\

& \rev{Linear SVM} & \rev{80/20} & \rev{.667} & \rev{.657} & \rev{.657} & \rev{.657} \\

& \rev{Linear SVM} & \rev{LOSO}  & \rev{.421} & \rev{.369} & \rev{.393} & \rev{.372} \\

\bottomrule

\end{tabular}

\end{table}

\begin{table*}[t]

\footnotesize
\centering

\caption{Top unimodal features for perceived discomfort classification under 80/20 and LOSO evaluation. For RF and GB, values denote model based feature importance scores. \rev{For Linear SVM, values denote normalized absolute coefficient based importance scores.}}

\label{tab:unimodal_feature_importance}

\renewcommand{\arraystretch}{1.15}

\begin{tabular}{lllp{0.54\linewidth}}

\toprule

\rowcolor[HTML]{D9ECE8}
\textbf{Modality} &
\textbf{Model} &
\textbf{Eval} &
\centering\arraybackslash \textbf{Top Features (importance)} \\

\midrule

IMU only & RF &
80/20 &
pitch\_mean (.29),
pitch\_dom\_freq (.15),
pitch\_skew (.14),
pitch\_spec\_entropy (.10),
pitch\_hjorth\_complexity (.07)
\\

& GB &
80/20 &
pitch\_mean (.27),
pitch\_dom\_freq (.26),
pitch\_std (.13),
pitch\_velocity\_mean (.13),
pitch\_spec\_entropy (.06)
\\

& \rev{Linear SVM} &
\rev{80/20} &
\rev{pitch\_spec\_entropy (1.00),
pitch\_mean (.83),
pitch\_dom\_freq (.62),
pitch\_hjorth\_mobility (.38),
pitch\_hjorth\_complexity (.34)}
\\

& RF &
LOSO &
pitch\_mean (.25),
pitch\_dom\_freq (.16),
pitch\_skew (.14),
pitch\_spec\_entropy (.09),
pitch\_kurt (.08)
\\

& GB &
LOSO &
pitch\_mean (.29),
pitch\_dom\_freq (.23),
pitch\_skew (.10),
pitch\_spec\_entropy (.09),
pitch\_kurt (.08)
\\

& \rev{Linear SVM} &
\rev{LOSO} &
\rev{pitch\_mean (.95),
pitch\_dom\_freq (.55),
pitch\_skew (.48),
pitch\_kurt (.43),
pitch\_std (.35)}
\\

\midrule

EEG only & RF &
80/20 &
ch3\_hjorth\_complexity\_mean (.18),
ch4\_mean\_mean (.14),
ch5\_beta\_rel\_mean (.13),
ch3\_beta\_rel\_mean (.13),
ch1\_ch3\_coh\_theta\_std (.13)
\\

& GB &
80/20 &
ch3\_beta\_rel\_mean (.27),
ch2\_hjorth\_complexity\_min (.16),
ch3\_hjorth\_complexity\_mean (.15),
ch5\_beta\_rel\_mean (.13),
ch5\_spec\_entropy\_max (.11)
\\

& \rev{Linear SVM} &
\rev{80/20} &
\rev{ch3\_hjorth\_complexity\_mean (1.00),
ch5\_beta\_rel\_mean (.96),
ch5\_alpha\_rel\_std (.93),
ch5\_hjorth\_complexity\_mean (.67),
ch3\_beta\_rel\_mean (.60)}
\\

& RF &
LOSO &
ch4\_mean\_mean (.15),
ch2\_hjorth\_mobility\_std (.13),
ch1\_ch3\_coh\_theta\_std (.13),
ch3\_hjorth\_complexity\_mean (.12),
ch5\_kurt\_mean (.12)
\\

& GB &
LOSO &
ch1\_ch3\_coh\_theta\_std (.19),
ch4\_mean\_mean (.19),
ch5\_kurt\_mean (.16),
ch4\_hjorth\_complexity\_mean (.13),
ch2\_alpha\_rel\_std (.13)
\\

& \rev{Linear SVM} &
\rev{LOSO} &
\rev{ch3\_skew\_max (.58),
ch4\_mean\_mean (.58),
ch2\_beta\_rel\_min (.51),
ch2\_hjorth\_mobility\_std (.49),
ch2\_tbr\_mean (.44)}
\\

\bottomrule

\end{tabular}

\end{table*}

\begin{table}[t]

\footnotesize
\centering

\caption{Top multimodal features for discomfort classification under 80/20 random split and LOSO evaluation. Values denote model specific feature importance measures. For RF, SVM, and GB, values denote model based feature importance scores. \rev{For Linear SVM, values denote normalized absolute coefficient based importance scores.}}

\label{tab:multimodal_feature_importance}

\renewcommand{\arraystretch}{1.15}

\begin{tabular}{llp{0.62\linewidth}}

\toprule

\rowcolor[HTML]{D9ECE8}
\textbf{Setting} &
\textbf{Model} &
\centering\arraybackslash \textbf{Top Features (importance)} \\

\midrule

80/20 &

RF &
OTE\_L-FpZ\_theta\_rel\_mean (.17),
RF-FpZ\_tbr\_min (.14),
OTE\_L-FpZ\_kurtosis\_min (.14),
OTE\_R-FpZ\_hjorth\_complexity\_max (.10),
RF-FpZ\_tbr\_std (.08)
\\

&

SVM &
OTE\_L-FpZ\_theta\_rel\_mean (.01),
OTE\_R-FpZ\_hjorth\_mobility\_max (.01),
pitch\_mean\_corr (.003),
RF-FpZ\_tbr\_min (.00),
OTE\_L-FpZ\_kurtosis\_min ($-.07$)
\\

&

GB &
OTE\_L-FpZ\_theta\_rel\_mean (.20),
pitch\_mean\_corr (.18),
RF-FpZ\_tbr\_min (.13),
OTE\_L-FpZ\_kurtosis\_min (.11),
OTE\_L-FpZ\_tbr\_std (.10)
\\

&

\rev{Linear SVM} &
\rev{OTE\_L-FpZ\_tbr\_std (1.00),
pitch\_mean\_corr (.97),
OTE\_L-FpZ\_hjorth\_mobility\_max (.90),
OTE\_L-FpZ\_theta\_rel\_mean (.48),
RF-FpZ\_tbr\_std (.42)}
\\

\midrule

LOSO &

RF &
pitch\_mean\_corr (.23),
OTE\_L-FpZ\_theta\_rel\_mean (.21),
LF-FpZ\_skewness\_max (.17),
OTE\_L-FpZ\_skewness\_max (.15),
OTE\_L-FpZ\_kurtosis\_mean (.12)
\\

&

SVM &
OTE\_R-FpZ\_theta\_rel\_mean (.06),
RF-FpZ\_skewness\_std (.02),
OTE\_L-FpZ\_theta\_rel\_mean (.02),
OTE\_L-FpZ\_kurtosis\_mean (.01),
coh\_OTE\_LR\_alpha\_mean (.01)
\\

&

GB &
pitch\_mean\_corr (.30),
OTE\_L-FpZ\_theta\_rel\_mean (.25),
LF-FpZ\_skewness\_max (.17),
RF-FpZ\_skewness\_std (.16),
OTE\_L-FpZ\_abr\_min (.16)
\\

&

\rev{Linear SVM} &
\rev{OTE\_L-FpZ\_kurtosis\_mean (.77),
OTE\_L-FpZ\_kurtosis\_min (.71),
OTE\_R-FpZ\_hjorth\_complexity\_max (.44),
OTE\_R-FpZ\_tbr\_std (.42),
RF-FpZ\_theta\_rel\_mean (.41)}
\\

\bottomrule

\end{tabular}

\end{table}
\normalsize
\subsection{Energy Consumption}
We quantify on-device computational cost by measuring 
each model's serialized size and inference latency. RF 
models achieve the highest classification accuracy but 
incur higher computational overhead, with inference 
latency of 13--15\,ms and memory footprints up to 
0.7\,MB. In contrast, SVM requires only 0.03\,ms per 
inference and 5.5\,KB of memory, approximately 400$\times$ 
lower latency than RF. These results indicate that 
lightweight classifiers are well suited for long-term 
on-device operation, and our prototype can support 
extended user monitoring with modest energy consumption 
(Table~\ref{tab:energy_consumption}).
\\\noindent\textbf{Key Takeaway:} All models run with low latency (13--15\,ms) and 
modest memory footprint, supporting real-time on-device 
deployment on resource-constrained wearable platforms.
\begin{table}[htbp]

\centering

\caption{Model footprint and runtime metrics for on-device inference. We report mean inference latency, 95th percentile latency (p95), and serialized model size for IMU, EEG, and multimodal classifiers under 80/20 and LOSO evaluation.}

\label{tab:energy_consumption}

\footnotesize

\setlength{\tabcolsep}{5pt}
\renewcommand{\arraystretch}{1.10}

\begin{tabular}{lllccc}

\toprule

\rowcolor[HTML]{D9ECE8}
\textbf{Modality} &
\textbf{Classifier} &
\textbf{Method} &
\textbf{Latency (ms)} &
\textbf{p95 (ms)} &
\textbf{Size (KB)} \\

\midrule

IMU & RF  & 80/20 & 13.916 & 14.964 & 67.08 \\
IMU & RF  & LOSO  & 12.885 & 13.590 & 523.53 \\
IMU & SVM & 80/20 & \textbf{.035} & \textbf{.062} & \textbf{5.31} \\
IMU & SVM & LOSO  & \textbf{.031} & \textbf{.035} & \textbf{5.89} \\
IMU & GB  & 80/20 & .074 & .082 & 116.96 \\
IMU & GB  & LOSO  & .074 & .093 & 113.44 \\

\midrule

EEG & RF  & 80/20 & 13.834 & 14.757 & 66.45 \\
EEG & RF  & LOSO  & 13.190 & 13.674 & 516.03 \\
EEG & SVM & 80/20 & \textbf{.032} & \textbf{.032} & \textbf{5.11} \\
EEG & SVM & LOSO  & \textbf{.039} & .104 & \textbf{5.50} \\
EEG & GB  & 80/20 & .074 & .105 & 131.87 \\
EEG & GB  & LOSO  & .076 & .097 & 128.50 \\

\midrule

Multimodal & RF  & 80/20 & 12.976 & 13.581 & 719.75 \\
Multimodal & RF  & LOSO  & 14.116 & 15.679 & 141.22 \\
Multimodal & SVM & 80/20 & \textbf{.028} & \textbf{.032} & \textbf{5.58} \\
Multimodal & SVM & LOSO  & \textbf{.028} & \textbf{.029} & \textbf{5.48} \\
Multimodal & GB  & 80/20 & .094 & .098 & 490.02 \\
Multimodal & GB  & LOSO  & .091 & .092 & 322.03 \\

\bottomrule

\end{tabular}

\normalsize

\end{table}

\normalsize

%% file: acmart-primary/UserFeasibilityMHCI_REvision.tex
\section{\rev{\system in the Wild: A Preliminary Real-Time Comparative Study}}
\label{para:pilotstudy}
\rev{We conduct a preliminary exploratory study to compare 
two real-time alert policies: posture-only (P-only) and 
cognition-aware (P+C), as described in 
Section~\ref{sec:algorithmposturecognition}. We recruit 
thirteen participants, including five from the earlier 
data collection and eight newly recruited. We explicitly 
evaluate system behavior on participants whose data was 
not used during model development (P6--P13) to assess 
cross-user robustness~\cite{wang2024ubiphysio,
cen2022exploring}. Each participant completes both 
conditions in a randomized order to control for order 
effects.} Each session begins with a 1-minute baseline 
phase for calibration, followed by a 3-minute numerical calculation task. Calibration is repeated before every session because participants may change posture or move between tasks, making a continuous baseline unreliable \cite{donaldson2021within}.
Participants receive visual and haptic alerts and their 
task scores are logged. When an alert is triggered, a 
corrective message and guided stretching 
video\footnote{\url{https://www.youtube.com/watch?v=Q12nIfVCpdU}} 
are sent. Participants adjust their posture freely at 
their own discretion. At the end of each condition, 
participants complete a modified System Usability Scale 
(SUS) and an interruptibility survey, both rated on a 
7-point Likert scale. The full questionnaire items are 
in Appendix~\ref{app:questionnaire}.

\begin{figure*}[t]
\centering
\begin{subfigure}[b]{0.48\textwidth}
\includegraphics[width=\textwidth]{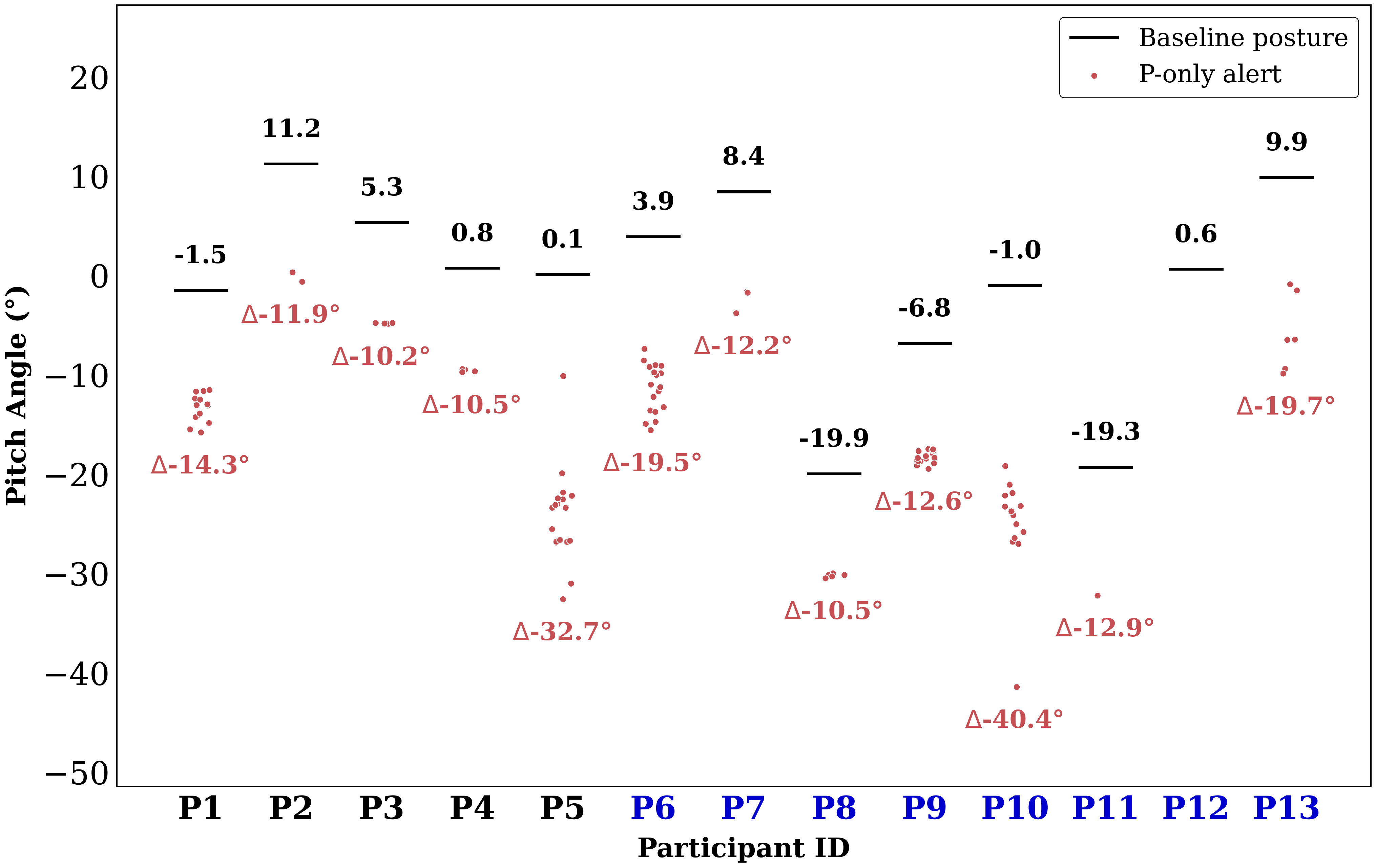}
\caption{\textbf{P-only Alert policy}.}
\label{fig:ponly_pitch}
\end{subfigure}
\hfill
\begin{subfigure}[b]{0.48\textwidth}
\includegraphics[width=\textwidth]{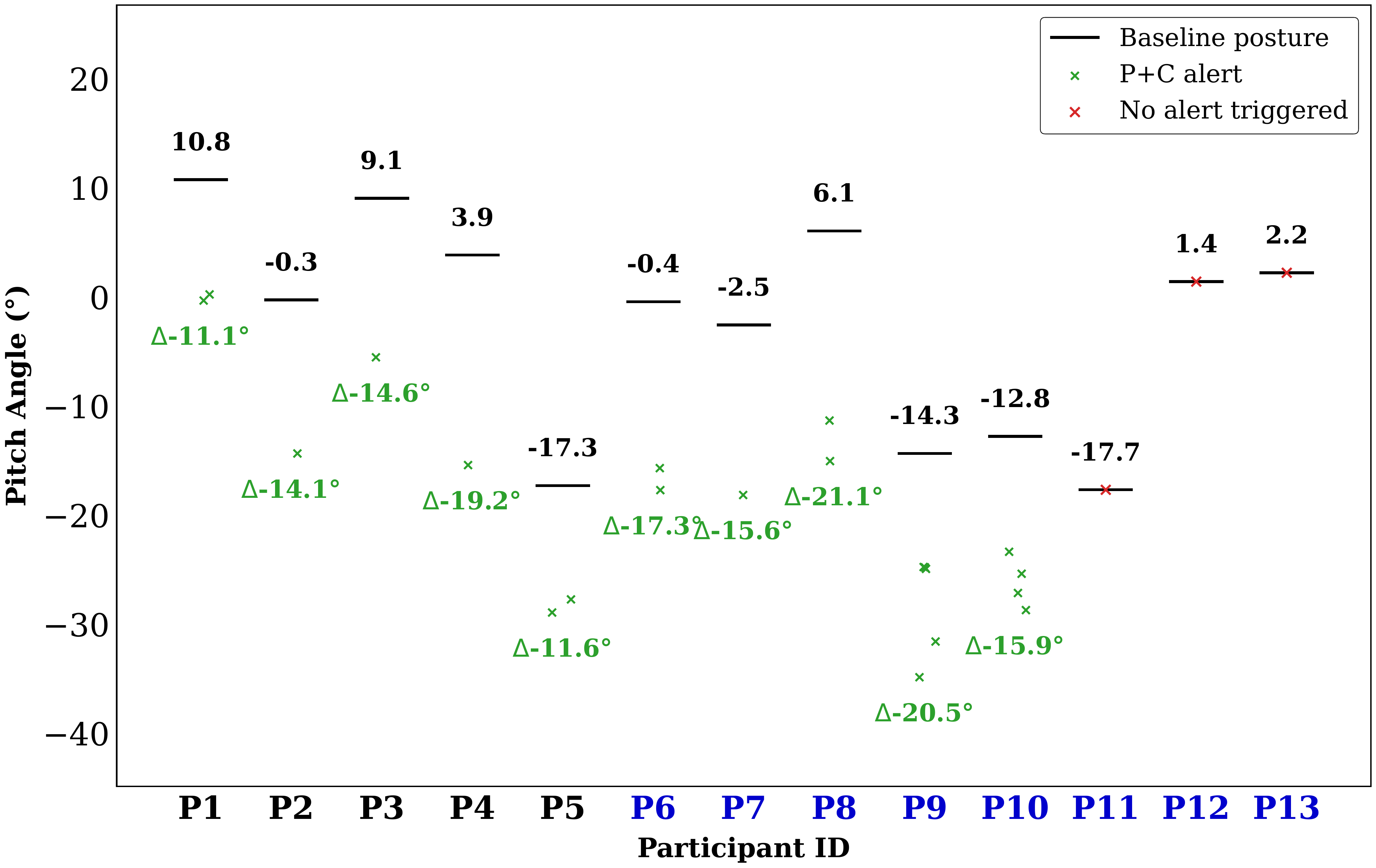}
\caption{\textbf{P+C Alert policy.}}
\label{fig:pc_pitch}
\end{subfigure}
\caption{\rev{Comparative alert and postural deviation analysis. The P+C policy triggers significantly fewer alerts by considering cognitive load compared to the P-only alert policy. P6 to P13 are unseen participants.}}
\end{figure*}

\paragraph{\textbf{\rev{Experimental Results}}}
\rev{From Table~\ref{tab:alert_comparison} and 
Figures~\ref{fig:ponly_pitch} and \ref{fig:pc_pitch}, 
alerts decrease from 9.6 (P-only) to 1.8 (P+C) across 
$N=13$ participants, corresponding to an 81\% reduction 
consistent across known and unseen groups. For example, 
alerts decrease from 18 to 2 for P6 and from 16 to 1 for 
P9. P+C alerts consistently exceed the pitch deviation 
threshold, confirming that the reduction reflects improved 
timing rather than relaxed criteria. P13 shows zero P+C 
alerts, indicating selective suppression when 
interruptibility conditions are not met. The mean rest 
probability ranges from .50 to .77, confirming that alerts 
are delivered during lower cognitive load periods. 
Consistent trends for unseen participants (P6--P13) 
confirm generalizability beyond participant-specific 
fitting.}

\begin{table}[t]
\centering
\caption{\rev{Per-participant alert statistics for 
posture-only (P-only) and posture + cognition (P+C) 
policies. \textbf{\# Alerts}: total alerts per session. 
\textbf{Mean sustain time}: average duration (s) from 
alert onset to user dismissal. \textbf{Mean rest 
probability}: estimated probability of low cognitive load 
at alert delivery (P+C only). P13 receives no P+C alerts 
because sustained bad posture does not coincide with an 
interruptible cognitive state.}}

\label{tab:alert_comparison}
\footnotesize
\renewcommand{\arraystretch}{1.12}

\resizebox{\linewidth}{!}{

\begin{tabular}{lcccccc}

\toprule

\rowcolor[HTML]{D9ECE8}
\textbf{Participant} &
\multicolumn{2}{c}{\textbf{P-only}} &
\multicolumn{3}{c}{\textbf{P+C Policy}} \\

\cmidrule(lr){2-3}
\cmidrule(lr){4-6}

\rowcolor[HTML]{D9ECE8}
&
\# \textbf{Alerts} &
\textbf{Mean sustain time (s)} &
\textbf{\# Alerts }&
\textbf{Mean sustain time (s)} &
\textbf{Mean rest probability} \\

\midrule

P1 (known)   & 15 & 2.20 & 2 & 2.83 & .61 \\
P2 (known)   & 5  & 5.41 & 1 & 5.72 & .76 \\
P3 (known)   & 4  & 2.47 & 1 & 2.97 & .70 \\
P4 (known)   & 4  & 3.36 & 1 & 5.73 & .77 \\
P5 (known)   & 18 & 1.96 & 2 & 3.25 & .72 \\

\midrule

P6 (unseen)  & 18 & 3.25 & 2 & 5.42 & .70 \\
P7 (unseen)  & 7  & 2.72 & 1 & 7.96 & .50 \\
P8 (unseen)  & 3  & 4.97 & 2 & 2.87 & .75 \\

\rev{P9 (unseen)}  & \rev{16} & \rev{1.58} & \rev{1} & \rev{3.01} & \rev{.75} \\

\rev{P10 (unseen)} & \rev{14} & \rev{1.96} & \rev{7} & \rev{2.19} & \rev{.67} \\

\rev{P11 (unseen)} & \rev{1}  & \rev{2.13} & \rev{2} & \rev{2.57} & \rev{.75} \\

\rev{P12 (unseen)} & \rev{14} & \rev{2.26} & \rev{1} & \rev{4.35} & \rev{.52} \\

\rev{P13 (unseen)} & \rev{6}  & \rev{2.30} & \rev{0} & \rev{--}   & \rev{--}   \\

\bottomrule

\end{tabular}

}

\end{table}
\normalsize

\begin{table*}[t]
\centering
\small

\caption{\revdb{Participant-level alert exposure and normalized posture response outcomes for five follow-up participants (P9--P13). \textbf{Response rate} is the normalized posture response, computed as corrected alerts divided by total issued alerts. \textbf{Mean pitch deviation at alert} is the mean absolute deviation of head pitch from each participant's calibrated neutral baseline at alert onset, $|\text{pitch}_{\text{before}}-\text{baseline}|$. Alert counts differ across participants and policies because P-only triggers whenever sustained poor posture crosses the threshold, whereas P+C additionally requires alignment with a low-cognitive-load window or the 30\,s override. Response rate normalises for these unequal exposures. Full per-alert pitch and recovery records are in Appendix~\ref{app:per_alert_details}.}}
\label{tab:per_alert}

\vspace{-4pt}
\setlength{\tabcolsep}{2pt}
\renewcommand{\arraystretch}{1.12}

\resizebox{\textwidth}{!}{%
\begin{tabular}{@{}l c c c c c c c c@{}}
\toprule
\rowcolor[HTML]{D9ECE8}
& \multicolumn{4}{c}{\textbf{\revdb{P-only}}}
& \multicolumn{4}{c}{\textbf{\revdb{P+C Policy}}} \\
\cmidrule(lr){2-5}\cmidrule(lr){6-9}

\rowcolor[HTML]{D9ECE8}
\textbf{\revdb{Participant}}
& \textbf{\revdb{\shortstack{\#\\Alerts}}}
& \textbf{\revdb{\shortstack{\#\\Corrected}}}
& \textbf{\revdb{\shortstack{Response\\rate}}}
& \textbf{\revdb{\shortstack{Mean pitch\\dev.\ ($^\circ$)}}}
& \textbf{\revdb{\shortstack{\#\\Alerts}}}
& \textbf{\revdb{\shortstack{\#\\Corrected}}}
& \textbf{\revdb{\shortstack{Response\\rate}}}
& \textbf{\revdb{\shortstack{Mean pitch\\dev.\ ($^\circ$)}}} \\
\midrule
 
\revdb{P9}
& \revdb{16} & \revdb{16} & \revdb{100.0\%} & \revdb{11.44}
& \revdb{1}  & \revdb{1}  & \revdb{100.0\%} & \revdb{10.73} \\
 
\revdb{P10}
& \revdb{14} & \revdb{1}  & \revdb{7.1\%}   & \revdb{24.08}
& \revdb{7}  & \revdb{7}  & \revdb{100.0\%} & \revdb{11.30} \\
 
\revdb{P11}
& \revdb{1}  & \revdb{1}  & \revdb{100.0\%} & \revdb{12.92}
& \revdb{2}  & \revdb{2}  & \revdb{100.0\%} & \revdb{12.59} \\
 
\revdb{P12}
& \revdb{14} & \revdb{5}  & \revdb{35.7\%}  & \revdb{16.00}
& \revdb{1}  & \revdb{1}  & \revdb{100.0\%} & \revdb{17.46} \\
 
\revdb{P13}
& \revdb{6}  & \revdb{4}  & \revdb{66.7\%}  & \revdb{15.62}
& \revdb{0}  & \revdb{0}  & \revdb{N/A$^\dagger$} & \revdb{--} \\
 
\midrule
\multicolumn{1}{@{}l}{\textbf{\revdb{\shortstack[l]{Participant-level Mean\\(N=5)}}}}
& \textbf{\revdb{10.2}} & \textbf{\revdb{5.4}}  & \textbf{\revdb{61.9\%}}
& \textbf{\revdb{16.01}}
& \textbf{\revdb{2.2}}  & \textbf{\revdb{2.2}}  & \textbf{\revdb{100.0\%$^\dagger$}}
& \textbf{\revdb{13.02$^\dagger$}} \\
 
\bottomrule
\end{tabular}
}
\vspace{-3pt}
\begin{flushleft}
\footnotesize
\revdb{$^\dagger$ P13 receives no P+C alert during the session and is excluded from P+C response-rate and mean pitch-deviation averages.}
\end{flushleft}

\end{table*}

\paragraph{\revdb{\textbf{Tradeoff between interruptibility and ergonomic correction.}}}
\revdb{Across all 13 participants, P+C issues 23 alerts compared with 125 under P-only, corresponding to an 81.6\% reduction in interruption frequency (Table~\ref{tab:alert_comparison}). To examine whether this reduction preserves ergonomic benefit, we conduct a post-hoc study for five unseen participants (P9--P13). Table~\ref{tab:per_alert} reports participant-level alert counts, normalized response rates, and pitch deviations, along with full per-alert pitch and recovery records in Appendix~\ref{app:per_alert_details}. Despite heterogeneous alert counts, a consistent pattern of response rate shows that P+C substantially reduces alert overall while maintaining ergonomic benefits.}

\revdb{\textbf{Participant-Level Response Rate Analysis}: To assess ergonomic benefit and interruptibility, we use the normalized mean response rate as the primary proxy metric. Here, response rate is computed as corrected alerts divided by total issued alerts for each participant. From Table~\ref{tab:per_alert}, P+C reduces the mean participant-level alert count from 10.2 to 2.2 per session, corresponding to a 78.4\% reduction in alert exposure. The mean participant-level response rate increases from 61.9\% under P-only to 100.0\% under P+C. Across this subset, P+C consistently achieves \textbf{equal or higher response rates} per alert while delivering substantially fewer interruptions. This pattern holds across different participant profiles. Participants with low P-only responsiveness benefit most. P10 responds to only 1 of 14 P-only alerts (7.1\%) but to all 7 P+C alerts (100.0\%), consistent with alert fatigue under posture-threshold triggering. Similarly, P12 improves from 5 of 14 P-only alerts (35.7\%) to 1 of 1 P+C alerts (100.0\%). Participants who already respond consistently under P-only also benefit through reduced alert burden. P9 and P11 already achieve 100.0\% response under P-only. In these cases, P+C provides no additional gain in response rate, but it reduces alert exposure; for example, P9's alerts decrease from 16 to 1 with no loss in correction outcome. P11 receives two P+C alerts, both triggered by the 30,s override, confirming that the override remains active when posture does not recover within the permitted duration. P13 receives no P+C alert because the participant's poor-posture episodes did not persist long enough to trigger the override and did not coincide with a low-cognitive-load interruptible window.}

\revdb{\textbf{Variation in Alert Exposure and Participant Response}:
In Table~\ref{tab:per_alert}, alert counts and response rates differ
across participants for three reasons. First, participants differ in
neutral posture, sitting habits, pain sensitivity, musculature, and
tolerance for sustained flexion. These individual differences determine
how often each participant enters poor posture and how long each
participant remains in it. Identical alert criteria therefore expose
participants to different alert frequencies. Second, the two policies
trigger alerts under different conditions. P-only issues an alert upon
detecting sustained non-neutral posture. P+C applies the same posture
criterion together with an evaluation of estimated task-induced
cognitive load. P+C delays alerts during high-load periods unless poor
posture persists for $T_{\text{cont}} = 30$\,s. A participant sustaining
high cognitive load throughout the task therefore receives fewer P+C
alerts than a participant with moderate or variable load, provided the
override never triggers. The fixed override also interacts with
individual posture behavior. Participants recovering from poor posture
before the 30\,s window may receive few or no P+C alerts. Participants
remaining in poor posture beyond 30\,s receive a P+C alert even during
high cognitive load, because the override limits prolonged uncorrected
posture. Third, participants respond to alerts differently. Some
participants ignore frequent alerts under P-only. This pattern is
consistent with alert fatigue~\cite{ancker2017effects}. The same
participants respond reliably to the less frequent and better-timed
alerts under P+C. Other participants correct posture consistently under
both policies.}
\begin{figure}[t]
    \centering
    \includegraphics[width=0.7\columnwidth]{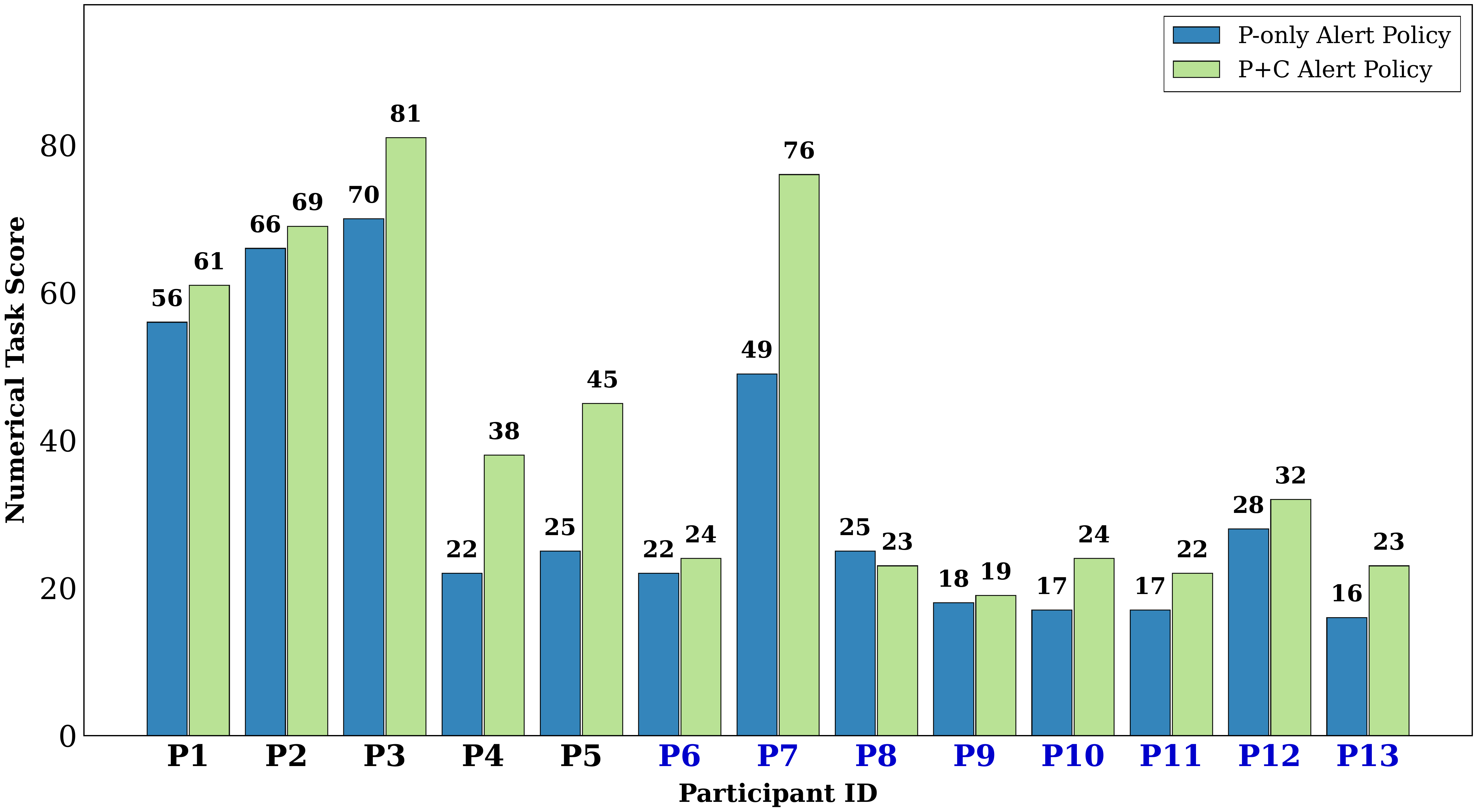}
    \caption{\rev{Comparison of arithmetic task performance between the posture-only (P-only) and cognition-aware (P+C) alert policy.}}
    \Description{Bar chart comparing arithmetic task scores under the posture-only and cognition-aware alert policies across participants.}
    \label{fig:task_performance_comparison}
\end{figure}

\textbf{Override Rationale:} Our choice of a fixed continuous bad-posture override parameter ($T_{\text{cont}} = 30$\,s) is motivated by ergonomics and musculoskeletal research linking risk to posture angle, exposure duration, repetition, and recovery profile, rather than brief isolated  deviations~\cite{santos2025efficacy,bernard1997musculoskeletal,szeto2002field,straker2009relationships,dong2022association}. EMG-based studies similarly show that muscle fatigue manifests as gradual changes in EMG features over sustained postures, on time scales from 120\,s~\cite{igarashi2016effect} to 15 minutes~\cite{guo2021weak}. We therefore treat 30\,s as a practical design bound that limits uncorrected exposure rather than a clinical cutoff, and it can be tightened for users with weaker musculature or pre-existing pain. In our design, cognition-aware alerts preserve focus during high-engagement periods, and the timeout override ensures that prolonged poor posture still triggers intervention when no interruptible state occurs, providing a bounded tradeoff between interruption reduction and ergonomic safety.

\revdb{In summary, alert count alone is an insufficient basis for comparing policies because it does not indicate whether alerts actually lead to ergonomic corrections. Therefore, we use the normalized response rate to capture this trade-off. For our P+C policy, the normalized response rate reveals a consistent trend of maintaining or improving posture responses relative to the P-only policy, while also reducing alert fatigue.}
\paragraph{\rev{\textbf{User experience and interruptibility analysis}.}} \rev {Our preliminary within-subjects evaluation ($N = 13$) shows improved user experience under the P+C policy. The mean System Usability Scale (SUS) increases to $74.7 \pm 9.1$ compared to $52.3 \pm 13.3$ for P-only ($t(12) = 7.56, p < .001$). Perceived interruptibility also improves, with lower disruption scores under P+C (5.28 $\pm$ 0.85) than P-only (2.82 $\pm$ 0.75), yielding a large effect size (Cohen’s $d = 2.07$). These results indicate that context-aware alerting reduces perceived interruption while maintaining task engagement. Given the limited sample size and short session duration, we treat these findings as preliminary. Qualitative feedback further suggests that participants found the P+C system better aligned with task demands and less intrusive during high cognitive load.}

\paragraph{\rev{\textbf{Effectiveness Observation.} Figure \ref{fig:task_performance_comparison} shows improved task performance under the P+C alert policy compared to the P-only baseline. Out of 13 participants, 12 achieved higher arithmetic task scores under the P+C condition. The mean score increases from 33.2 ($SD = 18.9$) under P-only to 41.3 ($SD = 21.8$) under P+C, corresponding to an average improvement of 8.2 points. These results indicate that cognition-aware alert scheduling supports task performance by reducing poorly timed interruptions.}}

\rev{These observations are directionally consistent with prior work on intelligent interruption in HCI, suggesting that context-aware alert scheduling can improve intervention timing and reduce perceived disruption during active tasks \cite{chen2022predicting,goyal2017intelligent}. Our results motivate future longitudinal studies to examine long-term adherence, broader user preference, and the design space of cognition-aware intervention strategies across more diverse real-world settings.}

\noindent\textbf{Key Takeaway}: Cognition-aware alerting significantly reduces alert frequency while improving user experience compared to posture-only feedback. By deferring alerts during high cognitive load and intervening at receptive moments, the system delivers fewer but better-timed interventions without compromising posture correction. Participants consistently perceived cognition-aware alerts as less disruptive, more acceptable, and better aligned with sustained productivity.

\normalsize

%% file: acmart-primary/Discussions_new.tex
\section{Discussion}
Our work focuses on intelligent, adaptive interventions, an often overlooked aspect of posture assistants. Existing systems typically trigger corrections based on minor posture deviations, which can be disruptive and impractical for long-term use. We address this by introducing a cognition-aware alert policy that considers \rev{task-induced cognitive load} in addition to posture.\\
\noindent \textbf{Necessity of Multimodal Sensing:} IMU data reliably captures posture and correlates with physical discomfort, while EEG provides insight into \rev{task-induced cognitive load}. \system combines these modalities by using IMU to detect posture deviations and EEG to time alerts during less disruptive moments. Our real-time study shows that this integration improves user experience by reducing alert burden and increasing perceived usability. \rev{Our findings align with prior work on posture-aware systems \cite{baer2022posture}, which report that frequent threshold-based alerts can lead to alert fatigue and reduced compliance. We show that incorporating cognitive context significantly reduces unnecessary interruptions while maintaining ergonomic benefits.}\\
\noindent \textbf{Enhancing User Experience and Productivity:} The proposed policy reduces alerts by 81\% while maintaining ergonomic effectiveness. This suggests that avoiding interruptions during high-focus periods helps preserve productivity. The model generalizes well to unseen participants, indicating that alerts are both timely and meaningful. These results highlight that ergonomic feedback should be interruptibility-aware rather than purely threshold-based. Instead of enforcing constant correction, our approach focuses on preventing prolonged harmful posture, making the system more practical and user-friendly.
\rev{This is consistent with prior findings in interruptibility research \cite{adamczyk2004if,obuchi2016investigating}, implying that interventions should be delivered during periods of low cognitive load, when users are more receptive to behavioral correction.}

\subsection{Practicality and Limitations}
\label{para:limitations}
\rev{While our results demonstrate the feasibility of 
cognition-aware ergonomic assistance, several limitations 
merit discussion. First, long-term wearability, social 
acceptability, and signal robustness of consumer-grade 
wearable EEG in unconstrained environments remain open 
challenges. \rev{The Frenz Brainband is well suited for 
detecting broad \rev{task-induced cognitive load} contrasts 
such as Rest versus Active~\cite{knierim2025advancing}, 
but fine-grained cognitive state distinctions and 
source-level analyses typically require higher-density 
channels~\cite{heib2021cognitive}. Our system is better characterized as suitable for interruptibility-aware 
alerting rather than precise neurological assessment.} 
Second, our study focuses on short-term controlled tasks 
and future work is needed to evaluate long-term behavioral 
change and adherence in real-world deployments. \rev{Third, 
controlled sessions are sufficient to elicit EEG-detectable 
cognitive load~\cite{nirabi2025cognitive,xiong2020pattern} 
and postural degradation~\cite{igarashi2016effect}, but 
gradual postural drift during longer unstructured work 
sessions may follow different dynamics. Future longitudinal 
studies should examine how cognitive load and posture 
co-evolve over extended naturalistic work periods.} 
Finally, our \rev{task-induced cognitive load} model 
simplifies mental workload into coarse categories, 
sufficient for interruptibility-aware alerting, but richer 
cognitive models may further improve personalization at 
the cost of additional complexity and privacy concerns.}
\subsection{Future Directions}
\label{para:futuredirections}
Future development of \system will incorporate 
recommendations from end users and ergonomics specialists, 
including exploring smart glasses or helmet-based 
prototypes for robust skin contact and improved signal 
fidelity. Our current reliance on traditional machine 
learning presents limitations as feature selection is 
labor-intensive and requires domain expertise. We will 
therefore investigate deep learning architectures for 
automated feature extraction from raw sensor streams, 
better suited to scaling with larger datasets and 
capturing subtle temporal patterns. Future iterations may 
also support more nuanced activity recognition using 
large-scale representation learning to distinguish task 
types, adapt alert preferences, and account for 
collaborative work sessions. \rev{Our pilot evaluation in 
Section~\ref{para:pilotstudy} is a preliminary case study 
rather than a definitive user study. The small sample 
($N{=}13$), short session duration, and single baseline 
comparison limit the conclusions that can be drawn about 
generalizability and long-term effectiveness. Additionally, 
longer naturalistic work sessions may exhibit slower and 
more variable posture drift than our shortened session for 
each posture. }\revdb{In addition, the P+C policy exposes a limitation of the current fixed-threshold design: a single set of cooldown and override parameters cannot adapt to individual postural habits or cognitive engagement patterns. Future work should therefore investigate personalized timeout and deferral parameters—such as user-specific override thresholds and alert deferral windows—learned from longitudinal observations of each user’s posture patterns, work environment, and behavioral responses to feedback. Such personalization may enable the system to better balance ergonomic protection and interruptibility across users with different sitting habits, musculature, pain sensitivity, and tolerance for sustained flexion. Longer-term, in-the-wild studies are needed to examine how cognitive load, posture adaptation, and user responses evolve over extended real-world use, and to establish more robust evidence for the effectiveness of cognition-aware alerting policies.}
We also plan to monitor correct neck exercise 
performance~\cite{taylor2013posture} and extend focus 
beyond musculoskeletal health to other tech-related risks 
such as eye strain. Current solutions such as Apple's 
Screen Distance\footnote{\url{https://support.apple.com/en-us/105007}} 
issue disruptive alerts that are often ignored. We aim to 
develop cognition-aware and context-sensitive interventions 
to mitigate eye strain risk and integrate additional 
sensing modalities such as ambient sensors, cameras, and 
audio to capture environmental factors including workspace 
lighting and viewing distance.

%% file: acmart-primary/Conclusion.tex
\section{Conclusion}
\rev{In this work, we present \system, a head-worn wearable system that integrates IMU-based posture sensing with EEG-based \rev{task-induced cognitive load} estimation to enable cognition-aware ergonomic interventions. Our results show that incorporating timing into alert delivery is more effective than threshold-based approaches, reducing alert frequency by 81\% while improving posture correction by 38\%. These findings shows that modeling holistic user state enables fewer, better-timed interventions without sacrificing ergonomic effectiveness. Future work will evaluate long-term adherence and real-world performance through longitudinal deployments.The code and supplementary materials for this work are available at \url{https://github.com/umassos/ErgoAssist}.}

%% file: acmart-primary/Appendix.tex
\appendix
\clearpage
\section{Participant Demographics}
\label{app:participants}

Table~\ref{tab:participants1} summarizes the demographic 
details of all 24 participants including age, gender, 
device usage patterns, and vision range.

\begin{center}
\footnotesize
\captionof{table}{\textbf{Participant demographics.}}
\label{tab:participants1}

\renewcommand{\arraystretch}{1.08}

\begin{tabular}{c c c c c c}
\toprule
\rowcolor[HTML]{D9ECE8}
\textbf{Participant} &
\textbf{Age} &
\textbf{Gender} &
\textbf{Laptop use} &
\textbf{Other electronics use} &
\textbf{Vision range} \\
\midrule
P1  & 18 to 25 & Female & > 8 hrs & 3 to 5 hrs & -4.00, -3.75 \\
P2  & 26 to 35 & Male   & > 8 hrs & 5 to 8 hrs & 6/6 \\
P3  & 26 to 35 & Male   & > 8 hrs & 1 to 3 hrs & 6/6 \\
P4  & 26 to 35 & Female & > 8 hrs & 1 to 3 hrs & 6/6 \\
P5  & 36 to 45 & Male   & 1 to 3 hrs & 1 to 3 hrs & 6/6 \\
P6  & 26 to 35 & Female & > 8 hrs & 3 to 5 hrs & 6/6 \\
P7  & 26 to 35 & Male   & > 8 hrs & 1 to 3 hrs & 6/6 \\
P8  & 26 to 35 & Male   & > 8 hrs & 3 to 5 hrs & -0.25, -0.25 \\
P9  & 26 to 35 & Male   & 5 to 8 hrs & 1 to 3 hrs & 6/6 \\
P10 & 26 to 35 & Male   & > 8 hrs & > 8 hrs & 6/6 \\
P11 & 26 to 35 & Female & 3 to 5 hrs & 5 to 8 hrs & -2.5, -2.5 \\
P12 & 26 to 35 & Female & 5 to 8 hrs & 1 to 3 hrs & 6/6 \\
P13 & 18 to 25 & Female & 5 to 8 hrs & 3 to 5 hrs & 6/6 \\
P14 & 18 to 25 & Female & > 8 hrs & 1 to 3 hrs & 6/6 \\
P15 & 36 to 45 & Male   & 5 to 8 hrs & 1 to 3 hrs & 2/2.5 \\
P16 & 26 to 35 & Male   & 5 to 8 hrs & 1 to 3 hrs & -3/-2 \\
P17 & 18 to 25 & Male   & 5 to 8 hrs & 3 to 5 hrs & 6/6 \\
P18 & 18 to 25 & Male   & 1 to 3 hrs & 5 to 8 hrs & -2/-2 \\
P19 & 26 to 35 & Male   & > 8 hrs & 3 to 5 hrs & 6/6 \\
P20 & 26 to 35 & Male   & > 8 hrs & 1 to 3 hrs & 6/6 \\
P21 & 26 to 35 & Male   & > 8 hrs & 1 to 3 hrs & -3/-3 \\
P22 & 26 to 35 & Male   & 5 to 8 hrs & 5 to 8 hrs & 0.5/0.5 \\
P23 & 26 to 35 & Male   & 3 to 5 hrs & 5 to 8 hrs & 4/4 \\
P24 & 26 to 35 & Male   & > 8 hrs & 1 to 3 hrs & 6/6 \\
\bottomrule
\end{tabular}

\end{center}

\normalsize

\section{Questionnaire Items}
\label{app:questionnaire}
\paragraph{\textbf{Subjective Workload: NASA-TLX}}
Participants rate the following dimensions from 1 (very 
low) to 7 (very high):
\begin{itemize}
    \item \textbf{Mental Demand:} How much mental and 
    perceptual activity was required?
    \item \textbf{Physical Demand:} How much physical 
    activity was required?
    \item \textbf{Temporal Demand:} How much time pressure 
    was felt?
    \item \textbf{Performance:} How successful were you 
    in accomplishing the task goals?
    \item \textbf{Effort:} How hard did you work to achieve 
    your level of performance?
    \item \textbf{Frustration:} How irritated or stressed 
    did you feel during the task?
\end{itemize}

\paragraph{\textbf{Musculoskeletal Discomfort: CMDQ}}
Participants rate discomfort from 1 (no discomfort) to 7 
(extreme discomfort) for the following body regions: Neck, 
Upper back, Arms, Eyes.

\paragraph{\textbf{Modified System Usability Scale (SUS)}}
\begin{enumerate}
    \item I think that I would like to use this app 
    frequently for my work.
    \item I found the app unnecessarily complex.
    \item I thought the alerts were easy to understand 
    and follow.
    \item I think that I would need the support of a 
    technical person to be able to use this app.
    \item I thought there was too much inconsistency in 
    the alert timing.
    \item I would imagine that most people would learn to 
    use the app very quickly.
    \item I found the alerts very awkward to use during 
    my tasks.
    \item I felt very confident using the app to manage 
    my posture.
    \item I needed to learn a lot of things before I could 
    get going with this app.
\end{enumerate}

\clearpage
\section{Detailed Per-Alert Posture Correction Records}
\label{app:per_alert_details}

\begin{center}
\scriptsize
\setlength{\tabcolsep}{3pt}
\renewcommand{\arraystretch}{0.90}

\captionof{table}{\rev{Per-alert posture correction for unseen participants (P9--P13). \textbf{Baseline pitch}: neutral pitch from calibration. \textbf{Pitch before/after alert}: head pitch at alert onset and after correction. \textbf{Recovery time}: seconds from alert to correction; 0.00 = immediate, dashes = no correction within 30\,s. \textbf{Total corrected}: alerts resulting in correction out of total issued. Baseline pitch values differ across P-only and P+C sessions because personalized calibration is recomputed independently before each session.}}
\label{tab:per_alert_details}
\vspace{-6pt}

\resizebox{\textwidth}{!}{%
\begin{tabular}{c l c r r r r c}

\toprule
\rowcolor[HTML]{D9ECE8}
\textbf{\rev{Participant}} &
\textbf{\rev{Condition}} &
\textbf{\rev{Alert \#}} &
\textbf{\rev{Baseline pitch (°)}} &
\textbf{\rev{Pitch before alert (°)}} &
\textbf{\rev{Pitch after alert (°)}} &
\textbf{\rev{Recovery time (s)}} &
\textbf{\rev{Total corrected}} \\
\midrule

\multirow{17}{*}{\rev{P9}}
& \rev{P-only} & \rev{1} & \multirow{16}{*}{\rev{-6.83}} & \rev{-17.65} & \rev{-16.65} & \rev{0.00}  & \multirow{16}{*}{\rev{16/16}} \\
& \rev{P-only} & \rev{2} & & \rev{-19.44} & \rev{-16.76} & \rev{22.33} & \\
& \rev{P-only} & \rev{3} & & \rev{-18.69} & \rev{-16.76} & \rev{12.27} & \\
& \rev{P-only} & \rev{4} & & \rev{-17.93} & \rev{-16.76} & \rev{1.50}  & \\
& \rev{P-only} & \rev{5} & & \rev{-17.46} & \rev{-16.76} & \rev{7.15}  & \\
& \rev{P-only} & \rev{6} & & \rev{-18.55} & \rev{-16.62} & \rev{0.00}  & \\
& \rev{P-only} & \rev{7} & & \rev{-17.57} & \rev{-16.59} & \rev{20.20} & \\
& \rev{P-only} & \rev{8} & & \rev{-18.32} & \rev{-16.59} & \rev{9.77}  & \\
& \rev{P-only} & \rev{9} & & \rev{-17.60} & \rev{-16.59} & \rev{0.00}  & \\
& \rev{P-only} & \rev{10} & & \rev{-18.41} & \rev{-16.62} & \rev{12.55} & \\
& \rev{P-only} & \rev{11} & & \rev{-19.11} & \rev{-16.62} & \rev{2.36}  & \\
& \rev{P-only} & \rev{12} & & \rev{-18.66} & \rev{-16.73} & \rev{10.96} & \\
& \rev{P-only} & \rev{13} & & \rev{-18.88} & \rev{-16.73} & \rev{0.44}  & \\
& \rev{P-only} & \rev{14} & & \rev{-18.16} & \rev{-16.79} & \rev{10.70} & \\
& \rev{P-only} & \rev{15} & & \rev{-18.35} & \rev{-16.79} & \rev{0.69}  & \\
& \rev{P-only} & \rev{16} & & \rev{-17.49} & \rev{-16.76} & \rev{0.00}  & \\
& \textbf{\rev{P+C}} & \textbf{\rev{1}} & \textbf{\rev{4.39}} & \textbf{\rev{-6.34}} & \textbf{\rev{-1.71}} & \textbf{\rev{3.02}} & \textbf{\rev{1/1}} \\
\midrule

\multirow{21}{*}{\rev{P10}}
& \rev{P-only} & \rev{1}  & \multirow{14}{*}{\rev{-0.98}} & \rev{-27.00} & \rev{-10.94} & \rev{2.13}  & \multirow{14}{*}{\rev{1/14}} \\
& \rev{P-only} & \rev{2}  & & \rev{-41.41} & \rev{--}      & \rev{--}    & \\
& \rev{P-only} & \rev{3}  & & \rev{-26.77} & \rev{--}      & \rev{--}    & \\
& \rev{P-only} & \rev{4}  & & \rev{-24.12} & \rev{--}      & \rev{--}    & \\
& \rev{P-only} & \rev{5}  & & \rev{-21.04} & \rev{--}      & \rev{--}    & \\
& \rev{P-only} & \rev{6}  & & \rev{-19.16} & \rev{--}      & \rev{--}    & \\
& \rev{P-only} & \rev{7}  & & \rev{-25.79} & \rev{--}      & \rev{--}    & \\
& \rev{P-only} & \rev{8}  & & \rev{-26.41} & \rev{--}      & \rev{--}    & \\
& \rev{P-only} & \rev{9}  & & \rev{-25.01} & \rev{--}      & \rev{--}    & \\
& \rev{P-only} & \rev{10} & & \rev{-23.72} & \rev{--}      & \rev{--}    & \\
& \rev{P-only} & \rev{11} & & \rev{-23.19} & \rev{--}      & \rev{--}    & \\
& \rev{P-only} & \rev{12} & & \rev{-23.25} & \rev{--}      & \rev{--}    & \\
& \rev{P-only} & \rev{13} & & \rev{-21.88} & \rev{--}      & \rev{--}    & \\
& \rev{P-only} & \rev{14} & & \rev{-22.13} & \rev{--}      & \rev{--}    & \\
& \textbf{\rev{P+C}} & \textbf{\rev{1}} & \multirow{7}{*}{\textbf{\rev{15.83}}} & \textbf{\rev{4.03}}  & \textbf{\rev{5.99}}  & \textbf{\rev{5.23}}  & \multirow{7}{*}{\textbf{\rev{7/7}}} \\
& \textbf{\rev{P+C}} & \textbf{\rev{2}} & & \textbf{\rev{3.72}}  & \textbf{\rev{9.12}}  & \textbf{\rev{12.19}} & \\
& \textbf{\rev{P+C}} & \textbf{\rev{3}} & & \textbf{\rev{4.64}}  & \textbf{\rev{9.12}}  & \textbf{\rev{2.19}}  & \\
& \textbf{\rev{P+C}} & \textbf{\rev{4}} & & \textbf{\rev{5.51}}  & \textbf{\rev{6.01}}  & \textbf{\rev{1.48}}  & \\
& \textbf{\rev{P+C}} & \textbf{\rev{5}} & & \textbf{\rev{4.28}}  & \textbf{\rev{6.01}}  & \textbf{\rev{2.89}}  & \\
& \textbf{\rev{P+C}} & \textbf{\rev{6}} & & \textbf{\rev{3.92}}  & \textbf{\rev{15.86}} & \textbf{\rev{14.41}} & \\
& \textbf{\rev{P+C}} & \textbf{\rev{7}} & & \textbf{\rev{5.64}}  & \textbf{\rev{11.86}} & \textbf{\rev{4.40}}  & \\
\midrule

\multirow{3}{*}{\rev{P11}}
& \rev{P-only} & \rev{1} & \rev{-19.28} & \rev{-32.20} & \rev{-29.26} & \rev{0.00} & \rev{1/1} \\
& \textbf{\rev{P+C}} & \textbf{\rev{1}} & \multirow{2}{*}{\textbf{\rev{-1.26}}} & \textbf{\rev{-15.67}} & \textbf{\rev{-10.66}} & \textbf{\rev{2.99}} & \multirow{2}{*}{\textbf{\rev{2/2}}} \\
& \textbf{\rev{P+C}} & \textbf{\rev{2}} & & \textbf{\rev{-12.03}} & \textbf{\rev{-5.74}}  & \textbf{\rev{2.18}} & \\
\midrule

\multirow{15}{*}{\rev{P12}}
& \rev{P-only} & \rev{1}  & \multirow{14}{*}{\rev{9.43}} & \rev{-0.67}  & \rev{-0.50}  & \rev{0.00}  & \multirow{14}{*}{\rev{5/14}} \\
& \rev{P-only} & \rev{2}  & & \rev{-13.04} & \rev{--}      & \rev{--}    & \\
& \rev{P-only} & \rev{3}  & & \rev{-10.52} & \rev{--}      & \rev{--}    & \\
& \rev{P-only} & \rev{4}  & & \rev{-10.83} & \rev{--}      & \rev{--}    & \\
& \rev{P-only} & \rev{5}  & & \rev{-5.79}  & \rev{--}      & \rev{--}    & \\
& \rev{P-only} & \rev{6}  & & \rev{-5.32}  & \rev{--}      & \rev{--}    & \\
& \rev{P-only} & \rev{7}  & & \rev{-4.90}  & \rev{-0.56}   & \rev{29.52} & \\
& \rev{P-only} & \rev{8}  & & \rev{-5.06}  & \rev{-0.56}   & \rev{19.80} & \\
& \rev{P-only} & \rev{9}  & & \rev{-4.11}  & \rev{-0.56}   & \rev{9.05}  & \\
& \rev{P-only} & \rev{10} & & \rev{-2.83}  & \rev{-0.56}   & \rev{0.00}  & \\
& \rev{P-only} & \rev{11} & & \rev{-8.90}  & \rev{--}      & \rev{--}    & \\
& \rev{P-only} & \rev{12} & & \rev{-7.27}  & \rev{--}      & \rev{--}    & \\
& \rev{P-only} & \rev{13} & & \rev{-7.27}  & \rev{--}      & \rev{--}    & \\
& \rev{P-only} & \rev{14} & & \rev{-5.51}  & \rev{--}      & \rev{--}    & \\
& \textbf{\rev{P+C}} & \textbf{\rev{1}} & \textbf{\rev{-1.96}} & \textbf{\rev{-19.42}} & \textbf{\rev{-7.64}} & \textbf{\rev{4.36}} & \textbf{\rev{1/1}} \\
\midrule

\multirow{7}{*}{\rev{P13}}
& \rev{P-only} & \rev{1} & \multirow{6}{*}{\rev{9.88}} & \rev{-1.48}  & \rev{-0.11}  & \rev{0.00}  & \multirow{6}{*}{\rev{4/6}} \\
& \rev{P-only} & \rev{2} & & \rev{-0.87}  & \rev{--}      & \rev{--}    & \\
& \rev{P-only} & \rev{3} & & \rev{-9.37}  & \rev{--}      & \rev{--}    & \\
& \rev{P-only} & \rev{4} & & \rev{-6.46}  & \rev{0.06}    & \rev{22.84} & \\
& \rev{P-only} & \rev{5} & & \rev{-6.43}  & \rev{0.06}    & \rev{13.12} & \\
& \rev{P-only} & \rev{6} & & \rev{-9.85}  & \rev{0.06}    & \rev{3.09}  & \\
& \textbf{\rev{P+C}} & \textbf{\rev{0}} & \textbf{\rev{6.21}} & \textbf{\rev{--}} & \textbf{\rev{--}} & \textbf{\rev{--}} & \textbf{\rev{0/0}} \\

\bottomrule
\end{tabular}
}

\end{center}

\normalsize